\documentclass[11pt]{article}

\usepackage[letterpaper, margin=0.8in]{geometry}

\usepackage{amsmath, amsfonts, amssymb}
\usepackage{slashed}
\usepackage{cite}
\usepackage[vcentermath]{youngtab}
\usepackage{empheq}
\usepackage{booktabs}
\usepackage{mathrsfs}
\usepackage{xcolor}
\usepackage[parsep]{collref}
\usepackage{cancel}
\usepackage[format=hang,labelfont={bf}]{caption}
\usepackage{notes2bib}
\usepackage{hyperref}
\hypersetup{
	unicode,	% use with \texorpdfstring
	colorlinks,
	citecolor=blue,
	linkcolor=[rgb]{0,.3,.5},
	urlcolor=[rgb]{0,.3,.5}, %cyan,% pdftex
	bookmarksopen=true,
	bookmarksopenlevel=\maxdimen,
        linktocpage,
}

\newcommand {\cA}{{\cal A}}
\newcommand {\cB}{{\cal B}}
\newcommand {\cC}{{\cal C}}
\newcommand {\cD}{{\cal D}}
\newcommand {\cE}{{\cal E}}
\newcommand {\cF}{{\cal F}}

\newcommand {\cH}{{\cal H}}

\newcommand {\cJ}{{\cal J}}
\newcommand {\cK}{{\cal K}}

\newcommand {\cM}{{\cal M}}
\newcommand {\cN}{{\cal N}}

\newcommand {\cP}{{\cal P}}

\newcommand {\cR}{{\cal R}}

\newcommand {\cT}{{\cal T}}

\newcommand {\cV}{{\cal V}}
\newcommand {\cW}{{\cal W}}

\newcommand{\ra}{{\mathrm a}}
\newcommand{\rb}{{\mathrm b}}
\newcommand{\rc}{{\mathrm c}}
\newcommand{\rd}{{\mathrm d}}

\newcommand{\rn}{{\mathrm n}}
\newcommand{\rp}{{\mathrm p}}

\newcommand{\bd}{{\dot{\beta}}}

\newcommand{\hm}{{\hat{m}}}
\newcommand{\hn}{{\hat{n}}}

\newcommand{\ha}{{\hat{a}}}
\newcommand{\hb}{{\hat{b}}}
\newcommand{\hc}{{\hat{c}}}
\newcommand{\hd}{{\hat{d}}}
\newcommand{\he}{{\hat{e}}}

\newcommand{\hC}{{\hat{C}}}

\newcommand{\bm}[1]{\mbox{\boldmath$#1$}}
\newcommand{\veps}{\varepsilon}
\newcommand{\eps}{{\epsilon}}
\newcommand{\eol}{\notag \\}

\newcommand{\ul}{\underline}
\newcommand{\pa}{\partial}

\DeclareMathOperator{\sdet}{sdet}

\newcommand{\rmm}{{\rm m}}

\newcommand{\Lie}{\mathbb L}

\newcommand{\ba}{{\overline a}}
\newcommand{\bb}{{\overline b}}
\newcommand{\bc}{{\overline c}}
\renewcommand{\bd}{{\overline d}}

\newcommand{\ua}{{\ul a}}
\newcommand{\ub}{{\ul b}}
\newcommand{\uc}{{\ul c}}
\newcommand{\ud}{{\ul d}}
\newcommand{\ue}{{\ul e}}

\newcommand{\um}{{\ul m}}
\newcommand{\un}{{\ul n}}

\newcommand{\hmu}{{\hat\mu}}

\newcommand{\tA}{{\tiny \rm A}}
\newcommand{\tB}{{\tiny \rm B}}
\newcommand{\tC}{{\tiny \rm C}}
\newcommand{\tM}{{\tiny \rm M}}
\newcommand{\tN}{{\tiny \rm N}}
\newcommand{\tP}{{\tiny \rm P}}

\newcommand{\drep}[1]{\mathtt{#1}}

\newcommand{\proj}{{\Big\vert}_{\tiny \texttt{proj}}}
\newcommand{\tinyproj}{{\vert}_{\tiny \texttt{proj}}}

\newcommand{\tQ}{\tilde Q}
\newcommand{\wtM}{\widetilde M}
\newcommand{\wtK}{\widetilde K}
\newcommand{\wtL}{\widetilde L}

\newcommand{\tX}{\widetilde X}
\newcommand{\tnabla}{\widetilde\nabla}
\newcommand{\zcV}{\mathring \cV}
\newcommand{\zD}{\mathring D}
\newcommand{\zcD}{\mathring \cD}

\newcommand{\zJ}{\mathring \cJ}

\newcommand{\g}[1]{\mathrm{#1}}

\newcommand{\opbraket}[2]{\langle #1 \vert #2 \rangle}
\newcommand{\varV}{v}
\newcommand{\ext}{\widehat}

\makeatletter
\g@addto@macro\bfseries{\boldmath}
\makeatother

\numberwithin{equation}{section}

\begin{document}

\thispagestyle{empty}

\begin{flushright}\small
MI-TH-2136

\end{flushright}

%%%%%%%%%%%%%%%%%%%%%%%%%%%%%%%%%%%%%%%%%

\bigskip
\bigskip

\vskip 10mm

\begin{center}

{\Large{\bf Exploring the geometry of supersymmetric double field theory}}

\end{center}

%%%%%%%%%%%%%%%%%%%%%%%%%%%%%%%%%%%%%%%%

\vskip 6mm

\begin{center}

{\bf Daniel Butter}

\vspace{1.5ex}

{\em George P. and Cynthia W. Mitchell Institute \\for Fundamental
Physics and Astronomy \\
Texas A\&M University, College Station, TX 77843-4242, USA}\\

\vspace{2ex}
\centerline{\small \texttt{dbutter@tamu.edu}}

\end{center}

\vskip0.5cm

\begin{center} {\bf Abstract } \end{center}

\begin{quotation}\noindent
The geometry of $\cN=1$ supersymmetric double field theory is revisited in superspace.
In order to maintain the constraints on the torsion tensor, the 
local tangent space group of $\g{O}(D) \times \g{O}(D)$
must be expanded to include a tower of higher dimension generators.
These include a generator in the irreducible hook representation of the Lorentz group, 
which gauges the shift symmetry (or ambiguity) of the spin connection. 
This gauging is possible even in the purely bosonic theory, where it leads to a
Lorentz curvature whose only non-vanishing pieces are the physical ones: the
generalized Einstein tensor and the generalized scalar curvature.
A relation to the super-Maxwell$_\infty$ algebra is proposed.
The superspace Bianchi identities are solved up through dimension two, and the 
component supersymmetry transformations and equations of motion are explicitly 
(re)derived.
\end{quotation}

% \vfill \hfill Version: \today

\newpage

\tableofcontents

\newpage

%%%%%%%%%%%%%%%%%%%%%%%%%%%%%%%%%%%%%%%%%%%%%%%%%%%%%%%%%%%%%%%%%%%%%%%%%%%%%%%%%
\section{Introduction}
%%%%%%%%%%%%%%%%%%%%%%%%%%%%%%%%%%%%%%%%%%%%%%%%%%%%%%%%%%%%%%%%%%%%%%%%%%%%%%%%%

Double field theory (DFT) is a formulation of the massless sector of string theory
that makes $T$-duality manifest \cite{Siegel:1993xq,Siegel:1993th, Hull:2009mi,Hull:2009zb,Hohm:2010jy,Hohm:2010pp}. It describes the low energy sector of 
bosonic string theory (or the bosonic NS-NS sector of supersymmetric string theory)
and unifies the metric and two-form together into a generalized metric, which transforms
under generalized diffeomorphisms governed by the group $\g{O}(D,D)$.
In order to incorporate fermions and supersymmetry, one requires a frame formulation
\cite{Siegel:1993xq,Siegel:1993th,Hohm:2010xe, Jeon:2010rw, Jeon:2011cn}, and in this approach the generalized
vielbein is an element of $\g{O}(D,D)$ with a doubled tangent space group
$\g{O}(D-1,1) \times \g{O}(D-1,1)$. 
For further references and discussions of important applications of DFT,
we refer the reader to the reviews \cite{Aldazabal:2013sca,Berman:2013eva}.

The topic of this paper is the structure of supersymmetric double field theory,
and specifically how to deduce it from a doubled supergeometry.
At the component level, the structure of $\cN=1$ double field theory
was identified to second order in fermions in
\cite{Hohm:2011nu} and to all orders in \cite{Jeon:2011sq} (see also \cite{Coimbra:2011nw}).
Remarkably, the prescient early papers by Siegel \cite{Siegel:1993xq,Siegel:1993th} 
\emph{already} contained a superspace description, from which one should be
able to deduce supersymmetry transformations and equations of motion.
In principle this ought to be a straightforward task, and the original motivation
of this paper was simply to analyze this example as a warm-up to the more interesting
type II and exceptional cases.\footnote{$\cN=2$ supersymmetric double field theory 
(that is, for type II strings) has been constructed at the component level \cite{Jeon:2012hp},
extending the bosonic construction of \cite{Hohm:2011zr, Hohm:2011dv}. $\cN=2$ superspace
formulations have been given in \cite{Hatsuda:2014qqa,Hatsuda:2014aza,Cederwall:2016ukd}
(see also \cite{Bandos:2015cha, Park:2016sbw}).
Supersymmetric exceptional field theories have been built for $E_6$, $E_7$, and $E_8$
\cite{Godazgar:2014nqa,Musaev:2014lna,Baguet:2016jph} (with partial results for
$E_{11}$ \cite{Bossard:2019ksx}), but have been discussed in superspace only for 
the $E_7$ case \cite{Butter:2018bkl}. Note that an off-shell formulation of
11D superspace exists involving pure spinors 
\cite{Cederwall:2009ez,Cederwall:2010tn, Berkovits:2018gbq}, and it would be
interesting to explore whether a duality-covariant generalization is possible.}

Surprisingly, this is not as straightforward as it might seem. Aside from the technical aspects
of reducing a superspace formulation to a component one -- similar to more conventional
superspaces, albeit a bit more technically involved when working with
generalized super-diffeomorphisms -- one encounters a roadblock absent in conventional
superspaces or even in bosonic DFT. In order for the supervielbein to be an unconstrained
element of a supergroup -- the relevant one here being the orthosymplectic group
$\g{OSp}(D,D|2s)$ with $D=10$ and $s=16$ --
the local tangent space symmetry must be extended beyond
the doubled Lorentz group $\g{O}(D-1,1) \times \g{O}(D-1,1)$ familiar in the bosonic
theory. This is because the supervielbein carries a large number of degrees of freedom,
and only some of them can be eliminated using the usual Wess-Zumino gauge-fixing
conditions. To guarantee the remaining components drop out, an enhanced
local gauge symmetry is necessary.
This was implicit already in \cite{Siegel:1993th}, where it was used to eliminate
these additional fields almost immediately.
Our new observation is that consistency with the supersymmetry algebra implies that
this is just the first extension of a (possibly infinite) tower of local gauge
symmetries. Only the lowest few levels of these symmetries act on the supervielbein
via a tangent space rotation.

This extension turns out to not purely be an artifact of superspace.
It casts a shadow on the purely bosonic theory as well.
It is a fact of life that in the frame formulation of DFT, the
spin connection is not fully determined by the vanishing torsion conditions.
This can be restated in the following way: the purely left-handed and right-handed 
components of the spin connection are ambiguous under shifts in the irreducible 
hook representations, i.e.
\begin{align}\label{E:LambdaShift}
\delta \omega_{a b c} = \Lambda_{a|bc}~, \qquad
\delta \omega_{\overline{abc}} = \Lambda_{\ba|\overline{bc}}~,
\end{align}
where $\Lambda_{[a|bc]} = \Lambda^{b}{}_{|bc} = 0$ and similarly 
for $\Lambda_{\ba|\overline{bc}}$. Usually this is handled by requiring that
these representations of $\omega$
drop out of all physical quantities (actions, equations of motion, etc.).
However, in principle, there is no reason why one cannot demand that \eqref{E:LambdaShift}
be an honest local gauge invariance, which we impose to eliminate unphysical degrees of
freedom. After all, the doubled Lorentz group is introduced to eliminate the
unphysical degrees of freedom in the doubled vielbein. What we discover in
superspace is this new local symmetry is actually \emph{required} by closure of the
extended tangent space algebra. In other words, it transforms a ``bug'' of DFT 
into a ``feature."

This new local symmetry naturally implies others. This is because its gauging
necessitates the introduction of new gauge fields $h_{\ha, b|cd}$ and
$h_{\ha, \bb | \overline{cd}}$. In order to not introduce new degrees of freedom,
these gauge fields should be determined in terms of the generalized vielbein
in some way. What happens is that just as $\omega$ appears algebraically in the
torsion tensor, $h$ appears algebraically in the Lorentz curvature.
Imposing that the torsion tensor vanishes determines most of $\omega$, and similarly
imposing that some components of the Lorentz curvature tensor vanish determines
most of $h$. But because $h$ itself retains a residual shift symmetry analogous
to \eqref{E:LambdaShift}, the process may continue, with further connections 
and curvatures introduced at ever higher dimensions.
An important consequence of all of these additional symmetries is that the only
non-vanishing components of the generalized curvature tensors (torsion, Lorentz
curvature, etc.) that survive are the physical ones.

In attempting to make sense of this hierarchy, we have drawn inspiration from the work
of Pol\'a\v{c}ek and Siegel \cite{Polacek:2013nla}. They showed how one could incorporate
the spin connection into a larger \emph{megavielbein} by extending the doubled spacetime so that
the extended derivative $D_\cM = (M_{\ha \hb}, \pa_\hm, \wtM^{\ha\hb})$
includes the $\g{SO}(D-1,1) \times \g{SO}(D-1,1)$ Lorentz generator $M_{\ha \hb}$, 
along with a composite dual generator $\wtM^{\ha\hb}$, required so that
the extended metric $\eta_{\cM \cN}$ is invertible. In gauging the shift symmetry
\eqref{E:LambdaShift}, its descendants, and their superpartners, we find that 
the dual generators extending $\wtM^{\ha \hb}$ can be embedded in a version
of the super-Maxwell$_\infty$ algebra (see recent discussions in \cite{Gomis:2017cmt,Gomis:2018xmo}).
It is unclear to us if this fully characterizes the super-algebra or if a 
quotient must be taken.
We discuss some of these issues in the conclusion.

The paper is arranged as follows. We begin in section \ref{S:BosonicDFT} with
a discussion of the geometry of bosonic double field theory. This is largely
for review and to fix notation, but we elaborate on the shift symmetry
of the spin connection and how the introduction of the higher connection $h$
permits one to build a Lorentz curvature tensor with only physical components.
In section \ref{S:SDFT} we extend the formalism to superspace. This is primarily
a review of \cite{Siegel:1993th} but in modified language and formalism. Particular
attention is paid to how the tangent space group is used to eliminate unphysical
components of the vielbein. This is followed by section \ref{S:SGeo}, where we
explicitly analyze the generalized torsion and curvature tensors up through
dimension 2. It is here that we uncover the need for an extended local gauge
symmetry and its connection with the super-Maxwell$_\infty$ group. The component
supersymmetry transformations and equations of motion are derived in section
\ref{S:Components} and perfect agreement is found with \cite{Hohm:2011nu, Jeon:2011sq}.
As is typical for superspaces with maximal supersymmetry, the lower dimension ($\leq 1/2$) torsion constraints effectively place the theory on-shell.\footnote{The notable exception is
off-shell pure spinor superspace \cite{Cederwall:2009ez,Cederwall:2010tn, Berkovits:2018gbq},
but the connection to conventional supergeometry is obscured there.} 
Finally in section \ref{S:Discussion}, we discuss several open questions and possible
extensions.

Several technical appendices are included. 
Appendix \ref{A:Notation} summarizes the conventions we use.
Appendix \ref{A:PS} reviews the
Pol\'a\v{c}ek-Siegel (PS) formalism \cite{Polacek:2013nla}
for embedding connections into a megavielbein.
While we do not use such a megavielbein explicitly in the main body of the paper, the PS
formalism does provide a useful explanation for the transformation rules of the connections
and the construction of the curvature tensors when a generic group is gauged in DFT.
Appendix \ref{A:SMaxwell} gives the construction of the super-Maxwell$_\infty$ algebra
and its extension, which appears to play a role in building the local gauge group of $\cN=1$ DFT.

%%%%%%%%%%%%%%%%%%%%%%%%%%%%%%%%%%%%%%%%%%%%%%%%%%%%%%%%%%%%%%%%%%%%%%%%%%%%%%%%%
\section{Bosonic geometry of double field theory}\label{S:BosonicDFT}
%%%%%%%%%%%%%%%%%%%%%%%%%%%%%%%%%%%%%%%%%%%%%%%%%%%%%%%%%%%%%%%%%%%%%%%%%%%%%%%%%

Let us begin with a discussion of bosonic double field theory. This is well-studied
material over the last decade, so we will be relatively brief, highlighting only the
geometry of covariant derivatives, connections, etc. that we will be extending
to the supersymmetric case. The major difference with conventional formulations
is that we include two new connection-like fields. The first is the Pol\'a\v{c}ek-Siegel 
field that permits the construction of a generalized doubled 
Lorentz curvature \cite{Polacek:2013nla}.
The second is the new connection $h$ that transforms in the irreducible hook representation
of the doubled Lorentz group. This new connection gauges shift symmetries in the
spin connection. Together these fields are responsible for eliminating all but the
physical components of the doubled Lorentz curvature.

\subsection{Review of bosonic DFT geometry}
The material here mostly follows that of \cite{Siegel:1993th, Hohm:2010xe}
with only superficial alterations.

We begin with a bosonic space with a doubled set of coordinates $x^\hm$, $\hm = 1,\cdots, 2D$.
Typically these coordinates are denoted $x^M$ but we want to reserve the index $M$ 
for the supersymmetric case later on.
The space is furnished
with a constant invariant $\g{O}(D,D)$ metric $\eta^{\hm\hn}$ (and inverse $\eta_{\hm \hn}$),
with which we can raise (or lower) indices.
A generalized diffeomorphism of a vector $W^\hm$ with weight $w$ is given by
\begin{align}
\delta_\xi W^\hm = \Lie^{(w)}_\xi W^\hm := 
    \xi^\hn \pa_\hn W^\hm - W^\hn (\pa_\hn \xi^\hm - \pa^\hm \xi_\hn)
    + w \,\pa_\hn \xi^\hn\, W^\hm~.
\end{align}
These are $\g{O}(D,D) \times \mathbb R^+$ 
generalized Lie derivatives. Requiring closure of the algebra implies the
section condition,
\begin{align}
\eta^{\hm \hn} \pa_\hm \otimes \pa_\hn = 0
\end{align}
with the derivatives understood to act on either the same or different objects.
By construction, the weight-zero tensor $\eta$ is invariant under generalized diffeomorphisms.

There are two different coordinate systems one commonly encounters in DFT. The first is the
\emph{standard coordinate system}, with
\begin{align}
x^\hm = (x^\rmm, \tilde x_\rmm)~, \qquad \pa_\hm = (\pa_\rmm, \tilde \pa^\rmm)~, \qquad
\eta^{\hm \hn} =
\begin{pmatrix}
0 & \delta^\rmm{}_\rn \\
\delta_\rmm{}^\rn & 0
\end{pmatrix}
\end{align}
where $\rmm=1,\cdots, D$.
In toroidal compactifications of string theory, $x^\rmm$ corresponds to the center-of-mass 
coordinate of the string and $\tilde x_{\rmm}$ corresponds to the coordinate dual to the
winding momenta. The second is the \emph{left/right-moving coordinate system},
where the center-of-mass and winding mode coordinates are combined as
$x^m = \frac{1}{\sqrt 2} (x^\rmm + \eta^{\rmm \rn} \tilde x_{\rn})$ and
$x^{\bar m} = \frac{1}{\sqrt 2} (x^\rmm - \eta^{\rmm \rn} \tilde x_{\rn})$,
where $\eta^{\rmm \rn}$ is the $\g{SO}(D-1,1)$ metric. In this coordinate system,
\begin{align}
\eta^{\hm \hn} =
\begin{pmatrix}
\eta^{mn} & 0 \\
0 & -\eta^{\bar m \bar n}
\end{pmatrix}~,
\end{align}
where $\eta^{\bar m \bar n} = \eta^{mn}$.\footnote{This is potentially confusing
notation since $(\eta)^{\bar m \bar n} = -\eta^{\bar m \bar n}$. We \emph{never} employ
the former explicitly.}
Typically, one works in the standard coordinate system for the world indices but in the
left/right-moving basis for tangent space indices.

The metric of DFT is a symmetric tensor $\cH_{\hm \hn}$ subject to an invariance condition
\begin{align}
\cH^{\hm \hn} := \eta^{\hm \hm'} \eta^{\hn \hn'} \cH_{\hm' \hn'} = (\cH^{-1})^{\hm \hn}~.
\end{align}
The DFT metric may be
decomposed in the standard coordinate basis as
\begin{align}
\cH_{\hm \hn} =
\begin{pmatrix}
g_{\rm{mn}} - b_{\rm m k} g^{\rm k l} b_{\rm l n} & b_{\rm m k} g^{\rm k n} \\
- g^{\rm m k} b_{\rm k n} & g^{\rm m n}
\end{pmatrix}~, \qquad
\cH^{\hm \hn} =
\begin{pmatrix}
g^{\rm m n} & - g^{\rm m k} b_{\rm k n}\\
b_{\rm m k} g^{\rm k n} & g_{\rm mn} - b_{\rm m k} g^{\rm k l} b_{\rm l n}
\end{pmatrix}~.
\end{align}
In supersymmetric DFT, as in conventional supergravity, it is necessary to introduce a vielbein
along with its associated tangent space. Here, the vielbein is a field $V_\hm{}^\ha$
subject to two conditions:
\begin{align}
\cH_{\hm \hn} = V_\hm{}^\ha V_\hn{}^\hb \,\cH_{\ha \hb}~, \qquad
\eta_{\hm \hn} = V_\hm{}^\ha V_\hn{}^\hb \,\eta_{\ha \hb}~,
\end{align}
where $\cH_{\ha \hb}$ and $\eta_{\ha \hb}$ are fixed constant tensors.
The second condition implies that the vielbein is an $\g{O}(D,D)$ element, 
with its inverse given by
$V_\ha{}^\hm = \eta_{\ha \hb} \eta^{\hm \hn} V_\hn{}^\hb$.
In the standard parametrization of the tangent space group, we choose
\begin{align}
\cH_{\ha \hb} =
\begin{pmatrix}
\eta_{\rm a b} & 0 \\
0 & \eta^{\rm a b}
\end{pmatrix}~, \qquad
\eta_{\ha \hb} =
\begin{pmatrix}
0 & \delta_\ra{}^\rb \\
\delta^\ra{}_\rb & 0
\end{pmatrix}~, 
\end{align}
where $\eta_{\ra\rb}$ corresponds to the standard $\g{O}(D-1,1)$ metric.
These tensors are invariant under local infinitesimal tangent space transformations with
\begin{align}
\lambda_\ha{}^\hb &=
\begin{pmatrix}
\lambda_\ra{}^\rb & \tilde \lambda_{\ra\rb} \\
\tilde \lambda^{\ra\rb} & \lambda^\ra{}_\rb
\end{pmatrix}
\end{align}
where $\lambda$ and $\tilde\lambda$ are antisymmetric and their indices are raised
and lowered with $\eta_{\ra\rb}$.
Up to a local tangent space transformation, the vielbein can be chosen in an 
upper triangular gauge as
\begin{align}
V_\hm{}^\ha &=
\begin{pmatrix}
e_\rmm{}^\ra & b_{\rmm \rn} \,e_{\ra}{}^\rn \\
0 & e_\ra{}^\rmm
\end{pmatrix}~, \qquad
V_\ha{}^\hm =
\begin{pmatrix}
e_\ra{}^\rmm & -e_{\ra}{}^\rn\, b_{\rn \rmm}  \\
0 & e_\rmm{}^\ra
\end{pmatrix}~, \qquad
\end{align}

This is not the most convenient parametrization for either the string worldsheet or supersymmetry.
Instead, one may introduce left-handed indices $a,b,\cdots$
and right-handed indices $\ba,\bb,\cdots$, on which the tangent space group acts separately,
as $\g{O}(D-1,1)_L \times \g{O}(D-1,1)_R$. In this approach,
\begin{align}\label{E:BDFT.eta}
\eta_{\ha \hb} = 
\begin{pmatrix}
\eta_{a b} & 0 \\
0 & - \eta_{\overline{ab}}
\end{pmatrix}~, \qquad
\cH_{\ha \hb} = 
\begin{pmatrix}
\eta_{a b} & 0 \\
0 & \eta_{\overline{ab}}
\end{pmatrix}~,
\end{align}
where $\eta_{\overline{ab}} = \eta_{ab}$.
We choose to raise and lower barred indices with $\eta_{\overline{ab}}$
rather than $(\eta)_{\overline{ab}}$, so this means one must be careful when
decomposing a vector to specify whether the index is raised or lowered.
In the left/right tangent basis, the vielbein is given by
\begin{align}\label{E:CompDFT.eb}
V_\hm{}^\ha &= \frac{1}{\sqrt 2}
\begin{pmatrix}
e_\rmm{}^\ra + b_{\rmm \rn} e^{\ra \rn} & e_{\rmm}{}^\ra - b_{\rmm \rn} e^{\ra \rn} \\
e_\ra{}^\rmm & -e_\ra{}^\rmm
\end{pmatrix}~, \qquad
V_\ha{}^\hm = \frac{1}{\sqrt 2}
\begin{pmatrix}
e_\ra{}^\rmm & e_{\rmm \ra} + b_{\rmm \rn} e_\ra{}^\rn \\
e_{\ra}{}^\rmm & -e_{\rmm \ra} + b_{\rmm \rn} e_{\ra}{}^\rn
\end{pmatrix}~.
\end{align}

In either basis, the DFT vielbein transforms infinitesimally under 
generalized diffeomorphisms and $\g{O}(D-1,1)_L \times \g{O}(D-1,1)_R$ transformations as
\begin{align}
\delta V_\hm{}^\ha = \xi^\hn \pa_\hn V_\hm{}^\ha
    + V_\hn{}^\ha (\pa_\hm \xi^\hn - \pa^\hn \xi_\hm)
    - V_\hm{}^\hb \lambda_\hb{}^\ha~.
\end{align}

In addition to the DFT vielbein, we also need the dilaton.
We choose to treat the dilaton as a weight-1 scalar density, $\phi = e^{-2d}$, 
where $d$ is the usual DFT dilaton. $\phi$ transforms as
\begin{align}
\delta \phi = \xi^\hm \pa_\hm \phi + \pa_\hm \xi^\hm\, \phi
    = \pa_\hm(\xi^\hm \phi)~.
\end{align}
We trust this notation will not be too confusing. Later on, we will need to refer
to the standard (non-density) dilaton of supergravity, which we denote $\varphi$.
These are related by
\begin{align}
\phi = e^{-2d} = \det e_\rmm{}^\ra\, e^{-2 \varphi}~.
\end{align}

Together the dilaton and the vielbein can be
combined into a weight-1 vielbein,
which is an element of $\g{O}(D,D) \times \mathbb R^+$. However, it is more convenient
(not to mention conventional) to keep them as distinct fields.
The vielbein and the dilaton together can be used to construct two generalized fluxes
$F_{\ha \hb \hc}$ and $F_\ha$. These are given by (note the unconventional sign choices)
\begin{alignat}{3}
\Lie_{V_\ha} V_\hb{}^\hm &= -F_{\ha \hb}{}^\hc V_\hc{}^\hm
& \quad &\implies &\quad
F_{\ha \hb \hc} &= - 3 \,D_{[\ha} V_\hb{}^\hm \, V_{\hm \hc]}~, \eol
\Lie_{V_\ha}^{(+1)} \phi &= F_{\ha} \phi
& \quad &\implies &\quad
F_\ha &= D_\ha \log \phi + \pa_\hm V_\ha{}^\hm~.
\end{alignat}
The flattened derivatives $D_\ha := V_\ha{}^\hm \pa_\hm$ obey
\begin{align}
[D_\ha, D_\hb] = -F_{\ha \hb}{}^\hc D_\hc~, \qquad
D^\ha D_\ha = - F^\ha D_\ha~.
\end{align}
The unconventional choice of sign for the flux $F_{\ha \hb \hc}$ is to match our conventions
for the torsion tensor later on in superspace.

These flux tensors are covariant under generalized diffeomorphisms, but transform
inhomogeneously under Lorentz transformations. This latter issue can be rectified by
introducing by hand a spin connection $\omega_\hm{}_\ha{}^\hb$ transforming as
\begin{align}
\delta \omega_\hm{}_\ha{}^\hb = \Lie_\xi \omega_\hm{}_\ha{}^\hb
    + \pa_\hm \lambda_\ha{}^\hb
    + \lambda_\ha{}^\hc \omega_\hm{}_\hc{}^\hd
    - \omega_\hm{}_\ha{}^\hc \lambda_\hc{}^\hd~.
\end{align}
Because of the sign in $\eta_{\ha \hb}$ in \eqref{E:BDFT.eta}, one must be careful about
how $\omega$ is defined. We take
\begin{align}
\omega_{\hm \,\ha\hb} = (\omega_{\hm \,a b}, \,\,\omega_{\hm \,\overline{ab}})~, \qquad
\omega_{\hm \,\ha}{}^{\hb} = (\omega_{\hm \,a}{}^{b}, \,\,-\omega_{\hm \,\ba}{}^{\bb})~, 
\end{align}
and similarly for the Lorentz parameter $\lambda$.

Now the flux tensors may be modified to generalized torsion tensors $T_{\ha \hb \hc}$
and $T_\ha$ by replacing derivatives with covariant derivatives. In our conventions,
$\cD_\ha V_\hb = D_\ha V_\hb - \omega_{\ha \hb}{}^\hc V_\hc$, so that
\begin{alignat}{2}
T_{\ha \hb \hc} &:= - 3 \,\cD_{[\ha} V_\hb{}^\hm \, V_{\hm \hc]} 
    &&= F_{\ha \hb \hc} + 3\, \omega_{[\ha \hb \hc]}~, \\
T_\ha &:= \cD_\ha \log \phi + \cD_\hm V_\ha{}^\hm
    &&= F_\ha + \omega^\hb{}_{\hb \ha}~.
\end{alignat}
Setting to zero these torsions fixes the spin connection $\omega_{\ha\hb \hc}$ up to the irreducible
hook representations in the left and right Lorentz groups.
In other words, the spin connection is defined only up to local transformations
\begin{align}\label{E:BDFT.omegashift}
\delta \omega_{a b c} = \Lambda_{a|bc}~, \qquad
\delta \omega_{\overline{a b c}} = \Lambda_{\ba|\overline{bc}}
\end{align}
where $\Lambda_{a|bc}$ obeys (and similarly for $\Lambda_{\ba|\overline{bc}}$)
\begin{align}
\Lambda_{a|bc} = -\Lambda_{a|cb}~, \qquad
\Lambda_{[a|bc]} = 0~, \qquad \Lambda^b{}_{|bc} = 0~.
\end{align}

There is no direct analogue to the Riemann tensor in DFT. A close runner-up is
\begin{align}\label{E:Romega.naive}
R(\omega)_{\ha \hb \hc \hd} &:= 
    2 \,D_{[\ha} \omega_{\hb] \hc \hd}
    - 2 \,\omega_{[\ha |\hc}{}^{\he} \omega_{|\hb] \he \hd}
    + F_{\ha \hb}{}^\he \omega_{\he \hc \hd}
    + \frac{1}{2} \omega^\he{}_{\ha \hb} \,\omega_{\he \hc \hd}~,
\end{align}
which is covariant under diffeomorphisms but slightly non-covariant under Lorentz
transformations. Its Lorentz non-covariance is
\begin{align}\label{E:BDFT.Riem.nc}
\Delta_\lambda R(\omega)_{\ha \hb \hc \hd}
    = - \tfrac{1}{2} D^\he \lambda_{\ha \hb} \,\omega_{\he \hc \hd}
    + \tfrac{1}{2} D^\he \lambda_{\hc \hd} \,\omega_{\he \ha \hb}
\end{align}
where $\Delta_\lambda$ denotes the difference between the full Lorentz transformation
and the covariant part.
This means that while $R(\omega)_{a \bb \,c d}$ and $R(\omega)_{a \bb \,\overline{c d}}$ are
covariant, $R(\omega)_{a b \,c d}$, $R(\omega)_{ab \,\overline{cd}}$,
$R(\omega)_{\overline{a b} \,c d}$, and
$R(\omega)_{\overline{ab} \,\overline{cd}}$ are not.
Nevertheless, because \eqref{E:BDFT.Riem.nc} is pairwise antisymmetric in $\ha\hb$ and
$\hc \hd$, one may construct a symmetrized Riemann tensor
\begin{align}\label{E:RpairwiseSym}
\cR_{\ha \hb \hc \hd} = \frac{1}{2} R(\omega)_{\ha \hb \hc \hd} + \frac{1}{2} R(\omega)_{\hc \hd\ha \hb}
\end{align}
that is covariant under Lorentz transformations. This is the object that is usually
considered the analogue of the doubled Riemann tensor.

Unfortunately, $\cR_{\ha \hb \, \hc \hd}$ generally depends on undetermined parts
of the spin connection. Put another way, it is not a tensor under the shift
symmetry \eqref{E:BDFT.omegashift}. One finds that $R(\omega)$ and $\cR$ transform
as\footnote{We have exhibited the torsion tensor $T$ even though we have set it to zero.
This is because it will not vanish in the supersymmetric case.}
\begin{align}\label{E:BDFT.Romega.Lambda}
\delta_\Lambda R(\omega)_{\ha \hb \hc \hd}
    &= 2 \,\cD_{[\ha} \Lambda_{\hb]|\hc\hd}
    + T_{\ha \hb}{}^\he \Lambda_{\he|\hc\hd}
    + \tfrac{1}{2} \Lambda^\he{}_{|\ha \hb} \,\omega_{\he \hc \hd}
    - \tfrac{1}{2} \Lambda^\he{}_{|\hc \hd} \,\omega_{\he \ha \hb}~, \eol
\delta_\Lambda \cR_{\ha \hb \hc \hd}
    &= 
    \cD_{[\ha} \Lambda_{\hb]|\hc\hd}
    + \cD_{[\hc} \Lambda_{\hd]|\ha\hb}
    + \tfrac{1}{2} T_{\ha \hb}{}^\he \Lambda_{\he|\hc\hd}
    + \tfrac{1}{2} T_{\hc \hd}{}^\he \Lambda_{\he|\ha\hb}~.
\end{align}
Clearly, only a few components of $\cR_{\ha \hb \,\hc \hd}$ are $\Lambda$-invariant. These are
reduced further when one considers the torsion Bianchi identities, which read
\begin{subequations}\label{E:BDFT.Bianchi}
\begin{align}
\label{E:BDFT.Bianchi.a}
\cD_{[\ha} T_{\hb \hc \hd]} &= -\frac{3}{4} T_{[\ha \hb}{}^{\he} T_{\hc \hd] \he}
    + \frac{3}{2} \cR_{[\ha \hb \hc \hd]}~, \\
\label{E:BDFT.Bianchi.b}
2 \cD_{[\ha} T_{\hb]} &= -T_{\ha \hb}{}^\hc T_\hc - \cD^\hc T_{\hc \ha \hb}
    + \widehat \cR_{\ha \hb}~, \\
\label{E:BDFT.Bianchi.c}
\cD^\ha T_{\ha} &= -\frac{1}{2} T^\ha T_\ha + \frac{1}{12} (T_{\ha \hb \hc})^2
    - \frac{1}{2} \cR_{\ha \hb}{}^{\ha \hb}~,
\end{align}
\end{subequations}
where $\widehat \cR_{\ha \hb}$ is an antisymmetric tensor given by
\begin{align}\label{E:BDFT.hatR}
\widehat \cR_{\ha\hb} = 
    D^\hc \omega_{\hc \ha \hb}
    + F^\hc \omega_{\hc \ha \hb}
    + 2 R(\omega)_{\hd [\ha \hb]}{}^\hd~.
\end{align}
Setting $T=0$ above, one can show that the only invariant components of $\cR$ are
\begin{align}\label{E:BDFT.Rsurviving}
\cR = \cR_{ab}{}^{ab} = -\cR_{\overline{ab}}{}^{\overline{ab}}~, \qquad
\cR_{a \bb} = \cR_{a c \,\bb}{}^{c} = -\cR_{a \bc \,\bb}{}^{\bc} .
\end{align}

With these ingredients, a two-derivative action for DFT can be written down
in the absence of a cosmological constant:
\begin{align}
S = \int \rd^D x\,\rd^D \tilde x\, \phi \, \cR
\end{align}
where $\phi = e^{-2d}$ is the dilaton.
The two invariants \eqref{E:BDFT.Rsurviving} have a simple interpretation
in terms of this action.
The first is the equation of motion of the dilaton and
analogous to the Ricci scalar in general relativity.
The second is the analogue of the Einstein tensor, being the equation of motion 
of the generalized vielbein. 
These two objects correspond to two of the linearized invariants one can construct
in DFT, as noted already by Siegel in \cite{Siegel:1993th}. An additional
possible third invariant was noted by Siegel, with indices corresponding
to a mixed Riemann tensor $\cR_{ab\, \overline{cd}}$. We will comment briefly
about this object shortly.

\subsection{Extending the geometry}
Up until this point, we have mainly been reviewing the frame formulation of DFT.
Now we will begin to make some changes.
Recall that the spin connection $\omega$ was introduced to gauge the $\lambda$ gauge symmetry of
the vielbein. This had the effect of removing unphysical degrees of freedom; or to put
it another way, in considering only gauge-covariant quantities, we are guaranteed that
only physical degrees of freedom will appear.
In a similar manner, we can ask to gauge the $\Lambda$ shift symmetry of
the spin connection itself.

Naturally we should introduce new connections $h_{\hm\, b|cd}$ and $h_{\hm\, \bb|\overline{cd}}$
that transform covariantly under diffeomorphisms and Lorentz transformations, but under
$\Lambda$ gauge transformations as
\begin{align}\label{E:deltaLambda.h}
\delta_\Lambda h_{\hm\, b|cd} = \cD_\hm \Lambda_{b|cd}~, \qquad
\delta_\Lambda h_{\hm\, \bb|\overline{cd}} = \cD_\hm \Lambda_{\bb|\overline{cd}}~.
\end{align}
Evidently, we can eliminate a good portion of the $\Lambda$ transformation
in \eqref{E:BDFT.Romega.Lambda} by defining
\begin{align}
R(\omega, h)_{\ha \hb \,\hc \hd} &= R(\omega)_{\ha \hb \,\hc \hd}
    - h_{\ha\, \hb |\hc \hd}
    + h_{\hb\, \ha |\hc \hd}~.
\end{align}
Now one can check that under the $\Lambda$ transformation and the anomalous Lorentz
transformation,
\begin{align}\label{E:BDFT.R.Lambda}
(\delta_\Lambda + \Delta_\lambda) R(\omega,h)_{\ha \hb \,\hc \hd}
    &= \cT_{\ha \hb}{}^\he \Lambda_{\he|\hc\hd}
    + \tfrac{1}{2} (\Lambda^\he{}_{|\ha \hb} - D^{\he} \lambda_{\ha \hb}) \,\omega_{\he \hc \hd}
    - \tfrac{1}{2} (\Lambda^\he{}_{|\hc \hd} - D^{\he} \lambda_{\hc \hd})\,\omega_{\he \ha \hb}~.
\end{align}
The second and third terms are pairwise antisymmetric in $\ha \hb$ and $\hc \hd$, and so are eliminated when one builds $\cR_{\ha \hb \,\hc \hd}(\omega, h)$,
\begin{align}
(\delta_\Lambda + \Delta_\lambda) \cR(\omega,h)_{\ha \hb \,\hc \hd}
    &= \tfrac{1}{2} \cT_{\ha \hb}{}^\he \Lambda_{\he|\hc\hd}
    + \tfrac{1}{2} \cT_{\hc \hd}{}^\he \Lambda_{\he|\ha\hb}
\end{align}
Observe that the $\Lambda$ transformation acts covariantly here: it ``rotates''
the symmetrized curvature tensor back into the torsion tensor -- which in our case vanishes.

Rather than build $\cR_{\ha \hb\, \hc \hd}$ by brute-force pairwise symmetrizing the indices,
we can instead introduce yet another field that does this job automatically.
Pol\'a\v{c}ek and Siegel showed
that if one extends the space of generalized coordinates to include local Lorentz coordinates
$y^{ab}$ and $y^{\overline{ab}}$ (along with their duals to maintain the doubled geometry),
the spin connection is naturally incorporated into the vielbein \cite{Polacek:2013nla}.
One also gets for free a
new field, which we denote $p_{\ha \hb\,\hc\hd}$, which is pairwise antisymmetric and
with both pairs valued in the two Lorentz groups. While Pol\'a\v{c}ek and Siegel 
did not consider the connection $h$, it is straightforward to extend their analysis 
to include it, or even a completely arbitrary set of gauge connections.
We present this analysis in Appendix \ref{A:PS}.

The upshot is that one can directly build a covariant curvature tensor
\begin{align}
R(\omega, h, p)_{\ha \hb\, \hc \hd}
    = R(\omega)_{\ha \hb \,\hc \hd}
    - h_{\ha\, \hb |\hc \hd}
    + h_{\hb\, \ha |\hc \hd}
    - p_{\ha \hb\,}{}_{\hc \hd}
\end{align}
with $p$ transforming as
\begin{align}\label{E:delta.p}
(\delta_\Lambda + \Delta_\lambda) p_{\ha \hb\, \hc \hd} &=
    \tfrac{1}{2} (\Lambda^\he{}_{|\ha \hb} - D^{\he} \lambda_{\ha \hb}) \,\omega_{\he \hc \hd}
    - \tfrac{1}{2} (\Lambda^\he{}_{|\hc \hd} - D^{\he} \lambda_{\hc \hd})\,\omega_{\he \ha \hb}
\end{align}
so that only the torsion term survives in \eqref{E:BDFT.R.Lambda}.

We have just introduced a slew of new degrees of freedom, so we had better be able
to constrain the curvature tensor to eliminate them. First, we can entirely fix 
$p_{\ha \hb\, \hc \hd}$ by enforcing that $\cR_{\ha \hb \hc \hd}$ is pairwise 
symmetric where possible. That is, we can choose $p$ so that the following conditions hold:
\begin{subequations}
\begin{align}
R(\omega, h, p)_{a b, c d} &= R(\omega, h, p)_{c d, a b}~, \\
R(\omega, h, p)_{a b, \overline{c d}} &= R(\omega, h, p)_{\overline{c d}, a b}~, \\
R(\omega, h, p)_{\overline{a b}, \overline{c d}} &= R(\omega, h, p)_{\overline{c d}, \overline{a b}}~.
\end{align}
\end{subequations}
In doing so, the above curvatures each coincide with $\cR_{\ha \hb, \hc \hd}$.
Note however that
\begin{align}
\cR_{c d, a \bar b} = \cR_{a \bar b\, c d} 
    = \frac{1}{2} R(\omega,h,p)_{a \bar b, c d}~, \qquad
    R(\omega,h,p)_{c d, a \bar b} = 0 \,\,\,\text{(by definition)}
\end{align}
so it is not quite correct to claim $R(\omega,h,p)$ is the same as $\cR_{\ha \hb \hc \hd}$ defined in \eqref{E:RpairwiseSym}, although it will be true that
$\cR_{[\ha \hb \hc \hd]} = R(\omega,h,p)_{[\ha \hb \hc \hd]}$, so that the latter
can replace the former in the Bianchi identity. 

\emph{Henceforth, we use $\cR_{\ha \hb \hc\hd}$ as an alias for $R(\omega,h,p)_{\ha \hb \hc\hd}$.}

Next, one can fix a large portion of $h_{\hm \,\hb|\hc\hd}$ by
eliminating as many pieces of $\cR_{\ha \hb \,\hc \hd}$ as possible. 
This is a straightforward group theory question: what representations of
$h_{\ha\, \hb|\hc\hd}$ survive in the projection
$h_{[\ha\, \hb] |\hc \hd} + h_{[\hc\, \hd] |\ha \hb}$ ?
In terms of traceless Young tableaux, $h_{\ha\, \hb|\hc\hd}$ can be decomposed as
\begin{align}\label{E:YRep.h.a}
h_{a\, b|cd} &: \yng(1) \times \yng(2,1)
    = \yng(2,2) + \yng (3,1) + \yng(2,1,1) + \yng(1,1) + \yng(2) ~, \\
\label{E:YRep.h.b}
h_{\bar a\, b|cd} &: \overline{\yng(1)} \times \yng(2,1)
\end{align}
with similar expressions for $h_{a\, \bb |\overline{cd}}$ and 
$h_{\overline{a\, b}|\overline{cd}}$.
The pairwise symmetrized Riemann tensor decomposes as
\begin{align}
\label{E:YRep.R.a}
\cR_{a b\, c d} &: \yng(1,1) \times \yng(1,1) \, \Bigg\vert_{\rm sym} = 
    \yng(2,2) + \cancel{\yng(1,1,1,1)} + \yng(2) + \bullet~ \\
\label{E:YRep.R.b}
\cR_{\ba b\, c d} &: \overline{\yng(1)} \times \yng(1) \times \yng(1,1) = 
    \overline{\yng(1)} \times \Bigg(
    \yng(2,1) + \cancel{\yng(1,1,1)} + \yng(1) 
    \Bigg)
\end{align}
where we have crossed out the final representations that are killed by the
constraint $\cR_{[\ha \hb \hc \hd]} = 0$, which is the
Bianchi identity \eqref{E:BDFT.Bianchi.a} in the absence of 
torsion.\footnote{We emphasize that the equations \eqref{E:BDFT.Bianchi.a}
and \eqref{E:BDFT.Bianchi.c}
continue to hold with the new $\cR$ above, due to the symmetry properties of
$h$ and $p$. The curvature $\widehat \cR_{\ha \hb}$ in
\eqref{E:BDFT.Bianchi.b} develops a new interpretation,
which we discuss in Appendix \ref{A:PS}.}
(We do not list $\cR_{\overline{ab}\, cd}$ above because it is constrained
to vanish from this Bianchi identity.) 

Comparing \eqref{E:YRep.h.a} with \eqref{E:YRep.R.a}, it is clear
that the $\tiny\yng(2,2)$ and $\tiny\yng(2)$ representations of $h_{a \, b|cd}$ can
be used to fix the corresponding representations of $\cR_{a b\, cd}$ to zero.
Similarly, the single representation of $h_{\ba \, b|cd}$ can be used to fix 
that representation of $\cR_{\ba b\,cd}$. Similar comments pertain to the barred/unbarred
versions.

Taking into account the other Bianchi identities \eqref{E:BDFT.Bianchi.b}
and \eqref{E:BDFT.Bianchi.c} in the absence of torsion, it isn't hard to see
that we have eliminated all but the physical components of the Riemann tensor
given in \eqref{E:BDFT.Rsurviving}.
The remaining singlet of $\cR_{abcd}$ is related by \eqref{E:BDFT.Bianchi.c}
to the singlet of $\cR_{\overline{ab}\,\overline{cd}}$. Similarly, the $(1,1)$ tensor of
$\cR_{\ba b\, cd}$ is related to that of $\cR_{a \bb\, \overline{cd}}$ by
\eqref{E:BDFT.Bianchi.b}.
Recall there was a third linearized invariant discussed by Siegel in \cite{Siegel:1993th}.
This is nothing but the tensor $p_{a b\, \overline{cd}}$; its non-covariance \eqref{E:delta.p}
appears only at second order.

Naturally, the above process turns out to leave certain representations in $h$ unfixed,
specifically the $\tiny\yng (3,1) + \yng(2,1,1) + \yng(1,1)$ representations of
$h_{a\, b|cd}$ and their barred versions.
If we gauge these local symmetries, the process may continue.
In fact, it would seem we \emph{must} gauge these local symmetries in order for the
algebra to close. The reason is that the commutator of two local $\Lambda_{a|bc}$
transformations must close on a new generator (or vanish). From 
\eqref{E:deltaLambda.h} one can compute that
\begin{align}
[\delta_{\Lambda_2}, \delta_{\Lambda_1}] h_{a, b|cd}
    &= - 2 \, \Lambda_{[2}{}_{ab}{}^e \Lambda_{1]}{}_{ecd}
    + 2 \, \Lambda_{[2}{}_{ac}{}^e \Lambda_{1]}{}_{bde}
    - 2 \, \Lambda_{[2}{}_{ad}{}^e \Lambda_{1]}{}_{bce}~.
\end{align}
A laborious computation shows that the right-hand side involves only the
three representations $\tiny\yng (3,1) + \yng(2,1,1) + \yng(1,1)$.
In other words, closure of the algebra \emph{requires} that we now gauge the shift
symmetry in $h$. This seems likely to continue \emph{ad infinitum}, 
and the local gauge symmetry one generates seems to be closely related 
to the Maxwell$_\infty$ algebra. 

We elaborate on this further in Appendix \ref{A:SMaxwell} and in section \ref{S:SGeo.Detour}
for the supersymmetric case,
but let us briefly summarize the situation here. The Maxwell$_\infty$ algebra is
the free Lie algebra generated by $P_a$, combined with the Lorentz
generator \cite{Gomis:2017cmt,Gomis:2018xmo}.
Taking $P_a$ to have dimension 1, we define generators of positive dimension by
\begin{align}\label{E:BosonicMinf.1}
[P_a, P_b] = Y_{a b}~, \qquad
[P_a, Y_{b c}] = Y_{a,bc}~, \qquad \cdots
\end{align}
The generator $Y_{ab}$ is antisymmetric, and 
the generator $Y_{a,bc}$ lies in the reducible hook representation, since the
Jacobi identity of three $P$'s leads to $Y_{[a,bc]} = 0$. We will make a further
restriction for what we call the Maxwell$_\infty$ algebra that differs from
\cite{Gomis:2017cmt,Gomis:2018xmo}: we restrict $Y_{a,bc}$
to be traceless, and denote it $Y_{a|bc}$. Physically this can be understood if we
interpret $P_a$ in \eqref{E:BosonicMinf.1} to be the covariant derivative 
in a geometry with an on-shell Maxwell field strength. Then $Y_{ab}$ is $-\bm F_{ab}$
and then $Y_{a|bc}$ corresponds to its covariant derivative modulo the Bianchi
identity and the equation of motion.

Now it follows that at dimension 4,
\begin{align}
[Y_{a b}, Y_{c d}] &= Y_{ab,cd}~, \\
[P_a, Y_{b|cd}] &= Y_{a|b|cd} + \Big[
    \tfrac{3}{4} Y_{a b,c d} 
    + \tfrac{3}{8} \eta_{a c} Y_{b}{}^e{}_{,de}
    \Big]_{b|cd}~.
\end{align}
The first equation defines $Y_{ab,cd}$, so it is
in a reducible representation. In the second equation,
$Y_{a|b|cd}$ is in an irreducible representation,
which is symmetric in $ab$, antisymmetric in $cd$, traceless, and vanishing when
any three indices are antisymmetrized. The Jacobi identity guarantees these are
the only representations. In terms of traceless Young tableaux, these are
\begin{align}
Y_{a b, cd} : \yng(1,1) \wedge \yng(1,1) = \yng(2,1,1) + \yng(1,1) \qquad \qquad
Y_{a|b|cd} : \yng(3,1)
\end{align}
In terms of operators, $Y_{ab,cd}$ is $[\bm F_{ab}, \bm F_{cd}]$, so if we keep this
generator, we are really working with non-abelian Yang-Mills operators.
The second generator corresponds to two
symmetrized covariant derivatives of the field strength, modulo terms that involve
the Bianchi identity or the equation of motion.

The representations we are encountering match so far the representations
we seek for the extension of the local algebra of bosonic double field theory.
Unfortunately, the generators $Y$ we are discussing have the wrong dimension and their
algebra is not the same as the local algebra of $\Lambda$ transformations.
The solution is to recall in the Pol\'a\v{c}ek-Siegel formalism that for every
generator that we wish to gauge, we must introduce a dual generator that is 
paired with it. The $Y$'s discussed above are exactly these dual generators.
The algebra we are encountering can be embedded in the Pol\'a\v{c}ek-Siegel
extension of the Maxwell$_\infty$ algebra discussed above. 
The resulting infinite algebra involves generators (and similarly for barred
indices)
\begin{align}
\begin{tabular}{l|c|c|c|c|c|c|c|c|c}
dimension & $\cdots$ & $-2$ & $-1$ & $0$ & $+1$ & $+2$ & $+3$ & $+4$ & $\cdots$ \\ \hline
generator $\phantom{\Big|}$ & $\cdots$ & 
$L_{ab,cd}$, $L_{a|b|cd}$ & $K_{a|bc}$ & $M_{ab}$ & $P_a$ &
$\widetilde M^{ab}$ & $\widetilde K^{a|bc}$ &
$\widetilde L^{ab,cd}$, $\widetilde L^{a|b|cd}$ & $\cdots$
\end{tabular}
\end{align}
Up to normalizations, the $Y$ generators defined previously have been renamed
$\widetilde M$, $\widetilde K$, etc. to emphasize that they are the duals of
generators of non-positive dimension. We have denoted the shift generator of the $\omega$
connection by $K$ and the shift generators of the $h$ connection by $L$ for convenience.
We emphasize that $L_{ab,cd}$ is reducible.
For later use, we denote the subalgebra of non-positive generators by
\begin{align}
\ext{\g{SO}}(D-1,1) = \{ M_{a b}, K_{a|bc}, L_{ab,cd}, L_{a|b|cd}, \cdots\}~.
\end{align}

The generators and their duals are related
by an infinite dimensional invertible metric $\eta$, which is the metric
on the Pol\'a\v{c}ek-Siegel megaspace. ($P_a$ is its own dual.)
The metric $\eta$ is defined in the obvious way
by products of $\eta_{ab}$ with unit normalization. With this generator,
one can write the full algebra of generators $X_\cA$
schematically in terms of a commutator and
a symmetric metric,
\begin{align}
[X_\cA, X_\cB] = -f_{\cA \cB}{}^\cC X_\cC~, \qquad
\opbraket{X_\cA}{X_\cB} = \eta_{\cA \cB}
\end{align}
with the requirement that $f_{\cA \cB \cC} = f_{\cA \cB}{}^{\cD} \eta_{\cD \cC}$ 
is totally antisymmetric
(which follows by ad-invariance of the metric).
This allows one to use the algebra of positive dimension generators to ``reflect'' onto the
algebra of negative dimension generators. For example,
\begin{align}
[P_a, P_b] = -2 \wtM_{ab} \quad \implies \quad
[M_{c b}, P_a] = \eta_{ba} P_c - \eta_{c a} P_b~.
\end{align}
To avoid needless repetition, we will postpone further details until the supersymmetric case.

Another point to emphasize is that the kind of extension to $\omega$ that
we postulate here is very similar to how conformal gravity extends
Poincar\'e gravity. In conformal gravity, one introduces not just a spin connection
$\omega_{m \,a b}$ but also a Weyl connection $b_m$ to gauge dilatations.
The vanishing torsion condition determines $\omega_{m\, ab}$ but not $b_m$.
This suggests a new local symmetry, and indeed the new symmetry one encounters is
the special conformal transformation $K_a$, under which $b$ transforms linearly, 
$\delta b_m = e_m{}^a \Lambda_a$. One is forced to introduce a new gauge field $f_{m a}$,
which can be constrained to eliminate all parts of the Lorentz curvature \emph{except} for the
Weyl tensor. The behavior we are postulating for bosonic DFT is quite similar, except
that the local gauge symmetries do not terminate at $K$.

It may seem apparent to the reader that in the bosonic case there is no necessity
to introduce the local $\Lambda_{a|bc}$ symmetry or its corresponding gauge 
field $h_{\hm \, a|bc}$. This is true. But when we move on to the supersymmetric case, 
we will find that
the local tangent space symmetry should be extended beyond the doubled Lorentz group
and closure of those transformations will inevitably lead to the appearance of the
$\Lambda$ shift symmetry and its supersymmetric siblings.

%%%%%%%%%%%%%%%%%%%%%%%%%%%%%%%%%%%%%%%%%%%%%%%%%%%%%%%%%%%%%%%%%%%%%%%%%%%%%%%%%
\section{Double field theory in superspace}\label{S:SDFT}
%%%%%%%%%%%%%%%%%%%%%%%%%%%%%%%%%%%%%%%%%%%%%%%%%%%%%%%%%%%%%%%%%%%%%%%%%%%%%%%%%

\subsection{Elements of component supersymmetric DFT}
$\cN=1$ supersymmetric DFT was constructed independently by
Hohm and Kwak \cite{Hohm:2011nu} (to second order in fermions) 
and by Jeon, Lee, and Park \cite{Jeon:2011sq} (to all orders).
The conventions in these two approaches differ somewhat. We will more closely follow
Hohm and Kwak \cite{Hohm:2011nu}, although we will exchange the role of the left and
right Lorentz groups, along with a number of other minor alterations.

The extension to bosonic DFT involves introducing a gravitino $\Psi_\ba{}^\alpha$
and a dilatino $\chi_\alpha$. Roughly speaking, the gravitino is the superpartner
to the doubled vielbein and the dilatino is the superpartner to the dilaton.
The index $\alpha$ is a Majorana spinor index of $\g{SO}(9,1)_L$, and the gravitino
and dilatino are of opposite chirality. The vector index $\ba$ on the gravitino belongs
to $\g{SO}(9,1)_R$, so the gravitino transforms as a vector under one of the groups
and a spinor under the other. As usual, the supersymmetry parameter $\eps^\alpha$ 
is of the same chirality as the gravitino.

In our conventions, the supersymmetry transformation of the vielbein reads
\begin{align}\label{E:SDFT.comptrafo1}
\delta V_\hm{}^a &= V_\hm{}^{\bb} \, J_{\bb}{}^a~, \qquad
\delta V_\hm{}^\ba = V_\hm{}^{b} \, J_{b}{}^\ba~, \qquad
J_{\bb a} = -J_{a \bb} = -\kappa\, \eps^\alpha (\gamma_a)_{\alpha \beta} \Psi_\bb{}^\beta~,
\end{align}
and for the dilatino,
\begin{align}
\delta \phi &= -\kappa\, \phi \, \eps^\alpha \chi_\alpha~.
\end{align}
In these expressions, $\kappa$ is a dimensionless number that normalizes $\epsilon^\alpha$.
In order to match 10D $\cN=1$ supergravity with conventional normalizations, $\kappa$
involves an inconvenient factor of $\sqrt{2}$, so we have found it easier to leave this
number a variable for now. The fermions transform (to lowest order) as
\begin{align}\label{E:SDFT.comptrafo2}
\delta \Psi_\ba{}^\alpha = \cD_\ba \eps^\alpha + \cdots ~, \qquad
\delta \chi_\alpha = (\gamma^a)_{\alpha \beta} \cD_a \eps^\beta + \cdots~.
\end{align}
In contrast to Hohm and Kwak \cite{Hohm:2011nu}, $\cD_\ha$ is defined as $V_\ha{}^\hm \pa_\hm$
without a factor of $\sqrt{2}$.

A key feature of supersymmetric DFT is that both fermions are connections:
they each transform into a derivative of the supersymmetry parameter.
This implies that in any superspace formulation they ought to be encoded
in the supervielbein.

\subsection{Siegel's superspace DFT}
Let us now construct double field theory in $\cN=1$ superspace.
We begin with a review of Siegel's early construction \cite{Siegel:1993th},
with some minor modifications to conventions. While Siegel also treated heterotic
double field theory in the abelian limit, 
we restrict our attention to the case without vector multiplets for simplicity.
The basic details are also identical to Cederwall's discussion of double supergeometry
\cite{Cederwall:2016ukd}, except with half as many fermionic coordinates.

In analogy to conventional $\g{O}(D,D)$ double field theory, let us introduce
$\g{OSp}(D,D|2s)$ double field theory with $2D$ bosonic coordinates and $2s$ fermionic
ones. We will really only be concerned with $D=10$ and $s=16$.
Collectively, we denote these supercoordinates by $z^\cM$.
Superdiffeomorphisms have an $\g{OSp}(D,D|2s) \times\mathbb R^+$ structure, 
and act on a vector $W^\cM$ of weight $w$ as
\begin{align}
\Lie^{(w)}_\xi W^\cM = \xi^\cN \pa_\cN W^\cM 
    - \cW^\cN \Big(\pa_\cN \xi^\cM - \pa^\cM \xi_\cN (-1)^{nm}\Big)
    + w\, \pa_\cN \xi^\cN\, W^{\cM}\, (-1)^n 
\end{align}
where the grading $(-1)^{nm}$ is $-1$ if both $\cN$ and $\cM$ are fermionic and $+1$ otherwise.
An index $\cM$ is raised or lowered with the canonical orthosymplectic
element $\eta$ of $\g{OSp}(D,D|2s)$, with the usual NW-SE convention, i.e.
\begin{align}
W^\cM = \eta^{\cM \cN} W_\cN~, \qquad
W_\cM = W^\cN \eta_{\cN \cM}~.
\end{align}
The metric $\eta$ itself is graded symmetric, $\eta_{\cM \cN} = \eta_{\cN \cM} (-1)^{nm}$.
Because of the grading, $\eta_{\cM\cN}$ is not quite the inverse of $\eta^{\cM \cN}$;
instead, one finds
\begin{align}
\eta^{\cM \cP} \eta_{\cP\cN} = \delta_\cN{}^\cM  (-1)^{nm}~.
\end{align}
Closure of the algebra of superdiffeomorphisms requires the section condition
\begin{align}
\eta^{\cM \cN} \pa_\cN \otimes \pa_\cM = 0~.
\end{align}

In the standard coordinate basis, the coordinates are denoted
$z^\cM = (x^\rmm, \tilde x_\rmm, \theta^\mu, \tilde \theta_\mu)$ and $\eta$ is given by
\begin{align}
\eta^{\cM \cN} =
\begin{pmatrix}
0 & \delta^\rmm{}_\rn & 0 & 0 \\
\delta_\rmm{}^\rn & 0 & 0 & 0 \\
0 & 0 & 0 & \delta^\mu{}_\nu \\
0 & 0 & -\delta_\mu{}^\nu & 0
\end{pmatrix}~, \qquad
\eta_{\cM \cN} =
\begin{pmatrix}
0 & \delta_\rmm{}^\rn & 0 & 0 \\
\delta^\rmm{}_\rn & 0 & 0 & 0 \\
0 & 0 & 0 & \delta_\mu{}^\nu \\
0 & 0 & -\delta^\mu{}_\nu & 0
\end{pmatrix}~.
\end{align}
The section condition can then be solved by restricting fields and parameters to depend
only on $z^\tM = (x^\rmm, \theta^\mu)$, the coordinates of 10D $\cN=1$ superspace.
It will often be useful to collectively denote the doubled bosonic coordinates
by $x^\hm = (x^\rmm, \tilde x_\rmm)$ and the fermionic ones by $\theta^\hmu = (\theta^\mu, \tilde \theta_\mu)$. It will also be useful occasionally to use left/right-moving coordinates
$x^m$ and $x^{\bar m}$; these are defined exactly as in the bosonic case.
There are no fermionic analogues to the left or right-moving bosonic coordinates, 
partly because $\theta$ and $\tilde\theta$
have different engineering dimension. If $x$ and $\tilde x$ are both taken to have
dimension $-1$, and if $\theta$ has dimension $-\tfrac{1}{2}$ (to allow a conventional $\cN=1$ 
superspace to emerge), then the $\g{OSp}(D,D|2s)$ structure forces
$\tilde\theta$ to have dimension $-\tfrac{3}{2}$ \cite{Cederwall:2016ukd}.

The supervielbein is naturally taken as a weighted element of $\g{OSp}(D,D|2s) \times \mathbb R^+$.
As in the bosonic case, it is more convenient to split the supervielbein into an
$\g{OSp}(D,D|2s)$ element, which we denote $\cV_\cM{}^\cA$, and the 
superdilaton $\Phi$.\footnote{Siegel denoted the superdilaton by $\Phi^2$ so that
his $\Phi$ has weight $\tfrac{1}{2}$.}
These transform respectively as
\begin{subequations}\label{E:SDFT.VPhi.trafos}
\begin{align}
\label{E:SDFT.VPhi.trafos.a}
\delta \cV_\cM{}^\cA &= \Lie_\xi \cV_\cM{}^\cA
    = \xi^\cN \cV_\cM{}^\cA + \Big(\pa_\cM \xi^\cN - \pa^\cN \xi_\cM (-1)^{nm}\Big) \cV_\cN{}^\cA~, \\
\label{E:SDFT.VPhi.trafos.b}
\delta \Phi &= \Lie^{(+1)}_\xi \Phi
    = \xi^\cN \pa_\cN  \Phi
    + \pa_\cN \xi^\cN\,  \Phi\, (-1)^n~.
\end{align}
\end{subequations}
It is going to turn out that the superdilaton $\Phi$ differs
from the component dilaton $\phi = e^{-2d}$, so we have used a different name for it.
The index $\cA$ decomposes into a left sector $A$ and a right sector $\bar A$, with
\begin{align}
W^\cA = (W^A, W^{\bar A})~, \qquad 
W^A= (W^a, W^\alpha, W_\alpha)~, \qquad
W^{\bar A} = (W^{\bar a})~.
\end{align}
Siegel identified these sectors as the left and right-handed sectors of oscillators of
the affine Lie algebra of the superstring in the Hamiltonian framework. Here $a$ and $\bar a$
are the tangent space indices associated with the local $\g{O}(D-1,1)_L$ and $\g{O}(D-1,1)_R$ actions
on the bosonic DFT vielbein. In the supersymmetric case, $a$ is extended to include spinors
of both chiralities. The tangent space metric and its inverse are
\begin{align} \label{E:etaAB}
\eta^{\cA \cB} =
\left(\begin{array}{ccc|c}
\eta^{ab} & 0 & 0 & 0 \\
0 & 0 & \delta^\alpha{}_\beta & 0 \\
0 & -\delta_\alpha{}^\beta & 0 & 0 \\ \hline
0 & 0 & 0 & -\eta^{\overline{ab}} \phantom{\Big\vert} \\
\end{array}\right)~, \qquad
\eta_{\cA \cB} =
\left(\begin{array}{ccc|c}
\eta_{ab} & 0 & 0 & 0 \\
0 & 0 & \delta_\alpha{}^\beta & 0 \\
0 & -\delta^\alpha{}_\beta & 0 & 0 \\ \hline
0 & 0 & 0 & -\eta_{\overline{ab}}  \\
\end{array}\right)~,
\end{align}
where the horizontal and vertical lines emphasize
the split between the left and right sectors.
The metrics $\eta^{ab}$ and $\eta^{\overline{ab}}$ both describe $\g{SO}(D-1,1)$ 
(with mostly positive signature)
and are used to raise and lower their respective indices.
The condition that the supervielbein is a group element amounts to
\begin{align}
\cV_\cA{}^\cM = \eta^{\cM \cN} \cV_\cN{}^\cB \eta_{\cB \cA} \,(-1)^{am}~.
\end{align}
The right tangent space is $\g{O}(D-1,1)_R$ as in the bosonic theory, 
whereas the left tangent space must be a
subgroup of $\g{OSp}(D-1,1|2s)$. We will elaborate on the choice of this tangent space
momentarily.

Although it is natural to group together the left sector tangent space indices, we will
usually group bosonic indices together, so that $W^\cA = (W^a, W^\ba, W^\alpha, W_\alpha)$,
to be more in line with how we have ordered our coordinates.
Using the same rules for raising and lowering indices as before, one finds that
\begin{align}
W^\cA = (W^a, W^{\ba}, W^\alpha, W_\alpha) \quad \implies \quad
W_\cA = (W_a, -W_{\ba}, - W_\alpha, W^\alpha)~.
\end{align}
Therefore, when specifying explicit elements, one must be careful to emphasize whether the
indices are taken to be vectors or covectors because of the signs that are introduced above.

Just as one may use left/right coordinates $z^\cM = (x^m, x^{\bar m}, \theta^\mu, \theta_\mu)$
to match the left/right tangent space, one may occasionally wish to put the tangent space 
in its standard toroidal form, where
$W^\cA = (W^\ra, W_\ra, W^\alpha, W_\alpha)$, in line with the standard toroidal coordinates
$z^\cM = (x^\rmm, \tilde x_\rmm, \theta^\mu, \tilde \theta_\mu)$.
We use lower case Roman font indices $\ra, \rb, \cdots$ to denote the tangent frame
bosonic indices in the standard basis.

\subsection{Determining the tangent space}
\label{S:SDFT.Tangent}
In order to motivate how we will fix the tangent space in superspace, let's return to
bosonic DFT and discuss how one may go about determining its physical content.
We will take a more elaborate approach than is actually required, but this approach
will generalize more straightforwardly to superspace.

Working for the moment in the standard toroidal basis for both the coordinate and 
tangent space indices, the vielbein can be decomposed as
\begin{align}\label{E:BDFT.Decomp}
V_\hm{}^\ha =
\begin{pmatrix}
1 & b \\
0 & 1
\end{pmatrix}
\times
\begin{pmatrix}
e & 0 \\
0 & e^{-T}
\end{pmatrix}
\times
\begin{pmatrix}
1 & 0 \\
c & 1
\end{pmatrix}
\end{align}
where $b = b_{\rm mn}$, $e = e_{\rm m}{}^\ra$, and $c = c^{\ra \rb}$.
It is easy to check that this is an $\g{O}(D,D)$ element by computing
$V^{-1} = \eta V^T \eta$. This decomposition can be understood as arising by
first decomposing the generators $X_{\hm \hn}$ of $\g{O}(D,D)$ into 
$X^{\rm mn}$, $X_{\rm m}{}^\rn$, and $X_{\rm mn}$.
Our $\g{O}(D,D)$ and $\g{OSp}(D,D|2s)$ conventions can be found in Appendix \ref{A:Notation}.
The generators can be assigned levels $-1$, $0$, and $+1$ with respect to an outer automorphism
of the algebra. The above element then corresponds to a parametrization of the group element $V$ as
\begin{align}
V = 
\exp(\tfrac{1}{2} b_{\rm mn} X^{\rm nm}) \times 
\exp(a_\rmm{}^\rn X_\rn{}^\rmm) \times 
\exp(\tfrac{1}{2} c^{\rm mn} X_{\rm nm}) 
\end{align}
and taking $V$ to act to the left on the coordinate representation. One identifies
$e_\rmm{}^\ra = \exp(a)$.

In this parametrization, a general coordinate transformation
with $\pa^\rmm = 0$ transforms only $b_{\rmm \rn}$ and $e_\rmm{}^\ra$. To see this,
observe that the general coordinate transformation involves
\begin{align}
\cK_\hm{}^\hn := \pa_\hm \xi^\hn - \pa^\hn \xi_\hm =
\begin{pmatrix}
\pa_\rmm \xi^\rn & 2 \,\pa_{[\rmm} \xi_{\rn]} \\
0 & -\pa_\rn \xi^\rmm
\end{pmatrix}~,
\end{align}
which is an $\g{O}(D,D)$ generator at levels 0 and $-1$. Writing the
vielbein as $V = V_{-1} V_0 V_{+1}$ and the generator as
$\cK = \cK_{-1} + \cK_0$, a general coordinate
transformation then acts as
\begin{align}
\delta_{\rm g.c.} V = \cK V \quad \implies \quad 
\delta_{\rm g.c.} V_{-1} = \cK_{-1} V_{-1} + [\cK_0, V_{-1}] ~, \qquad
\delta_{\rm g.c.} V_0 = \cK_0 V_0~, \qquad
\delta_{\rm g.c.} V_{+1} = 0
\end{align}
which amount to the usual transformations
\begin{align}
\delta_{\rm g.c.} b_{\rm mn} = 2 \pa_{[\rmm} \xi_{\rn]} + \pa_\rmm \xi^\rp \,b_{\rp \rn} - b_{\rmm \rp} \,\pa_\rn \xi^\rp~, \qquad
\delta_{\rm g.c.} e_\rmm{}^\ra = \pa_\rmm \xi^\rn e_\rn{}^\ra ~, \qquad
\delta_{\rm g.c.} c^{\rm a b} = 0~.
\end{align}

The field $c^{\rm a b}$ itself can be set to zero using half of the
$\g{O}(D-1,1)_L \times \g{O}(D-1,1)_R$ tangent space symmetry. Recall that
$\mathfrak{so}(D-1,1)_L + \mathfrak{so}(D-1,1)_R$ is represented
in the standard toroidal basis as
\begin{align}
\lambda_\ha{}^\hb =
\begin{pmatrix}
\lambda_\ra{}^\rb & \tilde \lambda_{\ra\rb} \\
\tilde\lambda^{\ra\rb} & \lambda^\ra{}_\rb
\end{pmatrix}~, \qquad
\lambda = \frac{1}{2} (\lambda_L + \lambda_R)~, \qquad \tilde\lambda = \frac{1}{2} (\lambda_L - \lambda_R)
\end{align}
The indices $\ra,\rb,\cdots$ denote vectors of the diagonal $\g{SO}(D-1,1)$ transformation,
corresponding to antisymmetric $\lambda_{\ra\rb}$ above. The anti-diagonal transformations
are generated by antisymmetric $\tilde \lambda_{\ra\rb}$. (In both cases, we raise and
lower indices with the diagonal $\eta_{\ra\rb}$.) 
Now consider a double Lorentz transformation acting on the right of $V$.
The diagonal transformations, which lie at level 0, evidently transform $c$ as a 2-form, 
$e$ as a vector on the right,  and leave $b$ invariant, which justifies their index
structures. An infinitesimal $\tilde \lambda$ transformation is more complicated,
comprising a sum of levels $-1$ and $+1$. Because $c^{\ra\rb}$ is situated on the far 
right of the group element, lying at level $+1$, it can be eliminated immediately,
whereas $e_\rmm{}^\ra$ and $b_{\rm mn}$ will transform in much more complicated ways.
The explicit form of their transformations is irrelevant: we only need to know
that we can impose the gauge $c=0$. This gauge will then be undisturbed 
both by diagonal Lorentz transformations and by general coordinate transformations 
obeying $\pa^\rmm = 0$.

Returning to superspace, we introduce a decomposition of the
supervielbein in direct analogy to \eqref{E:BDFT.Decomp}.
Group the physical coordinates together so that
$z^\cM = (z^\tM, \tilde z_\tM) = (x^\rmm, \theta^\mu, \tilde x_\rmm, \tilde \theta_\mu)$ and 
similarly for the tangent space,
$W^\cA = (W^\tA, W_\tA) = (W^\ra, W^\alpha, W_\ra, W_\alpha)$. 
Here $z^{\tM}$ is the coordinate of 10D $\cN=1$ superspace and $\tA$ is its 
tangent superspace index. The supervielbein decomposes as
\begin{align}
\cV_\cM{}^\cA &=
\begin{pmatrix}
1 & B \\
0 & 1
\end{pmatrix}
\times
\begin{pmatrix}
E & 0 \\
0 & E^{-T}
\end{pmatrix}
\times
\begin{pmatrix}
1 & 0 \\
C & 1
\end{pmatrix}
\end{align}
where $B = B_{\tM\tN} (-)^n$, 
$E = E_\tM{}^\tA$, $E^{-T} = E_\tA{}^\tM (-1)^{am+a}$ and $C = C^{\tA\tB}$.
This corresponds to a level decomposition with the $\g{OSp}(D,D|2s)$ generator
$X_{\cM \cN}$ decomposed into the $\g{GL}(D|s)$ generator $X_\tM{}^\tN$ at level 0 and
generators $X_{\tM \tN}$ and $X^{\tM\tN}$ at levels $\pm1$.

The supervielbein $E_\tM{}^\tA$ and super-two-form $B_{\tM\tN}$ both play natural roles in 
10D $\cN=1$ superspace, and it is straightforward as in the bosonic case to show that
they transform under $\tilde z_\tM$-independent general coordinate transformations in the expected manner:
\begin{align}
\delta_{\rm g.c.} B_{\tM\tN} = 2\, \pa_{[\tM} \tilde \xi_{\tN]} 
    + \pa_\tM \xi^\tP B_{\tP \tN} - B_{\tM \tP} \,\pa_\tN \xi^\tP (-1)^{np}~, \qquad
\delta_{\rm g.c.} E_\tM{}^\tA = \pa_\tM \xi^\tN E_\tN{}^\tA
\end{align}
where we have taken $\xi_\cM = (\tilde \xi_\tM, \xi^\tM)$, or equivalently,
$\xi^\cM = (\xi^\tM, \tilde \xi_\tM (-)^m )$.
In analogy with the bosonic case, we would like to eliminate $C^{\tA\tB}$.
Because this involves not only $C^{\ra\rb}$ but also $C^{\ra \beta}$ and $C^{\alpha \beta}$,
we require an enhancement of the bosonic tangent group. Using current 
algebra arguments, Siegel made a choice that amounts to enhancing $\g{O}(D-1,1)_L$ 
to a subgroup of $\g{OSp}(D-1,1|2s)_L$  where the generator $\lambda_{A}{}^B$ is
subject to the following conditions 
\begin{align}
\lambda_\alpha{}^b = 0~, \qquad 
\lambda_\alpha{}^\beta = \frac{1}{4} \lambda^{ab} (\gamma_{ab})_\alpha{}^\beta~,
\end{align}
with $\lambda^{\alpha \beta}$, $\lambda^{ab}$, and $\lambda^{a \beta}$ unconstrained.
The second condition above amounts to the requirement that fermionic orthosymplectic
indices transform as spinors
under the $\g{SO}(D-1,1)$ subgroup of $\g{OSp}(D-1,1|2s)$.
In essence, the $\g{OSp}(D-1,1|2s)$ generators $M_{AB}$ are truncated to just the subalgebra
involving $M_{\alpha \beta}$, $M_{\alpha b}$, and 
$M'_{ab} = M_{a b} + \frac{1}{4} (\gamma_{ab})_\alpha{}^\beta M_\beta{}^\alpha$
corresponding to generators $\lambda^{\alpha \beta}$, $\lambda^{a \beta}$ and
$\lambda^{ab}$. As in the bosonic case, this is sufficient gauge freedom to 
eliminate $C^{\tA\tB}$.
Siegel went on in \cite{Siegel:1993th} to identify the generalized flux tensors (and torsion
tensors) of superspace and showed which constraints led to 10D $\cN=1$ superspace.

There is a subtlety in this approach when we compare with component supersymmetric DFT.
The component dilatino, which transforms 
non-tensorially under supersymmetry \eqref{E:SDFT.comptrafo2}, does not seem to have a 
natural interpretation as a component of a covariant torsion tensor, which is where it
arises in 10D $\cN=1$ superspace. For that reason, one would expect it to more naturally
arise in the supervielbein itself. But this appears in conflict with the simple 10D $\cN=1$
superspace reduction above, where it seems it must reside in a torsion tensor.

What seems to happen in Siegel's approach is that indeed the dilatino does appear as a 
dimension $\tfrac{1}{2}$ component of the torsion tensor. However, the torsion
tensor turns out to not actually be invariant under the $\lambda_A{}^B$ tangent space
transformation discussed above.
Instead, under a $\lambda_a{}^\beta$ transformation,
a dimension $\tfrac{1}{2}$ piece of the torsion tensor 
(namely, $\cT_{\alpha \beta}{}^\gamma$) transforms into
the dimension $0$ piece of the torsion tensor (namely,
$\cT_{\alpha \beta c} \propto (\gamma_c)_{\alpha\beta}$).
This transformation turns out to involve \emph{only} the spin-1/2 piece of $\lambda_a{}^\beta$.
This implies that $\cT_{\alpha \beta}{}^\gamma$ must carry a spin-1/2 field (identified
with the dilatino) that transforms inhomogeneously under the spin-1/2 part of
$\lambda_a{}^\beta$. It is essentially a compensator field for this symmetry. 
This anomalous tangent space symmetry is responsible for
imparting the non-tensorial supersymmetry transformation, as a SUSY transformation must
be accompanied by a compensating $\lambda$ gauge transformation to retain the gauge
$C^{\tA \tB} = 0$.

In formulating $\cN=1$ superspace DFT in detail, we have found it more useful to take a
different approach. We will eliminate the dilatino from the torsion tensor, fixing the
spin-1/2 part of the $\lambda_a{}^{\beta}$ transformation so that $\lambda_a{}^\beta$
is now purely spin-3/2. Now not all of $C^{\tA \tB}$
can be eliminated, and the piece that remains will be identified with the component
dilatino. While this obscures the reduction to 10D $\cN=1$ superspace (which will now require
a ``degauging'' to shift a field from the supervielbein to the torsion tensor),
it \emph{significantly} simplifies the superspace DFT
curvatures and streamlines the reduction to the component theory.\footnote{Perhaps the
most democratic approach would be to keep the full $\lambda_a{}^\beta$ symmetry and show 
how one can both reduce to 10D $\cN=1$ superspace and also to component supersymmetric DFT
depending on how the gauge is fixed.
It should be possible to take the formulation we present in this paper and ``regauge''
it to restore the full $\lambda_a{}^\beta$ symmetry by inserting a Stueckelberg field into
the torsion tensor.}

%%%%%%%%%%%%%%%%%%%%%%%%%%%%%%%%%%%%%%%%%%%%%%%%%%%%%%%%%%%%%%%%%%%%%%%%%%%%%%%%%
\section{Superspace geometry and its constraints}\label{S:SGeo}
%%%%%%%%%%%%%%%%%%%%%%%%%%%%%%%%%%%%%%%%%%%%%%%%%%%%%%%%%%%%%%%%%%%%%%%%%%%%%%%%%

Having introduced the supervielbein and its tangent space group, we can now explain
the superspace geometry and give the appropriate torsion constraints.

\subsection{Generalized fluxes and their constraints}
The supervielbein $\cV_\cM{}^\cA$ and the
superdilaton $\Phi$ transform under generalized diffeomorphisms as \eqref{E:SDFT.VPhi.trafos}.
From these fields, one can construct generalized 
fluxes $\cF_{\cA \cB \cC}$ and $\cF_\cA$ as
\begin{alignat}{3}
\Lie_{\cV_\cA} \cV_\cB{}^\cM &= -\cF_{\cA \cB}{}^\cC \cV_\cC{}^\cM
& \quad &\implies &\quad
\cF_{\cA \cB \cC} &= - 3 \,D_{[\cA} \cV_\cB{}^\cM \, \cV_{\cM \cC]}~, \eol
\Lie_{\cV_\cA}^{(+1)} \Phi &= \cF_{\cA} \Phi
& \quad &\implies &\quad
\cF_\cA &= D_\cA \log \Phi + \pa_\cM \cV_\cA{}^\cM (-)^{a m+m}~.
\end{alignat}
The flattened derivatives $D_\cA := \cV_\cA{}^\cM \pa_\cM$ here obey
\begin{align}
[D_\cA, D_\cB] = -\cF_{\cA \cB}{}^\cC D_\cC~, \qquad
D^\cA D_\cA = - \cF^\cA D_\cA~.
\end{align}
These flux tensors in turn obey the following Bianchi identities:
\begin{subequations}
\begin{align}
4 \,D_{[\cA} \cF_{\cB \cC \cD]} &= -3 \cF_{[\cA \cB|}{}^{\cE} \cF_{\cE |\cC \cD]}~, \\
2 D_{[\cA} \cF_{\cB]} &= -\cF_{\cA \cB}{}^\cC \cF_\cC - D^\cC \cF_{\cC \cA \cB}~, \\
D^\cA \cF_{\cA} &= -\frac{1}{2} \cF^\cA \cF_\cA - \frac{1}{12} \cF^{\cA \cB \cC} \cF_{\cC \cB \cA}~.
\end{align}
\end{subequations}

The fluxes themselves are not invariant under the tangent space group, which we denote by
$\cH$. This group leaves
the superdilaton invariant and acts on the supervielbein and fluxes as
\begin{align}
\delta \cV_\cA{}^\cM = \lambda_\cA{}^\cB \cV_{\cB}{}^\cM~, \quad
\delta \cF_{\cA \cB \cC} &= - 3 D_{[\cA} \lambda_{\cB \cC]} + 3 \lambda_{[\cA}{}^\cD \cF_{\cD| \cB \cC]}~, \qquad
\delta \cF_\cA = -D^\cB \lambda_{\cB \cA} - \cF^\cB \lambda_{\cB \cA}~.
\end{align}
The parameter $\lambda$ belongs to the group $\cH_L \times \g{SO}(D-1,1)_R$,
where $\cH_L$ is a subgroup of $\g{OSp}(D-1,1|2s)_L$.
As discussed in section \ref{S:SDFT.Tangent}, we restrict $\cH_L$ so that
the graded antisymmetric parameter $\lambda_{\cA \cB}$ obeys
\begin{align}
\lambda_{\alpha \,b} = \lambda_{\alpha \beta} = 0~, \qquad
\lambda_\alpha{}^\beta = \frac{1}{4} \lambda_{ab} (\gamma^{ab})_\alpha{}^\beta~, \qquad
(\gamma^a)_{\alpha \beta} \lambda_a{}^\beta = 0~.
\end{align}
In other words, the group $\cH_L$ is generated by $\g{SO}(D-1,1)$ transformations
$\lambda_{ab}$, $\gamma$-traceless fermionic transformations $\lambda_a{}^\beta$,
and symmetric $\lambda^{\alpha\beta}$ transformations. We are interested in
$D=10$ and $s=16$, which means $\lambda^{\alpha\beta}$ is reducible: it corresponds
to a vector and a self-dual 5-form.

\begin{table}[t]
\centering
\renewcommand{\arraystretch}{1.5}
\begin{tabular}{c|l}
\toprule
dimension &  fluxes \\ 
\hline
$-\frac{1}{2}$ & $\cF_{\alpha \beta \gamma} = \cT_{\alpha\beta\gamma}$ \\ 
\hline
$0$ & $\cF_{\alpha \beta c} = \cT_{\alpha \beta c} $ \\
    & $\cF_{\alpha \beta \bc} = \cT_{\alpha \beta \bc}$ \\
\hline
$\frac{1}{2}$
    & $\cF_{\alpha b \bar c} = \cT_{\alpha b \bar c}$ \\
    & $\cF_{\alpha b c}$, $\cF_{\alpha \overline{bc}}$,
    $\cF_{\alpha \beta}{}^\gamma$, $\cF_\alpha$ \\
\hline
$1$
    & $\cF_{a b c}$, $\cF_{a b \bar c}$,  $\cF_{a \overline{bc}}$, $\cF_{\overline{abc}}$ \\
    & $\cF_{\bar a}{}_\beta{}^\gamma$, $\cF_{a}{}_\beta{}^\gamma$ \\
    & $\cF_a$, $\cF_{\ba}$ \\
\hline
$\tfrac{3}{2}$
    & $\cF_{a b}{}^\gamma$,
    $\cF_{a \bar b}{}^\gamma$,
    $\cF_{\overline{ab}}{}^\gamma$ \\
    & $\cF_{\alpha}{}^{\beta \gamma}$, $\cF^\alpha$ \\
\hline
$2$
    & $\cF_{a}{}^{\beta \gamma}$, $\cF_{\bar a}{}^{\beta \gamma}$ \\
\hline
$\tfrac{5}{2}$
    & $\cF^{\alpha \beta \gamma}$ \\
\bottomrule
\end{tabular}
\captionsetup{width=0.6\textwidth}
\caption{Generalized fluxes in superspace DFT. The lowest dimension fluxes
are covariant and may be identified as torsion tensors.}
\label{T:Fluxes}
\end{table}

The generalized fluxes are grouped by dimension in Table \ref{T:Fluxes}. 
Given the restrictions placed on the
tangent space, some of the fluxes of low dimension are already gauge invariant
and can be identified immediately with invariant torsion tensors $\cT$.
Let's review the constraints imposed by Siegel \cite{Siegel:1993th}:
\begin{itemize}
\item 
The dimension $-\tfrac{1}{2}$ tensor $\cT_{\alpha\beta\gamma}$ corresponds in 10D $\cN=1$ superspace
to the lowest dimension component $H_{\alpha\beta\gamma}$ of the 3-form field strength.
This is constrained to vanish in conventional superspace, and the same choice 
should be made here.
\item At dimension 0, one encounters $\cT_{\alpha \beta c}$ and $\cT_{\alpha \beta \bc}$.
The only sensible choice is to take $\cT_{\alpha \beta c} = \kappa\, (\gamma_c)_{\alpha\beta}$,
for $\kappa$ some numerical constant, and $\cT_{\alpha\beta \bc} = 0$.
These amount to the usual dimension 0 constraints,
$T_{\alpha \beta}{}^{\rc} \propto (\gamma^\rc)_{\alpha \beta}$ and 
$H_{\alpha \beta \rc} \propto (\gamma_\rc)_{\alpha\beta}$ in 10D $\cN=1$ superspace.
Our choice of $\kappa$ is left unspecified, but later on we will explain how to
choose it to recover conventional normalizations.

\item At dimension $\tfrac{1}{2}$, the torsion tensor $\cT_{\alpha b \bc}$ can be shown to
be $\gamma$-traceless in $\alpha b$ as a consequence of the flux Bianchi identities.
This implies that $\cT_{\alpha b \bc} \propto (\gamma_{b})_{\alpha\beta} \cW_\bc{}^\beta$
for some superfield $\cW_\bc{}^\beta$.
There is no covariant dimension $\tfrac{1}{2}$ field in component supersymmetric DFT that
$\cW_\bc{}^\beta$ should be identified with, so we choose to set this field 
to zero.\footnotemark

\end{itemize}

\footnotetext{In the case of heterotic DFT, Siegel 
identified $\cW_\bc{}^\beta$ with the gauginos of the vector multiplets when 
$\bc$ extended over the $n$ vector indices. In extending our approach to heterotic
DFT, we would still keep $\cW_\bc{}^\beta$
vanishing in all cases. The gaugino then would appear in the supervielbein as part of
$\Psi_{\ba}{}^\beta$, just as it does in component supersymmetric DFT \cite{Hohm:2011nu}.
The difference with Siegel's approach is that he includes an extra tangent space symmetry that 
allows one to remove the would-be gaugino from the supervielbein, just as with the dilatino.}

We summarize here the constraints we have explicitly imposed so far:
\begin{align}\label{E:FluxConstraints}
\cT_{\alpha\beta\gamma} = 0~, \qquad
\cT_{\alpha \beta c} = \kappa (\gamma_c)_{\alpha \beta}~, \qquad
\cT_{\alpha \beta \bc} = 0~, \qquad
\cT_{\alpha b \bc} = 0~.
\end{align}
The other geometric fluxes must be augmented by tangent space connections, as in
the bosonic case, if they are to be gauge covariant.

\subsection{Tangent space connections and torsion tensors}
\label{S:SGeo.b}
Following Siegel, let us introduce a tangent space connection. We denote it $\Omega_{\cA \cB \cC}$,
using the common superspace convention of denoting superfields with capital letters.
This is to avoid confusion with the component $\omega_{\ha \hb \hc}$ connection, which will
receive gravitino corrections. The reader is cautioned to not confuse the superspace
connection $\Omega$ with the anholonomy coefficient $D_\cA \cV_\cB{}^\cM \cV_{\cM \cC}$, 
which is frequently denoted $\Omega$ in the literature.

The connection $\Omega_{\cA \cB \cC}$ has indices $\cB \cC$ valued in 
$\cH_L \times \g{SO}(D-1,1)_R$.
A conventional $\cH_L \times \g{SO}(D-1,1)_R$ connection should transform as
\begin{align}\label{E:dOmega.expected}
\delta \Omega_{\cM \, \cA\cB}
    \stackrel{?}{=} \Lie_\xi \Omega_\cM{}_{\cA\cB}
    + \pa_\cM \lambda_{\cA \cB}
    - \Omega_{\cM \, \cA}{}^\cC \lambda_{\cC \cB}
    + \Omega_{\cM \, \cB}{}^\cC \lambda_{\cC \cA} (-)^{ab}
\end{align}
but we are going to find that this transformation rule is deformed by terms
explicitly involving the DFT vielbein $\cV_\cM{}^\cA$. These deformations arise because
we will be choosing the connections so that higher dimension torsion tensors vanish.
Let us see what happens explicitly as we analyze the first few cases.

Define the torsion tensors so that
\begin{alignat}{2}
\label{E:Torsion}
\cT_{\cA \cB \cC} &:= - 3 \,\cD_{[\cA} \cV_\cB{}^\cM \, \cV_{\cM \cC]}  & 
    &= \cF_{\cA \cB \cC} + 3\, \Omega_{[\cA \cB \cC]}~, \\
\label{E:DilTorsion}
\cT_\cA &:= \cD_\cA \log \Phi + \cD_\cM \cV_\cA{}^\cM & 
    &= \cF_\cA + \Omega^\cB{}_{\cB \cA}~.
\end{alignat}
At dimension $\tfrac{1}{2}$, there are four terms:
$\cT_{\alpha b c}$, $\cT_{\alpha \overline{bc}}$, $\cT_{\alpha \beta}{}^\gamma$, and $\cT_\alpha$.
Let's focus on the first two. It is easy to see from their definitions that they can be set
to zero by appropriately choosing $\Omega_{\alpha bc}$ and $\Omega_{\alpha \overline{bc}}$,
\begin{align}
\cT_{\alpha b c} := \cF_{\alpha b c} + \Omega_{\alpha b c} \equiv 0~, \qquad
\cT_{\alpha \overline{b c}} := \cF_{\alpha b c} + \Omega_{\alpha \overline{b c}} \equiv 0~.
\end{align}
The second constraint is sensible in the sense that the variation of
$\cF_{\alpha \overline{bc}}$ implies a conventional transformation for $\Omega_{\alpha \overline{bc}}$.
Explicitly,
\begin{align}
\delta \cF_{\alpha \overline{b c}} &= -D_\alpha \lambda_{\overline{bc}} 
    + \lambda_\alpha{}^\beta \cF_{\beta \overline{b c}}
    - \lambda_\bb{}^\bd \cF_{\beta \overline{d c}}
    - \lambda_\bc{}^\bd \cF_{\beta \overline{b d}} \quad \implies \quad \eol
\delta \Omega_{\alpha \overline{b c}} &= D_\alpha \lambda_{\overline{bc}} 
    + \lambda_\alpha{}^\beta \Omega_{\beta \overline{b c}}
    - \lambda_\bb{}^\bd \Omega_{\alpha \overline{d c}}
    - \lambda_\bc{}^\bd \Omega_{\alpha \overline{b d}} ~.
\end{align}
However, the first constraint leads to an anomalous $\kappa$-dependent term
\begin{align}
\delta \cF_{\alpha b c} &= -D_\alpha \lambda_{bc} 
    + \lambda_\alpha{}^\beta \cF_{\beta b c}
    + \lambda_b{}^d \cF_{\alpha d c}
    + \lambda_c{}^d \cF_{\alpha b d}
    + 2 \cF_{\alpha \gamma [b} \lambda_{c]}{}^\gamma \quad \implies \quad \eol
\delta \Omega_{\alpha b c} &=
     D_\alpha \lambda_{bc}
    + \lambda_\alpha{}^\beta \Omega_{\beta b c}
    + \lambda_b{}^d \Omega_{\alpha d c}
    + \lambda_c{}^d \Omega_{\alpha b d}
     - 2\kappa \,(\gamma_{[b})_{\alpha \gamma} \lambda_{c]}{}^\gamma ~,
\end{align}
where we have used the explicit form of $\cF_{\alpha \gamma b}$.

There is a very simple reason for this. The natural $\cH_L$ algebra action
on covariant derivatives $\cD_\cA$ is 
$\delta_\lambda \cD_\cA = \lambda_\cA{}^\cB \cD_\cB$. For the torsion tensor,
this means that
$\delta_\lambda \cT_{\cA \cB \cC} = 3 \, \lambda_{[\cA|}{}^\cD \cT_{\cD| \cB \cC]}$.
But this implies for a transformation involving the dimension $\tfrac{1}{2}$ parameter
$\lambda_a{}^\beta$, that $\cT_{\alpha b c}$ should transform into the dimension zero
$\cT_{\alpha \beta c}$. The latter is non-vanishing. Requiring the former to vanish is
inconsistent \emph{unless} the $\lambda$ transformations are modified. We will see that
(as for $\Omega_{\alpha b c}$), we can consistently modify the transformations of the 
$\Omega$ connections to accommodate this.

The other torsion tensors at dimension $\tfrac{1}{2}$ are
\begin{align}
\cT_{\alpha \beta}{}^\gamma
    = \cF_{\alpha \beta}{}^\gamma
    + 2 \,\Omega_{(\alpha \beta)}{}^\gamma~, \qquad
\cT_{\alpha} = \cF_\alpha - \Omega_{\beta \alpha}{}^\beta
\end{align}
These involve a spin connection that has already been defined, since
$\Omega_{\alpha \beta}{}^\gamma = \tfrac{1}{4} \Omega_{\alpha b c} (\gamma^{bc})_\beta{}^\gamma$.
Using the variation of $\Omega_{\alpha bc}$ defined above, the $\lambda$
transformation of $\cT_{\alpha \beta}{}^\gamma$, up to covariant terms involving $\lambda_{ab}$, becomes
\begin{align}
\Delta_\lambda \cT_{\alpha \beta}{}^\gamma
    = \kappa\, \delta_{(\alpha}{}^\gamma (\gamma^c \lambda_c)_{\beta)}
    - \frac{1}{2} \kappa\, (\gamma_b)_{\alpha \beta} (\gamma^b \gamma^c \lambda_c)^\gamma~.
\end{align}
For convenience, we have suppressed indices for sequential spinor contractions, so that
e.g. 
$(\gamma^b \gamma^c \lambda_c)^\gamma = 
    (\gamma^b)^{\gamma \delta} (\gamma^c)_{\delta \epsilon} \lambda_c{}^\epsilon$.
This variation cancels out when $\lambda_c{}^\gamma$ is $\gamma$-traceless. If this assumption
is not made, then one must introduce a spin-1/2 field to populate $\cT_{\alpha\beta}{}^\gamma$.
As we have already mentioned, this field would ultimately play the role of the dilatino.
As we have elected to keep the dilatino within the supervielbein, we will take
$\lambda_c{}^\alpha$ to be purely spin-3/2 without further comment.
It follows that $\Delta_\lambda \cT_{\alpha \beta}{}^\gamma$ vanishes.
Similar comments pertain to $\cT_\alpha$, whose variation is
$\Delta_\lambda \cT_\alpha = -\tfrac{9}{2} \kappa \,(\gamma^c \lambda_c)_{\alpha}$,
which similarly vanishes.

Ultimately we will require both $\cT_{\alpha \beta}{}^\gamma$ and $\cT_\alpha$
to vanish as a constraint on the geometry
(as there is no covariant dimension $\tfrac{1}{2}$ quantity at the component level to
identify them with), but for now, we will leave them unspecified.
At this point we have defined all dimension $\tfrac{1}{2}$ components
of $\Omega$.

Let's move on to dimension 1 torsions. There are four torsions that are familiar
from bosonic DFT and we constrain them in the same way:
\begin{alignat}{2}
\cT_{a b c} &= \cF_{a b c} + 3 \,\Omega_{[a b c]} \equiv 0~, &\qquad
\cT_{\overline{abc}} &= \cF_{\overline{abc}} + 3 \,\Omega_{[\overline{a b c}]} \equiv 0~, \eol
\cT_{\ba b c} &= \cF_{\ba b c} + \Omega_{\ba b c} \equiv 0~, &\qquad
\cT_{a \overline{bc}} &= \cF_{a \overline{bc}} + \Omega_{a \overline{bc}} \equiv 0~.
\end{alignat}
As in the bosonic case, this defines only the totally antisymmetric parts of
$\Omega_{abc}$ and $\Omega_{\overline{abc}}$.
It implies the gauge transformations
\begin{alignat}{2}
\Delta_\lambda \Omega_{[a b c]} &= D_{[a} \lambda_{b c]} + \lambda_{[a}{}^\gamma \Omega_{\gamma bc]}~, 
    &\quad
\Delta_\lambda \Omega_{[\overline{abc}]} &= D_{[\bar a} \lambda_{\overline{bc}]}~, \eol
\Delta_\lambda\Omega_{a \overline{bc}} &= D_{a} \lambda_{\overline{bc}} 
    + \lambda_{a}{}^\gamma \Omega_{\gamma \overline{bc}}~, &\quad
\Delta_\lambda \Omega_{\bar a bc} &= D_{\bar a} \lambda_{bc}~
\end{alignat}
where again we use $\Delta_\lambda$ to denote $\lambda$ transformations up to
covariant $\lambda_{ab}$ or $\lambda_{\overline{ab}}$ terms.

Two more dimension 1 torsions involve only the connections $\Omega_{\ba b c}$
and $\Omega_{\overline{a\,bc}}$:
\begin{align}
\cT_{\bar a} &= \cF_{\bar a} - \Omega^{\bb}{}_{\overline{ba}} \equiv 0~, \qquad
\cT_{\bar a}{}_\beta{}^\gamma
    = \cF_{\bar a}{}_\beta{}^\gamma + \Omega_{\bar a}{}_\beta{}^\gamma~.
\end{align}
The first can be set to zero by fixing the trace of the $\g{SO}(9,1)_R$ spin connection.
The second torsion cannot be constrained, but we can check that its non-covariant $\lambda$
transformation vanishes, $\Delta_\lambda \cT_{\bar a}{}_\beta{}^\gamma = 0$. Again,
this is consistent with it later being fixed to zero as a geometric constraint.

The remaining dimension 1 torsions both involve the additional $\cH_L$ connection
$\Omega_{\alpha b}{}^\gamma$:
\begin{align}
\cT_a = \cF_a + \Omega^b{}_{b a} - \Omega_\beta{}^\beta{}_a, \qquad
\cT_{a}{}_\beta{}^\gamma
    = \cF_a{}_\beta{}^\gamma + \Omega_a{}_\beta{}^\gamma
    - \Omega_\beta{}_a{}^\gamma~.
\end{align}
The first torsion can be related to the second by defining 
\begin{align}
\Omega^b{}_{ba} := -\cF_a - \cF_{a \beta}{}^\beta \quad \implies \quad
\cT_a = -\cT_{a \beta}{}^\beta
\end{align}
The second torsion can be used to define $\Omega_\alpha{}_b{}^\beta$ if we
suitably project onto the spin-3/2 part,
\begin{align}
\Omega_{\alpha b}{}^\beta := (\cF_{c \alpha}{}^\gamma + \Omega_{c \alpha}{}^\gamma) 
    (\delta_b{}^c \delta_\gamma{}^\beta - \tfrac{1}{10} (\gamma^c \gamma_b)_\gamma{}^\beta )
    \quad \implies \quad
\cT_{a}{}_\beta{}^\gamma = \tfrac{1}{10}\,(\gamma_a)^{\gamma \alpha} \cT_{\alpha \beta}
\end{align}
where $\cT_{\alpha\beta}$ has no particular symmetry property for its two indices.

Something a little bit subtle is occurring above. We are using $\Omega_{c a b}$ to define
$\Omega_{\alpha b}{}^\beta$ even though not all of $\Omega_{c a b}$ is actually defined.
Another way of saying this is that the definitions of $\Omega_{c a b}$ we have made allow
us to assign it a transformation
\begin{align}
(\delta_\Lambda + \Delta_\lambda) \Omega_{a b c} &= D_{a} \lambda_{b c} + \lambda_{a}{}^\gamma \Omega_{\gamma bc}
    - \tfrac{2}{9} \,\kappa\, \eta_{a [b} (\gamma_{c]})_{\alpha \beta} \lambda^{\alpha \beta}
    + \Lambda_{a|bc}
\end{align}
where $\Lambda_{a|bc}$ parametrizes the ambiguity in its definition. This same ambiguity
now appears in $\Omega_{\beta a}{}^\gamma$, which transforms as
\begin{align}
(\delta_\Lambda + \Delta_\lambda) \Omega_{\beta a}{}^\gamma &=
    D_\beta \lambda_a{}^\gamma
    + \Omega_\beta{}^{bc} \Big(
    \tfrac{1}{4} \lambda_a{}^\alpha (\gamma_{bc})_\alpha{}^\gamma
    - \eta_{a b} \lambda_c{}^\gamma
    \Big)
    \eol & \quad
%     - \kappa (\gamma_a)_{\beta \delta} \lambda^{\delta \gamma}
%     + \tfrac{1}{18} \kappa\, (\gamma_{a b})_\beta{}^\gamma (\gamma^b)_{\delta \epsilon} \lambda^{\delta \epsilon}
    + \kappa (\gamma_a)_{\beta \delta} \lambda^{\delta \gamma}
    - \tfrac{1}{18} \kappa\, (\gamma_{a b})_\beta{}^\gamma (\gamma^b)_{\delta \epsilon} \lambda^{\delta \epsilon}
    + \tfrac{1}{4} (\gamma^{b c})_\beta{}^\gamma \Lambda_{a|bc}~.
\end{align}
All dimension 1 components of $\Omega$ are now defined up to the irreducible
hook representation ambiguities $\Lambda_{a|bc}$ and $\Lambda_{\overline{a|bc}}$.

Next we turn to the dimension $\tfrac{3}{2}$ torsions:
\begin{subequations}
\begin{align}
\cT_{\overline{ab}}{}^\alpha &= \cF_{\overline{ab}}{}^\alpha + \Omega^\alpha{}_{\overline{ab}}~, \\
\cT_{a \bb}{}^\alpha &= \cF_{a \bb}{}^\alpha - \Omega_{\bb a}{}^\alpha~, \\
\cT_{a b}{}^\alpha &= \cF_{a b}{}^\alpha + \Omega^{\alpha}{}_{ab} + 2\, \Omega_{[a b]}{}^\alpha~, \\
\cT_{\alpha}{}^{\beta \gamma} 
    &= \cF_{\alpha}{}^{\beta \gamma}
    + \Omega_\alpha{}^{\beta \gamma} 
    + \tfrac{1}{2} (\gamma^{b c})_\alpha{}^{(\beta} \Omega^{\gamma)}{}_{bc}~, \\
\cT^\alpha
    &= \cF^\alpha + \Omega^b{}_{b}{}^\alpha
    + \tfrac{1}{4} (\gamma^{bc})_\beta{}^\alpha \Omega^\beta{}_{bc}
    - \Omega_\beta{}^{\beta \alpha}
\end{align}
\end{subequations}
The first four equations lead to natural definitions for the $\Omega$'s involved:
\begin{subequations}
\begin{alignat}{2}
\Omega^\alpha{}_{\overline{ab}} &:= -\cF_{\overline{ab}}{}^\alpha \quad &\implies \quad
\cT_{\overline{ab}}{}^\alpha &\equiv 0~, \\
\Omega_{\bb\, a}{}^\alpha &:= - \cF_{\bb  b}{}^\beta \Big(
    \delta_a{}^b \delta_\beta{}^\alpha - \tfrac{1}{10} (\gamma^b\gamma_a)_\beta{}^\alpha
\Big) \quad &\implies \quad
\cT_{a \bb}{}^\alpha &\equiv \tfrac{1}{10} \,(\gamma_{a})^{\alpha \beta} \cT_{\beta \bb}~, \\
\Omega^\alpha{}_{ab} &:=
    - \cF_{a b}{}^\alpha - 2\, \Omega_{[a b]}{}^\alpha \quad &\implies \quad
\cT_{a b}{}^\alpha &\equiv 0    ~, \\
\Omega_\alpha{}^{\beta \gamma} &:=
    - \cF_{\alpha}{}^{\beta \gamma}
    - \tfrac{1}{2} (\gamma^{b c})_\alpha{}^{(\beta} \Omega^{\gamma)}{}_{bc}
\quad &\implies \quad
\cT_{\alpha}{}^{\beta \gamma} &\equiv 0
\end{alignat}
\end{subequations}
Remarkably, there is no constraint on $\Omega_{a b}{}^\gamma$, which remains
\emph{completely undetermined}. These choices lead to the transformation rules
\begin{subequations}
\begin{align}
(\delta_\Lambda + \Delta_\lambda) \Omega^\alpha{}_{ab} &=
    D^\alpha \lambda_{ab}
    + \lambda^{\alpha \gamma} \Omega_{\gamma a b}
    + \lambda^{\alpha c} \Omega_{c a b}
    - 2 \,\Lambda_{[a,b]}{}^\alpha~, \\
(\delta_\Lambda + \Delta_\lambda) \Omega_{a b}{}^\beta &=
    D_a \lambda_b{}^\beta + \lambda_a{}^\gamma \Omega_{\gamma b}{}^\beta
    + \Omega_a{}^{cd} \Big(
        \tfrac{1}{4} \,\lambda_b{}^\gamma (\gamma_{cd})_\gamma{}^\beta
        - \eta_{b c} \lambda_d{}^\beta
    \Big)
    + \Lambda_{a, b}{}^\beta~, \\
(\delta_\Lambda + \Delta_\lambda) \Omega_\alpha{}^{\beta \gamma} &=
    D_\alpha \lambda^{\beta \gamma}
    + \tfrac{1}{2} \Omega_{\alpha b c}\, \lambda^{\delta (\beta} (\gamma^{b c})_\delta{}^{\gamma)}
    + 2 \,\Omega_{\alpha c}{}^{(\gamma} \lambda^{c \beta)}
    + (\gamma^{b c})_\alpha{}^{(\beta} \Lambda_{b,c}{}^{\gamma)}
\end{align}
\end{subequations}
where $\Lambda_{a,b}{}^\beta$ is $\gamma$-traceless in the last two indices.
This parametrizes the total ambiguity in the definition of
$\Omega_{a b}{}^\beta$. While it could be used to absorb all the other terms in 
$\delta \Omega_{a b}{}^\beta$, we have chosen to define $\delta \Omega_{a b}{}^\beta$
to match the patterns that have been exhibited so far. 

The remaining torsions at dimension 2 and dimension $\tfrac{5}{2}$ are
\begin{align}
\cT_{a}{}^{\beta \gamma} = \cF_a{}^{\beta \gamma} + \Omega_a{}^{\beta \gamma} 
    - 2 \,\Omega^{(\beta}{}_a{}^{\gamma)}~, \qquad
\cT_{\ba}{}^{\beta \gamma} = \cF_\ba{}^{\beta \gamma} + \Omega_\ba{}^{\beta \gamma} ~, \qquad
\cT^{\alpha \beta \gamma} = \cF^{\alpha \beta \gamma} + 3 \,\Omega^{(\alpha \beta \gamma)}~.
\end{align}
These suggest the definitions
\begin{subequations}
\begin{alignat}{2}
\Omega_a{}^{\beta \gamma} &:= -\cF_a{}^{\beta \gamma} + 2 \,\Omega^{(\beta}{}_a{}^{\gamma)}
\quad &\implies \quad
\cT_a{}^{\beta \gamma} &\equiv 0~, \\
\Omega_\ba{}^{\beta \gamma} &:= -\cF_\ba{}^{\beta \gamma}
\quad &\implies \quad
\cT_\ba{}^{\beta \gamma} &\equiv 0~, \\
\Omega^{(\alpha \,\beta \gamma)} &:= -\tfrac{1}{3} \cF^{\alpha \beta\gamma}
\quad &\implies \quad
\cT^{\alpha \beta \gamma} &\equiv 0~.
\end{alignat}
\end{subequations}
The last definition leaves the (irreducible) hook representation in $\Omega^{\alpha\, \beta\gamma}$
unfixed. These results lead to the transformation rules
\begin{subequations}
\begin{align}
(\delta_\Lambda + \Delta_\lambda)  \Omega_\ba{}^{\beta \gamma}
    &= D_\ba \lambda^{\beta \gamma}
    + \tfrac{1}{2} \Omega_{\ba b c} \lambda^{\delta (\beta} (\gamma^{bc})_\delta{}^{\gamma)}
    + 2 \,\Omega_{\ba c}{}^{(\gamma} \lambda^{c \beta)}~, \\
(\delta_\Lambda + \Delta_\lambda) \Omega_a{}^{\beta \gamma}
    &= D_a \lambda^{\beta \gamma}
    + \lambda_a{}^\alpha \Omega_\alpha{}^{\beta \gamma}
    + \tfrac{1}{2} \Omega_{a b c} \lambda^{\delta (\beta} (\gamma^{bc})_\delta{}^{\gamma)}
    + 2 \,\Omega_{a c}{}^{(\gamma} \lambda^{c \beta)}
    + 2 \,\Lambda^{(\beta}{}_{,b}{}^{\gamma)}~, \\   
(\delta_\Lambda + \Delta_\lambda) \Omega^{\alpha \beta \gamma}
    &= D^\alpha \lambda^{\beta \gamma}
    + \lambda^{\alpha \cD} \Omega_\cD{}^{\beta \gamma}
    + \tfrac{1}{2} \Omega^\alpha{}_{b c} \lambda^{\delta (\beta} (\gamma^{bc})_\delta{}^{\gamma)}
    + 2 \,\Omega^\alpha{}_{c}{}^{(\gamma} \lambda^{c \beta)}
    + \Lambda^{\alpha|\beta \gamma}~, \\
(\delta_\Lambda + \Delta_\lambda) \Omega^\alpha{}_b{}^\gamma
    &= D^\alpha \lambda_b{}^\gamma
    + \lambda^{\alpha \cD} \Omega_{\cD b}{}^\gamma
    + \Omega^{\alpha c d} \Big(
        \tfrac{1}{4} \lambda_b{}^\beta (\gamma_{cd})_\beta{}^\gamma
        - \eta_{b c} \lambda_d{}^\gamma
    \Big)
    + \Lambda^\alpha{}_{,b}{}^\gamma~,
\end{align}
\end{subequations}
where the hook representation
$\Lambda^{\alpha|\beta\gamma}$ (with $\Lambda^{(\alpha|\beta\gamma)} = 0$)
parametrizes the ambiguity in $\Omega^{\alpha \,\beta \gamma}$.

This completes our analysis of the tangent space $\Omega$ connections. Up to
the ambiguities parametrized by the $\Lambda$ parameters, we have determined each
of these connections. Their transformation rules can now be summarized in a uniform
way, where we take $\Omega_{\cM \cA \cB} := \cV_\cM{}^\cC \Omega_{\cC \cA \cB}$:
\begin{subequations}\label{E:OmegaTrafo}
\begin{align}
(\delta_\Lambda + \Delta_\lambda) \Omega_{\cM \,\overline{bc}}
    &= \pa_\cM \lambda_{\overline{bc}}
    + \cV_\cM{}^\ba \Lambda_{\ba|\overline{bc}}~, \\
(\delta_\Lambda + \Delta_\lambda) \Omega_{\cM \,bc}
    &= \pa_\cM \lambda_{bc}
    - 2\kappa\, \cV_\cM{}^\alpha \,(\gamma_{[b})_{\alpha \gamma} \lambda_{c]}{}^\gamma
    + \cV_\cM{}^{a} \Big(
    - \tfrac{2}{9} \kappa \,\eta_{a [b} (\gamma_{c]})_{\beta \gamma} \lambda^{\beta \gamma}
    + \Lambda_{a|bc}
    \Big)
    \eol & \quad
    - 2\, \cV_\cM{}_\alpha \Lambda_{[b,c]}{}^\alpha~, \\
(\delta_\Lambda + \Delta_\lambda) \Omega_{\cM \,b}{}^\gamma
    &= \pa_\cM \lambda_b{}^\gamma
    + \Omega_\cM{}^{c d} \Big(
        \tfrac{1}{4} \lambda_b{}^\alpha (\gamma_{cd})_\alpha{}^\gamma
        - \eta_{b c} \lambda_d{}^\gamma
    \Big)
    \eol & \quad
    + \cV_\cM{}^\alpha \Big(
        \kappa (\gamma_b)_{\alpha \delta} \lambda^{\delta \gamma}
        - \tfrac{1}{18} \kappa (\gamma_{b c})_\alpha{}^\gamma (\gamma^c)_{\delta \epsilon} \lambda^{\delta \epsilon}
        + \tfrac{1}{4} (\gamma^{cd})_\alpha{}^\gamma \Lambda_{b|cd}
    \Big)
    \eol & \quad
    + \cV_\cM{}^a \Lambda_{a,b}{}^\gamma
    + \cV_\cM{}_\alpha \Lambda^\alpha{}_{,b}{}^\gamma~, \\
(\delta_\Lambda + \Delta_\lambda) \Omega_\cM{}^{\beta \gamma}
    &= \pa_\cM \lambda^{\beta \gamma}
    + \tfrac{1}{2} \Omega_\cM{}_{b c} \lambda^{\delta (\beta} (\gamma^{bc})_\delta{}^{\gamma)}
    + 2 \,\Omega_\cM{}_{c}{}^{(\gamma} \lambda^{c \beta)}
    \eol & \quad
    + \cV_\cM{}^\alpha (\gamma^{b c})_\alpha{}^{(\beta} \Lambda_{b,c}{}^{\gamma)}
    + 2 \,\cV_{\cM}{}^a \Lambda^{(\beta}{}_{,a}{}^{\gamma)}
    + \cV_{\cM \alpha} \Lambda^{\alpha|\beta \gamma}~.
\end{align}
\end{subequations}
The key features of these transformations are as follows. The terms involving $\lambda$
that are $\kappa$-independent are nothing more than the expected $\g{OSp}(9,1|32)$
transformations given in \eqref{E:dOmega.expected}. These include the inhomogeneous
$\pa_\cM \lambda$ terms and the terms where $\Omega$ is transformed into another $\Omega$.
The remaining terms, which are $\kappa$-dependent or involve the additional $\Lambda$ terms,
correspond to rotations of $\Omega$ into the supervielbein. This look like shift symmetries
when $\Omega_\cM$ is written with a tangent space index.

At this point, there seems no reason to take the $\Lambda$ parameters seriously as local
transformations. So far they are only parametrizing our ignorance of the $\Omega$ connections.
But we will now check that the $\Lambda$ transformations are actually required for closure
of the gauge transformations.
To condense the notation, we will employ BRST terminology, and give the algebra
of gauge transformations with the nilpotent BRST operator $\delta$ acting on (graded) anti-commuting
parameters $\lambda$, so that, for $[\delta_1, \delta_2] = \delta_{12}$, the closure
of $\g{SO}(9,1)_L$ transformations
\begin{align}
\lambda_{12}{}_{a b} = - \lambda_1{}_{a}{}^c \lambda_2{}_{c b} 
    + \lambda_2{}_{a}{}^c \lambda_1{}_{c b} 
\end{align}
becomes a more compact expression
\begin{align}
\delta \lambda_{ab} = - \lambda_{a}{}^c \lambda_{c b}~.
\end{align}
The other tangent space algebraic relations are encoded in
\begin{align}
\delta \lambda_a{}^\beta =
    - \lambda_{a}{}^c \lambda_{c}{}^{\beta}
    - \lambda_{a}{}^{\gamma} \lambda_\gamma{}^\beta,\qquad
\delta \lambda^{\alpha \beta}
    = \lambda^{b}{}^{\alpha} \lambda_{b}{}^{\beta}
    - 2 \lambda^{\gamma (\alpha} \lambda_\gamma{}^{\beta)}~.
\end{align}
These expressions are just the explicit decomposition of
$\delta \lambda_{\cA \cB} = -\lambda_\cA{}^\cC \lambda_{\cC \cB}$.
More interesting are the closure conditions on the additional $\Lambda$ symmetries. 
For example, we find for $\Lambda_{a|bc}$ that
\begin{align}
\delta \Lambda_{a|bc} &=
2\kappa \,\lambda_{a}{}^{\alpha} (\gamma_{[b})_{\alpha \beta} \lambda_{c]}{}^{\beta}
+ \tfrac{2}{9}\kappa \,\eta_{a [b} (\gamma_{c]})_{\alpha \beta} \lambda^{d}{}^{\alpha} \lambda_{d}{}^{\beta}
- \lambda_{a}{}^{d} \Lambda_{d| b c}
- \lambda_{b}{}^{d} \Lambda_{a| d c}
- \lambda_{c}{}^{d} \Lambda_{a| b d}~.
\end{align}
This enforces that the closure of $\lambda_a{}^\alpha$ transformations requires a
$\Lambda_{a|bc}$ transformation.
Similarly, for the other $\Lambda$ parameters we find (writing $\Delta$ to suppress
the obvious $\lambda_a{}^b$ contributions to these terms)
\begin{subequations}
\begin{align}
\Delta \Lambda_{b\,a}{}^\beta &=
    \kappa (\gamma_{a})_{\alpha \gamma} \lambda^{\beta \alpha} \lambda_{b}{}^{\gamma} 
    - \tfrac{1}{9} \kappa\, (\gamma_{a})_{\alpha \gamma} \lambda^{\alpha \gamma} \lambda_{b}{}^{\beta} 
    + \tfrac{1}{9} \kappa \,\eta_{b a}\, (\gamma_{c})_{\alpha \gamma} \lambda^{\alpha \gamma} \lambda_{c}{}^{\beta} 
    \eol & \quad
    - \tfrac{1}{18} \kappa (\gamma^{c})_{\alpha \gamma} (\gamma_{b c})_{\delta}{}^{\beta} 
        \lambda^{\alpha \gamma} \lambda_{a}{}^{\delta} 
    - \tfrac{1}{18} \kappa\, (\gamma^{c})_{\alpha \gamma} (\gamma_{a c})_{\delta}{}^{\beta} 
        \lambda^{\alpha \gamma} \lambda_{b}{}^{\delta} 
    \eol & \quad
    + \tfrac{1}{4} \Lambda_{b| c d} (\gamma^{c d})_{\alpha}{}^{\beta} \lambda_{a}{}^{\alpha} 
    + \tfrac{1}{4} \Lambda_{a| c d} (\gamma^{c d})_{\alpha}{}^{\beta} \lambda_{b}{}^{\alpha} 
    - \Lambda_{b| a c} \lambda^{c}{}^{\beta}~, \\[1ex]
\Delta \Lambda^\alpha{}_a{}^\beta &=
-\kappa \,\lambda^{\alpha \gamma} (\gamma_{a})_{\gamma \delta} \,\lambda^{\delta\beta} 
+ \tfrac{1}{18} \kappa  
    \lambda^{\alpha \epsilon} (\gamma_{a b})_{\epsilon}{}^{\beta} (\gamma^{b})_{\gamma \delta} \lambda^{\gamma \delta} 
    \eol & \quad
    + \Lambda_{b a}{}^{\beta} \lambda^{b}{}^{\alpha} 
    + \Lambda_{a b}{}^{\alpha} \lambda^{b}{}^{\beta} 
    - \Lambda_{b a}{}^{\alpha} \lambda^{b}{}^{\beta} 
    - \tfrac{1}{2} \Lambda_{b c}{}^{\alpha} \lambda_{a}{}^{\gamma}  (\gamma^{b c})_{\gamma}{}^{\beta} 
    \eol & \quad
    + \tfrac{1}{4} \Lambda_{a| b c} \,\lambda^{\alpha \gamma} \,(\gamma^{b c})_{\gamma}{}^{\beta} ~, \\[1ex]
\Delta \Lambda^{\gamma|\alpha \beta} &=
- \Lambda_{a b}{}^{\gamma} (\gamma^{a b})_{\delta}{}^{(\alpha} \lambda^{\beta) \delta} 
+ \Lambda_{a b}{}^{(\alpha} (\gamma^{a b})_{\delta}{}^{\beta)} \lambda^{\gamma \delta} 
+ 2 \,\Lambda^{\gamma}{}_{a}{}^{(\alpha} \lambda^{a}{}^{\beta)}
- 2 \,\Lambda^{(\alpha}{}_{a}{}^{\beta)} \lambda^{a}{}^{\gamma} ~.
\end{align}
\end{subequations}
The point is that each of the additional $\Lambda$ transformations becomes necessary
for closure when $\kappa \neq 0$. Moreover, we find no additional symmetry
constraints on these parameters aside from $\gamma$-tracelessness of $\Lambda_{a \,b}{}^\beta$
and $\Lambda^\alpha{}_b{}^\beta$ on their $b\beta$ indices, and the vanishing of the totally
symmetric part of $\Lambda^{\alpha|\beta\gamma}$.

\begin{table}[t]
\centering
\renewcommand{\arraystretch}{1.5}
\begin{tabular}{c|c|c}
\toprule 
dimension & conventional constraints & remaining torsions\\ \hline
$-\frac{1}{2}$ &
$-$ & 
$\cT_{\alpha\beta\gamma} = 0$ \\
\hline
$0$ &
$-$ &
$\cT_{\alpha \beta c} = \kappa (\gamma_c)_{\alpha\beta}$ \,\,
$\cT_{\alpha \beta \bc} = 0$ \\
\hline
$\frac{1}{2}$
    & $\cT_{\alpha b c} = \cT_{\alpha \overline{bc}} = 0$
    & $\cT_{\alpha b \bc} = 0$ \\
    && $\cT_{\alpha \beta}{}^\gamma$, $\cT_\alpha$ \\
\hline
$1$ &
$\cT_{\ha \hb \hc} = \cT_\ha = 0$ & $\cT_\ba{}_\beta{}^\gamma$ \\
 & $\cT_{a}{}_\beta{}^\gamma = \tfrac{1}{10}\,(\gamma_a)^{\gamma \alpha} \cT_{\alpha \beta}$ & 
\\ \hline
$\tfrac{3}{2}$ &
$\cT_{\overline{ab}}{}^\alpha = \cT_{a b}{}^\alpha = \cT_{\alpha}{}^{\beta \gamma} = 0$ 
    & $\cT^\alpha$ \\
   & $\cT_{a \bb}{}^\gamma = \tfrac{1}{10}\, (\gamma_a)^{\gamma \alpha} \cT_{\alpha \bb}$ 
\\\hline
$2$ & 
$\cT_{\ha}{}^{\beta \gamma} = 0$ &
$-$ 
\\ \hline
$\tfrac{5}{2}$ &
$\cT^{\alpha \beta \gamma} = 0$ &
$-$ \\
\bottomrule
\end{tabular}
\captionsetup{width=0.8\textwidth}
\caption{Conventional and physical constraints on torsion.
Conventional constraints arise from
a specific choice of $\Omega_{\cA \cB \cC}$. Some of the remaining torsions 
have additional physical constraints imposed.
}
\label{T:TConstraints}
\end{table}

We will need to say a bit more about the structure of these additional transformations,
but for now let us return to comment on the torsions that have been set to zero.
These are summarized in Table \ref{T:TConstraints}.
In the middle column, we have listed all the constraints that are purely conventional,
meaning they arise from a specific choice of the $\Omega$ connection. The torsions in
the right column remain. Some of these have physical constraints imposed to eliminate
covariant objects in the torsion.
These are true constraints on the supervielbein, as opposed to conventional choices of $\Omega$.

The structure of the unfixed torsions can be understood by considering the
linearized supervielbein. Writing $\cV_\cM{}^\cA = \delta_\cM{}^\cA+ \varV_{\cM \cB} \,\eta^{\cB \cA} (-1)^b$
for $\varV_{\cB \cA}$ a graded antisymmetric quantity, we can see that $\varV$ transforms as
$\delta\varV_{\cB \cA} = 2\,\pa_{[\cB} \xi_{\cA]} - \lambda_{\cB \cA}$
under linearized diffeomorphisms and tangent space transformations.
Using $\lambda$ as much as possible to fix $\varV$, we find the following surviving
components. First are the negative dimension fields, $\varV_{\beta \alpha}$,
$\varV_{\beta a}$, and $\varV_{\beta \ba}$,  and the 
zero-dimension field $\varV_\alpha{}^\beta$.
These transform as
\begin{gather}
\delta \varV_{\beta \alpha} = - 2\, \pa_{(\beta} \xi_{\alpha)}~, \quad
\delta \varV_{\beta a} = \pa_\beta \xi_a + \pa_a \xi_\beta~, \quad
\delta \varV_{\beta \ba} = -\pa_\beta \xi_\ba + \pa_\ba \xi_\beta~, \eol
\delta \varV_{\beta}{}^\alpha = \pa_\beta \xi^\alpha - \pa^\alpha \xi_\beta
    - \tfrac{1}{2} (\gamma^{cd})_\beta{}^\alpha \pa_c \xi_d~.
\label{E:linearizedHTrafo1}
\end{gather}
The remaining fields are analogous to what we find at the component level.
There is the zero-dimension bosonic DFT vielbein $\varV_{\bb a}$, the gravitino
$\varV_\bb{}^\alpha \equiv \Psi_\bb{}^\alpha$, and the dilatino
$\varV_b{}^\alpha \equiv \tfrac{1}{10} (\gamma_b)^{\alpha \beta} \chi_\beta$.
Their transformation rules are\footnotemark
\begin{align}
\delta \varV_{\bb a} = \pa_\bb \xi_a + \pa_a \xi_\bb ~, \qquad
\delta \Psi_\bb{}^\alpha = \pa_\bb \xi^\alpha + \pa^\alpha \xi_\bb~, \qquad
\delta \chi_\alpha = (\gamma^b)_{\alpha \beta} (\pa_b \xi^\beta - \pa^\beta \xi_b)~.
\end{align}
Finally, there is the linearized superdilaton $\Phi = 1 + \varphi$,
transforming as
\begin{align}
\delta \varphi = \pa_a \xi^a + \pa_\ba \xi^\ba - \pa_\alpha \xi^\alpha - \pa^\alpha \xi_\alpha~.
\end{align}

\footnotetext{When reducing to components
in section \ref{S:Components},
we will solve the fermionic pieces of the section condition by choosing $\pa^\alpha = 0$
to eliminate $\tilde\theta$ dependence.
Then the linearized gravitino and the dilatino acquire their usual transformations.
The physical linearized dilaton will presumably be the combination $\varphi + \varV_\alpha{}^\alpha$. The other low dimension fields in \eqref{E:linearizedHTrafo1} evidently can be set to
zero using a Wess-Zumino gauge transformation -- that is, by using the $\theta$-dependent 
parts of the diffeomorphisms to eliminate them.}

Now we can consider the possible one derivative invariants one can 
build out of the linearized vielbein and the dilaton.
These are in direct correspondence with the torsion tensors already identified.
The ones with negative or vanishing dimension are
\begin{align}
\cT_{\alpha \beta \gamma} = 3\, \pa_{(\alpha} \varV_{\beta \gamma)}~, \qquad
\cT_{\alpha \beta \hc} = 2 \,\pa_{(\alpha} \varV_{\beta) \hc} + \pa_\hc \varV_{\alpha \beta}~, \qquad
\cT_{\alpha b \bc} = \pa_\alpha \varV_{b \bc} - \pa_b \varV_{\alpha \bc} + \pa_\bc \varV_{\alpha b}~.
\end{align}
These we have already constrained in \eqref{E:FluxConstraints}.
The ones at dimension $\tfrac{1}{2}$ are
\begin{align}
\cT_{\alpha \beta}{}^\gamma = 2 \,\pa_{(\alpha} \varV_{\beta)}{}^\gamma + \pa^\gamma \varV_{\alpha \beta}
    + \pa_c \varV_{(\alpha d} (\gamma^{c d})_{\beta)}{}^\gamma~, \qquad
\cT_\alpha = \pa_\alpha \varphi
    - \pa^b \varV_{\alpha b}
    + \pa^\bb \varV_{\alpha \bb}
    + \pa_\alpha \varV_\beta{}^\beta
    + 2 \,\pa^\beta \varV_{\alpha \beta}~.
\end{align}
These must vanish as well,
as there are no covariant physical fields in supersymmetric DFT at this dimension,
and we will impose their vanishing in due course.
At dimension 1, we find
\begin{align}
\cT_{\ba \beta}{}^\gamma
    &= \pa_\ba \varV_\beta{}^\gamma + \pa^\gamma \varV_{\beta \ba} - \pa_\beta \Psi_\ba{}^\gamma
        + \tfrac{1}{2} (\gamma^{cd})_\beta{}^\gamma \pa_c \varV_{\ba d}~, \eol
\cT_{\alpha \beta} \equiv (\gamma^b)_{\alpha \gamma} \cT_{b \beta}{}^\gamma
    &= -\pa_\beta \chi_\alpha
    + (\gamma^a)_{\alpha \beta} \Big(
    \tfrac{1}{2} \pa_a \varV_\gamma{}^\gamma
    + \tfrac{1}{2} \pa_a \varphi
    + \pa^\gamma \varV_{\gamma a}
    - \tfrac{1}{2} \pa^\bb \varV_{\bb a}
    \Big) 
    \eol & \quad
    + (\gamma^a)_{\alpha \gamma} \Big(
    \pa^\gamma \varV_{\beta a}
    + \pa_a \varV_\beta{}^\gamma
    \Big)
\end{align}
These should vanish as well. Finally at dimension $\tfrac{3}{2}$, we encounter
\begin{subequations}
\begin{align}
\cT^\alpha &=
    (\gamma^a)^{\alpha \beta} \,\pa_a \chi_\beta
    - \pa^\bb \Psi_{\bb}{}^\alpha
    + \pa^\alpha \varphi
    + 2 \pa^\beta \varV_\beta{}^\alpha
    + \pa^\alpha \varV_\beta{}^\beta
    ~, \\
\cT_{\alpha \bb} \equiv (\gamma^a)_{\alpha \gamma} \cT_{a \bb}{}^\gamma &= 
    (\gamma^a)_{\alpha \beta} \,\pa_a \Psi_\bb{}^\beta
    - \pa_\bb \chi_\alpha
    - (\gamma^a)_{\alpha \beta} \pa^\beta \varV_{\bb a}~.
\end{align}
\end{subequations}
These correspond respectively to the linearized dilatino and gravitino equations of motion.
As the superspace geometry places DFT on-shell, these should vanish,
as we will show.

\subsection{A detour on tangent space symmetries and an extended algebra}
\label{S:SGeo.Detour}

Before moving on to analyze the torsion constraints and Bianchi identities in superspace,
we need to make a digression into the structure of the tangent space symmetries.
As we have discussed, the closure of the orthosymplectic tangent space symmetries
implies new local shift symmetries on the spin connection. But this conclusion was
motivated purely by requiring consistency of the constraints that we seek to impose. 
In this subsection, we instead give a more abstract characterization of
the algebra itself, motivated purely by closure of the Jacobi identity.
Subsequently, we will show that the gauging of this algebra leads to the
same transformations on $\Omega$, and implies the existence of higher connections.

Let us recall the normalization of the $\g{OSp}(9,1|32)_L$ generators $M_{AB}$ so that
\begin{align}\label{E:OSpAction}
\frac{1}{2} [\lambda^{BC} M_{CB}, P_A] = \lambda_A{}^B P_B
\end{align}
where $P_A = (P_a, Q_\alpha, \tQ^\alpha)$ are the supertranslation
generators and the commutator should be understood as a graded commutator.
The above is the conventional orthosymplectic action, but soon we will see it
is deformed in various ways. To begin with,
the only generators we are keeping are $M_{a b}$ (which includes a piece of
$(\gamma^{ab})_{\beta}{}^\alpha  M_{\alpha}{}^\beta$ to rotate fermionic
indices), $M_{\alpha b}$ (now restricted to be $\gamma$-traceless), and
$M_{\alpha \beta}$.
We use the above formula merely to fix normalizations of these operators.

The Lorentz generator acts as usual as
\begin{align}
[M_{c b}, P_a] &= \eta_{ba} P_c - \eta_{c a} P_b~, \eol{}
[M_{c b}, Q_\alpha] &= -\tfrac{1}{2} (\gamma_{c b})_\alpha{}^\beta Q_\beta~, \eol{}
[M_{c b}, \tQ^\alpha] &= -\tfrac{1}{2} (\gamma_{c b})^\alpha{}_\beta \tQ^\beta~.
\end{align}
The generators $P_A$ themselves obey the algebra
\begin{align}\label{E:DFT.Torsion}
\{Q_\alpha, Q_\beta\} = -\kappa \, (\gamma^c)_{\alpha \beta}\, P_c~, \qquad
[Q_\alpha, P_b] = -\kappa \, (\gamma_b)_{\alpha \gamma} \tQ^\gamma~.
\end{align}
As usual for supersymmetry, this describes a rigid (but not flat) supergeometry
with a constant fixed torsion tensor. The presence of this
torsion tensor has deep consequences for the algebra, which we are about to discover!

The tangent space generator at dimension $-\tfrac{1}{2}$ is the $\gamma$-traceless 
fermionic operator $M_{\alpha a}$. Its algebra with $P_A$ turns out to be
\begin{subequations}
\label{E:Malphab.P}
\begin{align}
\label{E:Malphab.P.a}
\{M_{\beta b}, Q_\alpha\} &= \kappa \, (\gamma^c)_{\beta \alpha} M_{bc} 
    - \tfrac{1}{10} \kappa (\gamma_b \gamma^{cd})_{\beta \alpha} M_{cd}~, \\
[M_{\beta b}, P_a] &= \eta_{b a} Q_{\beta} - \tfrac{1}{10} (\gamma_{b} \gamma_a)_\beta{}^\gamma Q_\gamma~, \\
\{M_{\beta b}, \tQ^\alpha\} &= -\delta_\beta{}^\alpha P_b
    + \tfrac{1}{10} (\gamma_b \gamma^c)_\beta{}^\alpha P_c~,
\end{align}
\end{subequations}
where in each expression the second term is required to project onto spin-3/2.
The second and third algebraic relations follow directly from \eqref{E:OSpAction}.
The first relation would be zero if we employed \eqref{E:OSpAction}. That this
\emph{cannot} be zero follows from the Jacobi identity
$[\{M_{\alpha a}, Q_\beta\}, Q_\gamma] +\cdots$
in the presence of the torsion term \eqref{E:DFT.Torsion}. One must make a $\kappa$-dependent
deformation and the expression \eqref{E:Malphab.P.a} does the job.
Similarly, at dimension $-1$, $M_{\alpha \beta}$ acts as
\begin{subequations}
\label{E:Malphabeta.P}
\begin{align}
[M_{\gamma \beta}, Q_\alpha] &= 
    -2 \kappa (\gamma^{b})_{\alpha (\beta} M_{\gamma) b}
    + \tfrac{1}{9} \kappa (\gamma^{b})_{\gamma \beta} M_{\alpha b}~, \\
[M_{\gamma \beta}, P_a] &= \tfrac{2}{9} \kappa (\gamma^b)_{\gamma \beta} M_{ba}~, \\
[M_{\gamma \beta}, \tQ^\alpha] &= 2 \,\delta^\alpha{}_{(\gamma} Q_{\beta)}~.
\end{align}
\end{subequations}
Now the first and second relations involve deformations and are required by closure
of the Jacobi identity.

These results dovetail with the explicit connection transformations. The three
generators $P_A = (Q_\alpha, P_a, \tQ^\alpha)$ are represented on superfields by
covariant derivatives
$\nabla_A = \cV_A{}^\cM \pa_\cM - \tfrac{1}{2} \Omega_{A}{}^{BC} M_{CB} + \cdots$.
The algebra $[M_{CB}, P_A]$ is then represented via $\delta_\lambda \nabla_A$.
Computing $\delta_\lambda$ using \eqref{E:OmegaTrafo} leads to the
\eqref{E:Malphab.P} and \eqref{E:Malphabeta.P}.

But we are not finished. Closure of the $M_{AB}$ algebra on the connections required
a new local symmetry, and this should appear at the algebraic level. This ought to be
a new generator at dimension $-1$ in the irreducible hook representation, which we
denote $K_{a|bc}$. Working backwards from the $\Lambda$ transformations on $\Omega$,
we conclude that
\begin{subequations}\label{E:KabcAction}
\begin{alignat}{2}
[K_{a|bc}, P_d] &= \tfrac{1}{2} \eta_{d a} M_{b c} \proj & \quad
    &= \tfrac{1}{3} \eta_{d a} M_{b c} 
    - \tfrac{1}{3} \eta_{d [b} M_{c] a}
    + \tfrac{1}{9} \eta_{a [b} M_{c] d}~, \\
[K_{a|bc}, Q_\alpha] &= -\tfrac{1}{4} (\gamma_{b c})_{\alpha}{}^{\beta} M_{\beta a}
    \proj  & \quad 
        &= -\tfrac{1}{4} (\gamma_{b c})_{\alpha}{}^{\beta} M_{\beta a}
        + \tfrac{1}{4} (\gamma_{[b c})_{\alpha}{}^{\beta} M_{\beta a]}
        + \tfrac{1}{18} \eta_{a [b} M_{\alpha c]}~, \\
[K_{a|bc}, \tQ^{\alpha}] &= 0
\end{alignat}
\end{subequations}
where we introduce the notation $\tinyproj$ to denote projection onto the appropriate
representation (in this case, the irreducible hoop representation $a|bc$)
implied by the left-hand side of the equation.
To verify the necessity of this generator at the algebraic level, consider
the anticommutator of $M_{\alpha a}$ with itself. This is given by
\begin{align}
\{M_{\alpha a}, M_{\beta b}\} &=
    - \eta_{a b} M_{\alpha \beta} \proj
    + \text{$\kappa$-dependent terms}
\end{align}
The leading term is expected from the orthosymplectic algebra.
That $\kappa$-dependent terms are required follows by considering
further commutation with $Q_\alpha$.
The result is
\begin{align}\label{E:MMdim1}
\{M_{\alpha a}, M_{\beta b}\} &=
    - \eta_{a b} M_{\alpha \beta}
    + 2\kappa (\gamma^{c})_{\beta \alpha} ( K_{b| a c} + K_{a| b c} ) \proj
    \eol 
    &=
    - \eta_{a b} M_{\alpha \beta}
    + \tfrac{1}{10} (\gamma_{a} \gamma_b)_{\alpha}{}^{\delta} M_{\delta \beta}
    + \tfrac{1}{10} (\gamma_b \gamma_a)_{\beta}{}^\delta M_{\alpha \delta }
    - \tfrac{1}{100} (\gamma_{a} \gamma^{c})_{\alpha}{}^{\delta}  (\gamma_{b} \gamma_{c})_\beta{}^\epsilon M_{\delta\epsilon}
    \eol & \quad
    + \tfrac{14}{5} \kappa (\gamma_{c})_{\alpha \beta} K_{(a| b) c}
    + \tfrac{2}{5} \kappa (\gamma_{b}{}^{c d})_{\alpha \beta} \, K_{(a| c) d} 
    - \tfrac{2}{5}\kappa (\gamma_{a}{}^{c d})_{\alpha \beta} \, K_{(b| c) d} ~.
\end{align}
The presence of $K_{a|bc}$, acting as \eqref{E:KabcAction}, follows from
the Jacobi identity.
In principle, one can derive all the required operators and their algebraic relations
by computing the free Lie superalgebra of successive applications of $M_{\alpha a}$ to itself
and dropping the representations not required by the Jacobi identity. 
Let us denote this extension of $\cH_L$ by
\begin{align}
\ext\cH_L = \{ M_{a b}, M_{\alpha b}, M_{\alpha \beta}, K_{a|bc}, K_{a,b\beta}, \cdots\}
\end{align}

Unfortunately, this approach is inefficient. Already in \eqref{E:MMdim1}, there are an additional
five representations that could have appeared on the right-hand side. 
Luckily, the direct approach of building the $\Omega$ connections and determining
how they must transform short-circuited this analysis. But that is hardly an elegant
solution because repeating it for higher levels would require that we introduce
gauge connections, curvatures, and constraints prior to first characterizing the algebra.
It would be far simpler if we could just start with an algebra that works and derive the required
connections and curvatures directly.

This is what we will now propose. Recall the work of Pol\'a\v{c}ek and Siegel \cite{Polacek:2013nla}, who motivated by the structure of current algebras on the string worldsheet, 
proposed (in the bosonic case) pairing the double Lorentz generator
$M_{\ha \hb}$  with a dual generator $\wtM^{\ha \hb}$. This mirrors
the pairing of $P_{\ha}$ with itself (and, in the supersymmetric case, $Q_\alpha$ with $\tQ^\alpha$).
This permits one to combine $P_\ha$, $M_{\ha\hb}$ and $\wtM^{\ha \hb}$ into a single
$X_A$ operator with an extended metric $\eta^{A B}$. The current algebra then implies
that the structure constants $f_{AB}{}^C$ in $[X_A, X_B] = -f_{A B}{}^C X_C$ must be
totally antisymmetric when lowered with $\eta$.
As we have discussed, this amounts to the implication
\begin{align}
[M_{c b}, P_a] = \eta_{ba} P_c - \eta_{c a} P_b \quad \implies \quad
[P_a, P_b] = -2 \wtM_{ab}
\end{align}
and the algebra closes on $P_a$, $M_{ab}$, and $\wtM^{ab}$.
Because $P_a$ has dimension 1, $\eta$ must pair an operator of dimension $\Delta$ 
with one of $2-\Delta$; we see above that $\wtM^{ab}$ is dimension 2.
This generator can be understood as the first non-trivial generator
in the Maxwell$_\infty$ algebra, which is the free Lie algebra beginning with the
bosonic generators $P_a$. It extends to higher dimensions, with
$[P_a, [P_b,P_c]]$ corresponding, for example, to a reducible hook
representation. (As discussed in section \ref{S:BosonicDFT}, for the bosonic
case, we would further impose tracelessness.)
However, one can truncate the resulting algebra
to include only $P$, $M$ and $\wtM$, and this coincides with the algebra
originally discussed in \cite{Polacek:2013nla}.

\begin{table}[t]
\centering
\begin{tabular}{c|c|c|c|c}
\toprule
dimension & operator & constraint & dual operator & dual dimension \\ \hline
1    & $P^a$ & $-$ & $P_a$ & 1 \\[1ex] \hline
3/2  & $\tQ^\alpha$ $\phantom{\Big\vert}$ & $-$       & $Q_\alpha$    & $1/2$ \\[1ex] \hline
2 & $\wtM^{ab}$ $\phantom{\Big\vert}$ & antisymmetric & $M_{ab}$      & $0$ \\[1ex] \hline
5/2 & $\wtM^{a\beta}$ $\phantom{\Big\vert}$  & $\gamma$-traceless   
    & $M_{\alpha b}$  & $-1/2$  \\[1ex] \hline
3 & $\wtM^{\alpha \beta}$ $\phantom{\Big\vert}$ & symmetric   & $M_{\alpha \beta}$ & $-1$ \\
  & $\wtK^{a|bc}$       & hook irrep & $K_{a|bc}$ &  \\[1ex] \hline
7/2 & $\wtK^{a,b\beta}$ $\phantom{\Big\vert}$  &  $\gamma$-traceless on $b\beta$
    & $K_{a,b\beta}$  & $-3/2$ \\[1ex] \hline
4 & $\wtK^{\alpha, b \beta}$ $\phantom{\Big\vert}$  &  $\gamma$-traceless on $b\beta$
    & $K_{\alpha,b\beta}$  & $-2$ \\
  & $\wtL^{ab,cd}$ $\phantom{\Big\vert}$ & pairwise antisymmetric
  & $L_{ab,cd}$& \\
  & $\wtL^{a|b|cd}$ $\phantom{\Big\vert}$ & $\drep{21000}$ irrep
  & $L_{a|b|cd}$& \\[1ex]\hline
$\vdots$ & $\vdots$ & $\vdots$ & $\vdots$ & $\vdots$ \\
\bottomrule
\end{tabular}
\captionsetup{width=0.8\textwidth}
\caption{Generators of super-Maxwell$_\infty$ through dimension 4 and their duals.
The generators at dimension zero and below form a closed algebra, which we conjecture
is the extension $\ext{\cH}_L$ required for $\cN=1$ super-DFT.}
\label{T:GeneratorsHL}
\end{table}

If we require the same doubling of generators in the supersymmetric case, we find 
that the generators of $\ext\cH_L$, which possess non-positive dimension
$\Delta = 0, -1/2, -1, \cdots$,
have ``reflections'' at positive dimension $2-\Delta$ that coincide with the generators of
a free Lie superalgebra that we denote \emph{super-Maxwell$_\infty$}.
We list these in Table \ref{T:GeneratorsHL} up through dimension 4.
This superalgebra extends the supersymmetry algebra \eqref{E:DFT.Torsion}.
For the details of its construction, we refer the reader to Appendix \ref{A:SMaxwell}.
To make complete contact, we must provide the translation table between the generators
in the appendix and the duals of generators already encountered. These are
\begin{gather}
Y_{a b} = -2 \wtM_{ab}~, \quad
Y_b{}^\beta = \wtM_b{}^\alpha~, \quad
Y^{\alpha \beta} = 2 \wtM^{\alpha \beta}~, \eol
Y_{a|bc} = - \wtK_{a|bc}~, \quad
Y_{a,b}{}^\beta = - \wtK_{a,b}{}^\beta~, \qquad
Y^\alpha{}_{,b}{}^\beta = -\wtK^\alpha{}_{,b}{}^\beta~, \qquad
Y^{\alpha|\beta \gamma} = - \wtK^{\alpha|\beta\gamma}~.
\end{gather}
The generators we have denoted by $\wtK$ are dual to the generators that
shift the tangent space connections $\Omega$. The connections associated to $K$
will be collectively denoted $H$. These pick up shift symmetries as well, which we
denote by $L$ and their duals by $\wtL$.
Following the pattern here, we will identify all higher generators with negative
signs so that the appropriate $Y$'s become $-\widetilde K$ or $-\widetilde L$.

Below we give some of the algebraic relations of super-Maxwell$_\infty$
up through dimension 4.
\begin{subequations}
\begin{alignat}{2}
& \text{(dimension 2)}  &\qquad 
[P_a, P_b] &= -2 \wtM_{a b}~,
\qquad
\{Q_\alpha, \tQ^\beta\} = -\frac{1}{2} (\gamma^{ab})_\alpha{}^\beta \wtM_{a b} \\[2ex]
% 
%%%%%%%%%%%%%%%%%%%%%%%%%%%%
& \text{(dimension 5/2)}  &\qquad 
[P_b, \tQ^\alpha] &= \wtM_b{}^\alpha~, \qquad
[Q_\alpha, \wtM_{ab}] = - \kappa (\gamma_{[a})_{\alpha \beta} \wtM_{b]}{}^\beta \\[3ex]
% 
%%%%%%%%%%%%%%%%%%%%%%%%%%%%
& \text{(dimension 3)}  &\qquad 
\{\tQ^{\alpha}, \tQ^{\beta}\} &= 2 \, \wtM^{\alpha \beta} \eol
&& [P_a, \wtM_{bc}] 
    &= \tfrac{1}{2} \wtK_{a|bc} 
    - \tfrac{2}{9} \kappa \,\eta_{a [b} (\gamma_{c]})_{\alpha \beta} \wtM^{\alpha \beta} \eol
&& \{Q_\alpha, \wtM_{b}{}^\beta\}
    &= - \tfrac{1}{4} (\gamma^{cd})_{\alpha}{}^\beta  \wtK_{b|cd}
    -2 \kappa (\gamma_b)_{\alpha \gamma} \wtM^{\gamma \beta}
    + \tfrac{1}{9} \kappa (\gamma_{b c})_\alpha{}^\beta (\gamma^c)_{\gamma \delta} \wtM^{\gamma\delta}
    \\[3ex]
% 
% 
%%%%%%%%%%%%%%%%%%%%%%%%%%%%
& \text{(dimension 7/2)}  &\qquad 
[\tQ^\alpha, \wtM_{ab}] 
    &= -\wtK_{[a,b]}{}^\alpha~, \qquad
[P_a, \wtM_{b}{}^\beta]
    = -\wtK_{a, b}{}^\beta~, \eol
&& [Q_\alpha, \wtM^{\beta \gamma}] &= -\tfrac{1}{2} (\gamma^{ab})_\alpha{}^{(\beta} \wtK_{a,b}{}^{\gamma)}~, \eol
&& [Q_\alpha, \wtK_{b|cd}] &= 2\kappa (\gamma_b)_{\alpha \gamma} \wtK_{[c,d]}{}^\gamma
    + 2 \kappa (\gamma_{[c|})_{\alpha \gamma} \wtK_{b,|d]}{}^\gamma \\[3ex]
%     
% 
%%%%%%%%%%%%%%%%%%%%%%%%%%%%
& \text{(dimension 4)}  &\qquad 
\{\tQ^\alpha, \wtM^{b \beta}\} &= -\wtK^{\alpha,b\beta}~, \qquad
[P_a, \wtM^{\beta\gamma}] = - \wtK^\beta{}_{,a}{}^\gamma~, \eol&&
[P_a, \wtK_{b|cd}]
    &= \wtL_{a|b|cd}
    + \tfrac{3}{4} \wtL_{ab,cd}
    + \tfrac{3}{8} \eta_{a c} \eta^{e f} \wtL_{be,df}
    \eol &&& \quad
    - \tfrac{3}{10} \kappa \, \eta_{a c} (\gamma_b)_{\alpha \beta} \wtK^\alpha{}_{,d}{}^\beta
    + \tfrac{7}{10} \kappa \, \eta_{a c} (\gamma_d)_{\alpha \beta} \wtK^\alpha{}_{,b}{}^\beta
    \Big\vert_{\texttt{proj}} ~, \eol&&
\{Q_\alpha, \wtK_{a,b}{}^\beta\}
    &= -\kappa (\gamma_a)_{\alpha \gamma} \wtK^\gamma{}_{,b}{}^\beta
    -\kappa  (\gamma_b)_{\alpha \gamma} \wtK^\gamma{}_{,a}{}^\beta
    -\kappa  (\gamma_b)_{\alpha \gamma} \wtK^\beta{}_{,a}{}^\gamma
    \eol &&& \quad
    + \tfrac{1}{9} \kappa \, (\gamma_{bc})_\alpha{}^\beta \, 
        (\gamma^c)_{\gamma \delta} \wtK^\gamma{}_{,a}{}^\delta
    + \tfrac{1}{4} (\gamma^{cd})_\alpha{}^\beta [P_a, \wtK_{b|cd}]
\end{alignat}
\end{subequations}
The above relations can be ``reflected'' to derive the action of the untilded generators.
We give some of these results below, grouped again by dimension.
\begin{subequations}
\begin{alignat}{2}
& \text{(dimension 0)}  &\qquad 
\{M_{\gamma c}, Q_\alpha\} &= -\kappa (\gamma^b)_{\gamma\alpha} M_{b c} \proj~, \eol&&
[M_{\alpha \beta}, P_a] &= -\tfrac{2}{9} \kappa\, (\gamma^b)_{\alpha\beta}\, M_{a b}~, \qquad
[K_{a|bc}, P_d] = \tfrac{1}{2} \eta_{a d} M_{bc} \proj~, \eol&&
\{K_{a,b \beta}, Q^\gamma\} &= - \delta_\beta{}^\gamma M_{ab} \proj \\[3ex]
% 
% 
%%%%%%%%%%%%%%%%%%%%%%%%%%%%%%%%%%%%%%%
& \text{(dimension -1/2)}  &\qquad 
[M_{\beta \gamma}, Q_\alpha] 
    &= -2\kappa (\gamma^b)_{\alpha \beta} M_{\gamma b}
    + \tfrac{1}{9} \kappa (\gamma^b)_{\beta \gamma} M_{\alpha b} \proj \eol&&
[K_{b|cd}, Q_\alpha] &=
    -\tfrac{1}{4} (\gamma_{c d})_\alpha{}^\beta M_{\beta b} \proj \eol&&
[K_{c,b\beta}, P_a]
    &= -\eta_{c a} M_{\beta b} ~, \qquad
[K_{\gamma, b\beta}, Q^\alpha] = -\delta_\gamma{}^\alpha M_{b\beta} \\[3ex]
% 
% 
%%%%%%%%%%%%%%%%%%%%%%%%%%%%%%%%%%%%%%%
& \text{(dimension -1)}  &\qquad 
\{K_{c,b\beta}, Q_{\alpha}\}
    &= -\tfrac{1}{2} (\gamma_{cb})_\alpha{}^\gamma M_{\gamma \beta}
    + 2\kappa (\gamma^d)_{\alpha \beta} K_{d|cb}
    - 2 \kappa (\gamma^d)_{\alpha \beta} K_{c|bd} \proj \eol&&
[L_{e|b|cd}, P_a] &= \eta_{ae} K_{b|cd} \proj~, \qquad
[L_{de,bc}, P_{a}] = \tfrac{3}{4} \eta_{a d} K_{e|bc}
    + \tfrac{3}{8} \eta_{e c} K_{d|ab} \proj~, \eol&&
[K_{\gamma, b\beta}, P_a] &= -\eta_{b a} M_{\gamma \beta}
    + \tfrac{3}{10} \kappa (\gamma^d)_{\gamma \beta} K_{d|ba}
    + \tfrac{3}{10} \kappa (\gamma^d)_{\gamma \beta} K_{b|ad} \proj
\end{alignat}
\end{subequations}
Other commutators can be determined from the Jacobi identity.

This Lie superalgebra, which is an extension of the super-Maxwell$_\infty$ algebra,
would seem to continue to infinity in both directions. A separate construction of
it is sketched in Appendix \ref{A:SMaxwell.Local}. Unlike the super-Maxwell$_\infty$
algebra itself, which possesses only positive generators, the extension above includes
negative dimensions. This would seem to make it difficult to consistently truncate
the algebra, although we know of no proof of this.
The algebraic relations implied
down through dimension $-1$ do coincide with what a direct construction of connections,
curvatures, etc. would give. But we leave it an open question as to whether the
algebra required is actually the full extension of super-Maxwell$_\infty$
discussed above.

\subsection{New connections, transformations, and curvatures}

Now that we have a proposal for the algebra we want to gauge (at least up through dimension 4
and down to dimension -2),
we may go about introducing connections and curvatures. The idea is to match the
existing results for the torsion tensor and $\Omega$ connection from section \ref{S:SGeo.b}
and to extend them by introducing new $H$ connections, the superspace analogues of
the $h$ connection discussed in section \ref{S:BosonicDFT}.

First, we adopt the condensed notation of Appendix \ref{A:PS}, here extended to the
superspace case. We denote the supertranslation generators by 
$P_\cA$, the  $\ext\cH_L \times \ext{\g{SO}(9,1)}_R$ generators by $X_\ua$, and their duals by $\tX^\ua$.
Following Appendix \ref{A:PS}, we use $\ua, \ub, \cdots$ here to denote
$\ext\cH = \ext{\cH}_L \times \ext{\g{SO}(9,1)}_R$ indices.
We hope this will cause no confusion with the tangent space indices $(a, \ba)$.
Relative to the previous subsection, we're now including the right-handed $P$
and the $\ext{\g{SO}(9,1)}_R$
generators, and their duals, which commute with $P_A$, $\ext{\cH}_L$ and its dual.
The rigid algebra can be written
\begin{subequations}\label{E:RigidPSAlgebra}
\begin{align}
[P_\cA, P_\cB] &= -f_{\cA \cB}{}^\cC P_\cC - \tX^\uc  f_{\uc \cA \cB}~, \\
[X_\ua, P_\cB] &= -f_{\ua \cB}{}^\cC P_\cC - f_{\ua \cB}{}^\uc \,X_\uc~, \\
[X_\ua, X_\ub] &= - f_{\ua \ub}{}^\uc X_\uc~, \\[2ex]
[\tX^\ua, \tX^\ub] &= - \tX^\uc \, f_\uc{}^{\ua \ub}~, \\
[P_\cA, \tX^\ub] &= 
    - \tX^\uc f_{\uc\cA}{}^\ub~, \\
[X_\ua, \tX^\ub] &= -\tX^\uc\, f_{\uc\ua}{}^\ub
    - f_{\ua}{}^{\ub}{}^\cC P_\cC
    - f_\ua{}^{\ub \uc} X_\uc~.
\end{align}
\end{subequations}
A more compact form is
\begin{align}\label{E:RigidPSAlgebra.Compact}
[X_{\widehat\cA}, X_{\widehat \cB}] = -f_{\widehat \cA \widehat \cB}{}^{\widehat \cC} X_{\widehat \cC}
\end{align}
for $X_{\widehat\cA} = (X_\ua, P_\cA, \tX^\ua)$ and where
$f_{\widehat \cA \widehat \cB \widehat \cC} = 
    f_{\widehat \cA \widehat \cB}{}^{\widehat \cD} \eta_{\widehat \cD \widehat \cC}$
is totally antisymmetric with $\eta_{\widehat \cA \widehat \cB}$ given by
\begin{align}
\eta_{\widehat \cA \widehat \cB} =
\begin{pmatrix}
0 & 0 & \delta_\ua{}^\ub \\
0 & \eta_{\cA \cB} & 0 \\
(-1)^{\ua \ub} \,\delta_\ub{}^\ua  & 0 & 0
\end{pmatrix}~.
\end{align}

Following the discussion in Appendix \ref{A:PS}, we introduce 
the dilaton $\Phi$,
the vielbein $\cV_\cM{}^\cA$,
connections $H_\cM{}^\ua$, and an additional graded antisymmetric field
$P^{\ua \ub}$. These transform under diffeomorphisms and gauge transformations
(with parameter $\Lambda^\ua$) as\footnote{For purposes of legibility, we have suppressed
gradings in these and subsequent expressions.}
\begin{subequations}\label{E:PSTrafos}
\begin{align}
\label{E:PSTrafos.Phi}
\delta \Phi &= \Lie_\xi \Phi~, \\
\label{E:PSTrafos.V}
\delta \cV_\cM{}^\cA &= \Lie_\xi \cV_\cM{}^\cA + \cV_\cM{}^\cB \Lambda^\uc f_{\uc \cB}{}^\cA~, \\
\label{E:PSTrafos.H}
\delta H_\cM{}^\ua
    &= \Lie_\xi H_\cM{}^\ua
    + \pa_\cM \Lambda^\ua
    + H_\cM{}^\ub \Lambda^\uc f_{\uc \ub}{}^\ua
    + \cV_\cM{}^\cB \Lambda^\uc f_{\uc \cB}{}^\ua~, \\
\label{E:PSTrafos.P}
\delta P^{\ua \ub}
    &= \xi^\cM \pa_\cM P^{\ua \ub}
    - \Lambda^\uc f_{\uc}{}^{\ua \ub}
    - 2 \,\Lambda^\uc P^{\ud [\ua} f_{\uc \ud}{}^{\ub]}
    - H^{\cM [\ua} \pa_\cM \Lambda^{\ub]}
    - \Lambda^\uc H^{\cD [\ua} f_{\cD \uc}{}^{\ub]}~.
\end{align}
\end{subequations}
The connection $H_\cM{}^\ua$ generalizes $\Omega_\cM{}^{\cA \cB}$ and the field $P^{\ua \ub}$
generalizes the one introduced by Pol\`a\v{c}ek and Siegel \cite{Polacek:2013nla}.
Using these ingredients, we can construct covariant derivatives
\begin{align}
\nabla_\cA &= \cV_\cA{}^\cM \pa_\cM - H_\cA{}^{\ub} X_\ub~, \qquad
\tnabla^\ua = H^{\cM \ua} \pa_\cM + (P^{\ua \ub} - \tfrac{1}{2} H^{\cM \ua} H_\cM{}^{\ub} ) X_\ub~.
\end{align}
These correspond to the curved extensions of $P_\cA$ and $\tX^\ua$.
Their algebra can again be written \eqref{E:RigidPSAlgebra.Compact},
but with some of the components of $f$ now becoming structure functions.
These are the four curvatures
$\cT_{\cA \cB \cC}$, 
$\cR_{\cA \cB}{}^{\uc}$,
$\cR_\cA{}^{\ub \uc}$, and
$\cR^{\ua \ub \uc}$, which appear in the curved algebra as
\begin{subequations}\label{E:CurvedPSAlgebra}
\begin{align}
[\nabla_\cA, \nabla_\cB] &= 
    - \cT_{\cA \cB}{}^\cC \nabla_\cC
    - \cR_{\cA \cB}{}^\uc X_\uc
    - \tnabla^\uc\, f_{\uc \cA \cB} 
    ~, \\
[X_\ua, \nabla_\cB] &= -f_{\ua \cB}{}^\cC \nabla_\cC - f_{\ua \cB}{}^\uc \,X_\uc~, \\
[X_\ua, X_\ub] &= - f_{\ua \ub}{}^\uc X_\uc~, \\[2ex]
[\tnabla^\ua, \tnabla^\ub] &= 
    - \tnabla^\uc \, f_\uc{}^{\ua \ub}
    - \cR^{\ua \ub \cC} \nabla_\cC
    - \cR^{\ua \ub \uc} X_\uc
    ~, \\
[\nabla_\cA, \tnabla^\ub] &=
    - \tnabla^\uc \,f_\uc{}_\cA{}^\ub
    - \cR_\cA{}^\ub{}^\cC \,\nabla_\cC 
    - \cR_\cA{}^\ub{}^{\uc} X_\uc~, \\
[X_\ua, \tnabla^\ub] &= 
    - \tnabla^\uc \,f_\uc{}_\ua{}^{\ub}
    - f_\ua{}^\ub{}^\cC \nabla_\cC
    - f_\ua{}^\ub{}^\uc X_\uc~.
\end{align}
\end{subequations}
The torsion tensor is given as
\begin{align}
\cT_{\cC\cB\cA} &= -3 \nabla_{[\cC} \cV_\cB{}^\cM \cV_{\cM \cA]}~, \qquad
\nabla_\cC \cV_\cB{}^\cM := D_\cC \cV_\cB{}^\cM + H_\cC{}^\ud f_{\ud \cB}{}^\cA \cV_\cA{}^\cM~.
\end{align}
The curvature tensor $\cR_{\cC \cB}{}^\ua$ is
\begin{align}
\cR_{\cC \cB}{}^\ua &= 
    2\, D_{[\cC} H_{\cB]}{}^\ua
    + \cF_{\cC \cB}{}^\cD H_\cD{}^\ua
    - H_\cB{}^\ub H_\cC{}^\uc f_{\uc \ub}{}^\ua
    - 2\, H_{[\cC}{}^{\ud} f_{\ud \cB]}{}^\ua
%     \eol & \quad
    - f_{\cC \cB \,\ud} \, \Big(P^{\ud \,\ua} + \tfrac{1}{2} H^{\cF \ud} H_\cF{}^\ua\Big)~.
\end{align}
The other curvature tensors $\cR_\cA{}^{\ub \uc}$ and $\cR^{\ua \ub \uc}$ correspond to
covariantizations of $\nabla_\cA P^{\ub \uc}$ and $\tnabla^{[\ua} P^{\ub \uc]}$ and are
discussed further in Appendix \ref{A:PS}. 
In addition to these, one has dilaton-dependent curvatures
\begin{subequations}
\begin{align}
\cT_\cA &= \nabla_\cA \log\Phi + \nabla_\cM \cV_\cA{}^\cM + f_{\cA \ub}{}^\ub~, \\
\cR^\ua &= 
    D^\cB H_\cB{}^\ua + \cF^\cB H_\cB{}^\ua
    - H^{\cB \uc} f_{\uc \cB}{}^\ua
    + P^{\ub \uc} f_{\uc \ub}{}^{\ua}
    + f_\ub{}^{\ub \ua}
\end{align}
\end{subequations}
where $\cF_\cA := D_\cA \log\Phi + \pa_\cM \cV_\cA{}^\cM$.

Using these expressions, we can now give explicit expressions for the curvatures. 
The transformation rules for $\Omega$ and the expressions for the torsion tensors
reproduce what we found by direct construction in section \ref{S:SGeo.b},
so we do not repeat
that discussion here. We turn our attention directly to the $\Omega$ curvatures.
We restrict attention to components of curvatures only through dimension 2,
which is sufficient for understanding the two-derivative DFT action.

First, we take the curvature $\cR_{\cA \cB\, \overline{cd}}$, which will have
the simplest structure:
\begin{subequations}
\begin{align}
\cR_{\alpha \beta\, \overline{cd}} &= R(\Omega)_{\alpha \beta \,\overline{cd}} \\[2ex]
\cR_{\alpha \overline{b}\, \overline{cd}} &= R(\Omega)_{\alpha \bb\, \overline{cd} } - H_{\alpha\, \bb | \overline{cd}} \\
\cR_{\alpha b \,\overline{cd}} &= R(\Omega)_{\alpha b\,\overline{cd} } \\[2ex]
\cR_{a \bb \, \overline{cd}} &= R(\Omega)_{a \overline{bcd}} - H_{a\, \bb | \overline{cd}} \\
\cR_{\overline{ab} \,\overline{cd}} &= R(\Omega)_{\overline{ab}\, \overline{cd}}
    - H_{\ba, \bb|\overline{cd}}
    + H_{\bb, \ba|\overline{cd}}
    - P_{\overline{ab} \,\overline{cd}} \\
\cR_{a b \,\overline{cd}} &= R(\Omega)_{a b \,\overline{cd}}
    - P_{a b \,\overline{cd}} \\[2ex]
\cR_\alpha{}^\beta{}_{\overline{cd}} &= 
    R(\Omega)_\alpha{}^\beta{}_{\overline{cd}}
    - \tfrac{1}{4} (\gamma^{ab})_\alpha{}^\beta P_{ab\,\overline{cd}}~.
\end{align}
\end{subequations}
There remaining pieces $\cR_a{}^\beta{}_{\overline{cd}}$,
$\cR_\ba{}^\beta{}_{\overline {cd}}$, and $\cR^{\alpha \beta}{}_{\overline {cd}}$
are higher dimension, so we ignore them.
In the above expressions, $R(\Omega)_{\cA \cB \cC \cD}$ is the naive curvature
built from $\Omega$, which is the superspace analogue of \eqref{E:Romega.naive},
\begin{align}
R(\Omega)_{\cA \cB \cC \cD} &:= 
    2 \,D_{[\cA} \Omega_{\cB] \cC \cD}
    - 2 \,\Omega_{[\cA |\cC}{}^{\cE} \Omega_{|\cB] \cE \cD}
    + \cF_{\cA \cB}{}^\cE \Omega_{\cE \cC \cD}
    + \frac{1}{2} \Omega^\cE{}_{\cA \cB} \,\Omega_{\cE \cC \cD}~.
\end{align}
The connections $H_{\cA \, \bb | \overline{cd}}$ are necessary to gauge the
$\Lambda_{\bb|\overline{cd}}$ shift symmetry and the fields $P$ are required to
cure the Lorentz non-covariance of $R(\Omega)_{\cA \cB \cC \cD}$ when both
$\cA \cB$ and $\cC \cD$ correspond to tangent space transformations.
All of this is a straightforward extension of the bosonic case
(except for $\cR_\alpha{}^\beta{}_{\overline{cd}}$).

The curvature $\cR_{\cA \cB\, cd}$ is already a good deal more complicated:
\begin{subequations}
\begin{align}
\cR_{\alpha \beta \, c d} &= R(\Omega)_{\alpha \beta \,c d} 
    + 4 \,\kappa (\gamma_{[c})_{\gamma (\beta} \Omega_{\alpha) d]}{}^\gamma~, \\
\cR_{\alpha \bb \,c d} &= R(\Omega)_{\alpha \bb \,c d} 
    - 2 \kappa \,(\gamma_{[c})_{\alpha \gamma} \,\Omega_{\bb \,d]}{}^\gamma \\
\cR_{\alpha b \,c d} &= R(\Omega)_{\alpha b \,c d} 
    + \tfrac{2}{9} \kappa \,\eta_{b [c} (\gamma_{d]})_{\gamma \delta} \Omega_{\alpha}{}^{\gamma \delta}
    - 2 \kappa \,(\gamma_{[c|})_{\alpha \gamma} \,\Omega_{b|d]}{}^\gamma
    - H_{\alpha\, b | cd} \\[2ex]
\cR_{\overline{a b} \,c d} &= R(\Omega)_{\overline{ab} \,cd}
    - P_{\overline{a b} \,cd}\\
\cR_{a \bb \,c d} &=
    R(\Omega)_{a \bb \,c d}
    - \tfrac{2}{9} \kappa \,\eta_{a [c} (\gamma_{d]})_{\alpha \beta} \Omega_{\bb}{}^{\alpha \beta}
    + H_{\bb \, a|cd} \\
\cR_{a b \,c d} &= R(\Omega)_{a b \,c d} 
    + \tfrac{4}{9} \kappa \, \Omega_{[a}{}^{\alpha \beta}\, \eta_{b] [c} (\gamma_{d]})_{\alpha\beta}
    - H_{a, b|cd}
    + H_{b, a|cd}
    - P_{a b\, c d} \\[2ex]
\cR_\alpha{}^\beta{}_{cd} &=
    R(\Omega)_\alpha{}^\beta{}_{cd} 
    + H_{\alpha\, c,d}{}^\beta
    - H_{\alpha\, d,c}{}^\beta
    - \tfrac{1}{4} (\gamma^{ab})_\alpha{}^\beta P_{ab\,cd}~.
\end{align}
\end{subequations}
The new $\kappa$-dependent modifications involving $\Omega_{\cA b}{}^\gamma$
and $\Omega_\cA{}^{\beta \gamma}$ are required to ensure covariance under
$\lambda_b{}^\gamma$ and $\lambda^{\beta \gamma}$ transformations.

The curvature $\cR_{\cA \cB}{}_c{}^\gamma$ does not have a bosonic counterpart.
Its structure is fairly intricate, but only the lowest few pieces have a sufficiently
small dimension to be relevant for us:
\begin{subequations}
\begin{align}
\cR_{\alpha \beta \,c}{}^\gamma &= R(\Omega)_{\alpha \beta \,c}{}^\gamma
    - 2 \kappa (\gamma_c)_{\delta (\beta } \Omega_{\alpha)}{}^{\delta \gamma}
    + \tfrac{1}{9} \kappa \,(\gamma_{c d})_{(\beta|}{}^\gamma (\gamma^d)_{\delta \epsilon} \Omega_{|\alpha)}{}^{\delta \epsilon}
    - \tfrac{1}{2} (\gamma^{de})_{(\beta}{}^\gamma\, H_{\alpha)\, c|de}~, \\
\cR_{\alpha \bb \,c}{}^\gamma &=
    R(\Omega)_{\alpha \bb \,c}{}^\gamma
    + \kappa (\gamma_c)_{\alpha \delta} \Omega_\bb{}^{\delta \gamma}
    + \tfrac{1}{4} (\gamma^{ef})_\alpha{}^\gamma \, H_{\bb \,c|ef}~, \\
\cR_{\alpha b \,c}{}^\gamma &=
    R(\Omega)_{\alpha b \,c}{}^\gamma
    + \kappa\, (\gamma_c)_{\alpha \delta} \Omega_{b}{}^{\delta \gamma}
    - \tfrac{1}{18} \kappa\, (\gamma_{c d})_{\alpha}{}^\gamma \, (\gamma^d)_{\delta \epsilon} \Omega_b{}^{\delta \epsilon}
    + \tfrac{1}{4} (\gamma^{ef})_\alpha{}^\gamma\, H_{b,c|ef}
    - H_{\alpha\, b ,c}{}^\gamma~.
\end{align}
\end{subequations}
Again, $\Omega_{\cA b}{}^\gamma$ and $\Omega_\cA{}^{\beta \gamma}$ contributions
are required for $\lambda$-covariance, and similarly for the $H$ connections.
In the last expression we find a new connection $H_{\cA\, b,c}{}^\gamma$, which
gauges the $\Lambda_{b,c}{}^\gamma$ shift symmetry of $\Omega_{b,c}{}^\gamma$.
While there are $P_{\cA \cB\,}{}_c{}^\gamma$ contributions at higher dimension
when $\cA \cB$ corresponds to a generator, none appear here.

For $\cR_{\cA \cB}{}^{\gamma \delta}$, only the lowest component
is relevant:
\begin{align}
\cR_{\alpha \beta}{}^{\gamma \delta} &= R(\Omega)_{\alpha \beta}{}^{\gamma \delta}
    - 2 \, H_{(\alpha\, b, c}{}^{(\gamma} (\gamma^{bc})_{\beta)}{}^{\delta)}
\end{align}

The remaining curvatures are the dilaton-dependent ones $\cR^\ua$. At dimension 2,
these consist of
\begin{subequations}
\begin{align}
\cR_{a b} & = 
    R(\Omega)_{ab}
    + 2 \kappa\, (\gamma_a)_{\alpha \beta} \Omega^\alpha{}_b{}^\beta
    - \tfrac{2}{9} \kappa\, (\gamma_{[a})_{\alpha \beta} \Omega_{b]}{}^{\alpha \beta}
    - H^c{}_{c|ab}
    + 2 H_{\gamma\, [a,b]}{}^\gamma~, \\
\cR_{\overline{ab}} &= R(\Omega)_{\overline{ab}}
    + H^{\bc}{}_{\bc|\overline{ab}}
\end{align}
\end{subequations}
where $R(\Omega)_{\cA \cB} = D^\cC \Omega_{\cC \cA \cB} + \cF^\cC \Omega_{\cC \cA \cB}$.

We do not give the $H$ transformations explicitly, but they can be derived 
from \eqref{E:PSTrafos.H}.
As with the $\Omega$ connections, we can use the $H$ connections and $P$ fields to impose
a number of constraints on $\cR$:
\begin{subequations}
\label{E:HP.Constraints}
\begin{alignat}{2}
\label{E:HP.Constraints.a}
H_{\alpha\, \bb|\overline{cd}} \quad &\implies &\quad \cR_{\alpha \overline{b}\,\overline{cd}} \,\Big\vert_{\bb |\overline{cd}} &= 0~, \\
\label{E:HP.Constraints.b}
H_{\alpha\, b|cd} \quad &\implies &\quad \cR_{\alpha b \,cd} \Big\vert_{b|cd} &= 0~, \\
\label{E:HP.Constraints.c}
H_{\ba \, b|cd} \quad &\implies &\quad \cR_{\ba b \,c d} \Big\vert_{b|cd} &= 0~, \\
\label{E:HP.Constraints.d}
H_{a \, \bb|\overline{cd}} \quad &\implies &\quad \cR_{a \bb \,\overline{c d}} \Big\vert_{\bb|\overline{cd}} &= 0~, \\
\label{E:HP.Constraints.e}
P_{a b\, \overline{cd}} \quad &\implies &\quad \cR_{a b \,\overline{cd}} &= \cR_{\overline{cd} \,a b}~, \\
\label{E:HP.Constraints.f}
H_{a,b|cd}~,\,\, P_{ab\, cd} \quad &\implies &\quad
\cR_{a b\, c d} &= \tfrac{1}{45} \eta_{a [c} \eta_{d] b}\, \cR + \cR_{[ab\, cd]}~, \\
\label{E:HP.Constraints.g}
H_{\ba,\bb | \overline{cd}}~, \,\,P_{\overline{ab}\, \overline{cd}} \quad &\implies &\quad
\cR_{\overline{a b}\, \overline{c d}} &= \tfrac{1}{45} \eta_{\ba [\bc} \eta_{\bd] \bb}\, \bar\cR
    + \cR_{[\overline{a b}\, \overline{c d}]}~, \\
\label{E:HP.Constraints.h}
H_{\alpha \,b,c}{}^\gamma \quad &\implies &\quad 
    \cR_{\alpha b\, c}{}^\gamma &= 0
\end{alignat}
\end{subequations}
Above, we have used
$\cR = \cR_{ab}{}^{ab}$ and $\bar \cR = \cR_{\overline{ab}}{}^{\overline{ab}}$.

While these constraints
fully determine $P$, there remain three irreps of $H_{a,b|cd}$ that are undetermined.
As in the bosonic case, these are the $\tiny\yng (3,1)$ component
corresponding to $L_{a|b|cd}$ 
and the  $\tiny\yng(2,1,1) + \tiny\yng(1,1)$ components corresponding to $L_{ab,cd}$.

\subsection{Solution of the Bianchi identities through dimension 2}

Our last task in superspace is to analyze the torsion Bianchi identities to determine
what the unfixed pieces of the $\cR(\Omega)$ curvatures are. Here we will again
restrict our attention through dimension 2.

The Bianchi identities for the torsion tensor read
\begin{subequations}
\label{E:BI.Torsion}
\begin{align}
\label{E:BI.Torsion.a}
0 &= \cB_{\cA \cB \cC \cD} \equiv \Big[4 \,\nabla_{\cA} \cT_{\cB \cC \cD}
    + 3 \,\cT_{\cA \cB}{}^\cE \cT_{\cE \cC \cD}
    - 6 \,\cR_{\cA \cB \cC \cD}
    \Big]_{[\cA\cB\cC \cD]}~,\\
\label{E:BI.Torsion.b}
0 &= \cB_{\cA \cB } \equiv \Big[2 \nabla_{\cA} \cT_{\cB} 
    + \cT_{\cA \cB}{}^\cC \cT_\cC
    + \nabla^\cC \cT_{\cC \cA \cB}
    - \cR_{\cA}{}^{\cD}{}_{\cD \cB}
    + \cR_{\cB}{}^{\cD}{}_{\cD \cA}
    - \cR_{\cA \cB}
    \Big]_{[\cA\cB]}~, \\
\label{E:BI.Torsion.c}
0 &= \cB \equiv \nabla^\cA \cT_\cA
    + \tfrac{1}{2} \cT^\cA \cT_\cA
    + \tfrac{1}{12} \cT^{\cA \cB \cC} \cT_{\cC \cB \cA}
    - \tfrac{1}{2} \cR^{\cA \cB}{}_{\cB \cA}
\end{align}
\end{subequations}
The torsion tensors are given in Table \ref{T:TConstraints}, where we have listed the
conventional and physical constraints. 

The dimension $\tfrac{1}{2}$ torsions that we
have not yet fixed are $\cT_{\alpha \beta}{}^\gamma$ and $\cT_\alpha$.
From the $\alpha\beta\gamma d$ component of
\eqref{E:BI.Torsion.a}, we can show that $\cT_{\alpha\beta}{}^\gamma$ must obey
$(\gamma_d)_{\delta (\gamma} \cT_{\alpha \beta)}{}^\delta  = 0$,
which implies
$\cT_{\alpha \beta}{}^\gamma = X_{(\alpha} \delta_{\beta)}{}^\gamma
    - \frac{1}{2} (\gamma^c)_{\alpha \beta} (\gamma_c)^{\gamma \delta} X_\delta
$
in terms of some covariant field $X_\alpha$. If we had not already accounted
for the dilatino in the supervielbein, this is where it would appear. To avoid
introducing more degrees of freedom, we fix this dimension-1/2 piece to vanish.
The $\alpha\beta\gamma \bd$ component of \eqref{E:BI.Torsion.a} trivially vanishes.
If we hadn't already set $\cT_{\alpha b \bc}$ to zero, we would have found that
$(\gamma^c)_{(\alpha \beta} \cT_{\gamma) c \,\bd} = 0$ implies that
$\cT_{\alpha b \bc} \propto (\gamma_b)_{\alpha \beta} \cW^\beta{}_{\bc}$
in terms of a gaugino superfield.
We reiterate that for heterotic DFT, if the gaugino were not contained within 
the supervielbein, this is where it would appear.
That leaves $\cT_{\alpha}$. Although it's plausible that we could deduce
a constraint on this by looking at dimension 1 Bianchi identities, we are
instead simply going to impose that it vanishes. This gives a full set of
vanishing dimension-$\tfrac{1}{2}$ physical constraints,
\begin{align}
\cT_{\alpha \beta}{}^\gamma = \cT_\alpha = \cT_{\alpha b \bc} = 0~.
\end{align}

At dimension 1, we first analyze various components of
\eqref{E:BI.Torsion.a}.
These rapidly yield constraints:
\begin{alignat}{3}
\cB_{\alpha \beta \overline{cd}} &= 0 
&\quad &\implies &\quad
\cR_{\alpha \beta \, \overline{cd}} &= 0~, \eol
\cB_{\alpha \beta c \bd} &= 0 
&\quad &\implies &\quad
\cT_{\ba \beta}{}^\gamma &= 0~, \eol
\cB_{\alpha \beta c d} &= 0 
&\quad &\implies &\quad
\cR_{\alpha \beta \,c d} &= 
    -\tfrac{2}{5} \kappa\, \cT_{\gamma (\alpha} \, (\gamma_{cd})_{\beta)}{}^\gamma ~.
\end{alignat}
Next using  $\cB_{\alpha \beta \gamma}{}^\delta=0$, one can show that
$\cT_{\alpha \beta}$ must be purely a vector. However, $\cB_{\alpha\beta}=0$
is nonzero if $\cT_{\alpha\beta}$ is purely vectorial. Thus we find that
\begin{align}
\cB_{\alpha \beta \gamma}{}^\delta = \cB_{\alpha\beta} = 0
\quad \implies \quad
\cR_{\alpha \beta \,c d} = 0~, \qquad
\cT_{a}{}_\beta{}^\gamma = \tfrac{1}{10}\,(\gamma_a)^{\gamma \alpha} \cT_{\alpha \beta} = 0~.
\end{align}
This is as expected: there are no covariant dimension 1 fields or curvatures
in supersymmetric DFT.

At dimension $\tfrac{3}{2}$, there are more terms to analyze but fewer spinor indices,
so the group theory is simpler. We rapidly find
\begin{alignat}{3}
\cB_{\alpha \overline{bcd}} &= 0 
&\quad &\implies &\quad
\cR_{\alpha [\bb\, \overline{cd}]} &= 0~, \eol
\cB_{\alpha b \overline{cd}} &= 0 
&\quad &\implies &\quad
\cR_{\alpha b\,\overline{cd}} &= 0~, \eol
\cB_{\alpha \bb c d} &= 0
&\quad &\implies &\quad
\cR_{\alpha \bb\,c d} &= \tfrac{1}{5} \kappa\, (\gamma_{c d})_\alpha{}^\beta \cT_{\beta \bb}~, \eol
\cB_{\alpha b c d} &= 0
&\quad &\implies &\quad
\cR_{\alpha [b\, c d]} &= 0~.
\end{alignat}
The component $\cT_{\beta \bb}$ of $\cT_{a \bb}{}^\gamma$ is the covariantized
gravitino field equation.
Using $\cB_{\alpha \beta \bc}{}^\delta=0$, one can show that it vanishes,
fixing $\cR_{\alpha \bb\,c d} = 0$ and $\cT_{a \bb}{}^\gamma = 0$. This places
the component theory on-shell.
Next we find that
\begin{alignat}{3}
\cB_{\alpha \beta c}{}^\delta &= 0 
&\quad &\implies &\quad
\cR_{\alpha \beta c}{}^\gamma = \tfrac{1}{2} (\gamma^{a b})_{(\beta}{}^\gamma \cR_{\alpha) c a b}~.
\end{alignat}
This is a powerful constraint because $\cR_{\alpha \beta c}{}^\gamma$ is $\gamma$-traceless
on $c\gamma$. Since $\cR_{\alpha [cab]} = 0$, we conclude that
$\cR_{\alpha cab}$ must be in the irreducible hook representation
in $c|ab$. However, we have already fixed that piece to vanish using $H_{\alpha c|ab}$ in \eqref{E:HP.Constraints.b}. Thus we conclude that
$\cR_{\alpha \beta c}{}^\gamma$ and $\cR_{\alpha c a b}$ both vanish entirely.
The remaining Bianchi identities immediately give
\begin{alignat}{3}
\cB_{\alpha b} &= 0 
&\quad &\implies &\quad
\cT^\alpha &= 0~, \eol
\cB_{\alpha \bb} &= 0 
&\quad &\implies &\quad
\cR_{\alpha \bb \, \overline{c d}} \,\eta^{\overline{bd}} &= 0 \quad \implies \quad
\cR_{\alpha \bb \, \overline{c d}} = 0~.
\end{alignat}
The first equation is the covariantized dilatino equation of motion.
The last implication in the second equation
follows because (with $\cR_{\alpha [\bb \, \overline{c d}]}$ eliminated)
only the irreducible hook representation remains, but this is
precisely the piece we can remove by choosing $H_{\alpha \, \bb | \overline{c d}}$ as in
\eqref{E:HP.Constraints.a}.
In this way we have eliminated all dimension $\tfrac{3}{2}$ components of torsion
and curvature.

Finally, we turn to the dimension 2 Bianchi identities. First, we analyze
the purely bosonic pieces of \eqref{E:BI.Torsion.a}:
\begin{alignat}{3}
\cB_{\overline{abcd}} &= 0 
&\quad &\implies &\quad
\cR_{[\overline{ab} \, \overline{cd}]} &= 0~, \eol
\cB_{abcd} &= 0 
&\quad &\implies &\quad
\cR_{[ab \, cd]} &= 0~, \eol
\cB_{a \overline{bcd}} &= 0 
&\quad &\implies &\quad
\cR_{a [\bb \, \overline{cd}]} &= 0~, \eol
\cB_{abc \bd} &= 0 
&\quad &\implies &\quad
\cR_{\bd [c \, a b]} &= 0 ~, \eol
\cB_{ab \overline{cd}} &= 0 
&\quad &\implies &\quad
\cR_{a b \, \overline{cd}} &= -\cR_{\overline{cd}\, a b}~.
\end{alignat}
The first and second equations work similarly. Together with the conventional
constraints \eqref{E:HP.Constraints.f} and \eqref{E:HP.Constraints.g}, they imply that
$\cR_{ab\,cd}$ and $\cR_{\overline{ab}\,\overline{cd}}$ are pure traces.
The third and fourth equations, together with \eqref{E:HP.Constraints.c} and \eqref{E:HP.Constraints.d} imply that $\cR_{a \bb\, c d}$ and $\cR_{a \bb\, \overline{cd}}$
each consist of only the $\tiny\yng(1) \times \overline{\yng(1)}$ representations. Finally
the last equation, coupled with \eqref{E:HP.Constraints.e} tells us that
$\cR_{a b\, \overline{cd}}$ vanishes. In summary, we have
\begin{subequations}
\begin{alignat}{2}
\cR_{a b\, c d} &= \tfrac{1}{45} \eta_{a [c} \eta_{d] b}\, \cR &\qquad
\cR_{\overline{a b}\, \overline{c d}} &= \tfrac{1}{45} \eta_{\ba [\bc} \eta_{\bd] \bb}\, \bar\cR \\
\cR_{a \bb \, c d} &= \tfrac{2}{9} \,\eta_{a [c} \cR^e{}_{\bb\, e d]}~, &\qquad
\cR_{a \bb \, \overline{cd}} &= \tfrac{2}{9} \eta_{\bb [\bd} \cR_{a \overline{e}\, \bc]}{}^{\overline e}~, \\
\cR_{a b\, \overline{cd}} &= \cR_{\overline{a b}\, cd} = 0~.
\end{alignat}
\end{subequations}
The remaining Bianchi identities from \eqref{E:BI.Torsion.a} (and using
\eqref{E:HP.Constraints.h}) are
\begin{alignat}{3}
\cB_{\alpha}{}^\beta{}_{\overline{cd}} &= 0 
&\quad &\implies &\quad
\cR_\alpha{}^\beta{}_{\overline{cd}}
    &= -\tfrac{1}{4} \cR_{\overline{cd}\, a b} (\gamma^{ab})_\alpha{}^\beta = 0~, \eol
\cB_{\alpha}{}^\beta{}_{c d} &= 0 
&\quad &\implies &\quad
\cR_{\alpha}{}^\beta{}_{cd}
    &= - \tfrac{1}{4} (\gamma^{ab})_{\alpha}{}^\beta \cR_{cd\, ab}
%     - 2 \,\cR_{\alpha [c d]}{}^\beta
    = - \tfrac{1}{180} (\gamma_{c d})_\alpha{}^\beta\, \cR~, \eol
\cB_\alpha{}^\beta{}_\gamma{}^\delta &= 0
&\quad &\implies &\quad
\cR_{\alpha \gamma}{}^{\beta \delta} &= - \cR_{(\alpha}{}^{(\beta}{}_{cd}\, (\gamma^{cd})_{\gamma)}{}^{\delta)}
    = \tfrac{1}{180} (\gamma^{cd})_{(\alpha}{}^{(\beta} (\gamma_{cd})_{\gamma)}{}^{\delta)}\, \cR~, \eol
\cB_{\alpha}{}^\beta{}_{c \bd} &= 0 
&\quad &\implies &\quad
\cR_{\alpha \bd\, c}{}^\beta &= \tfrac{1}{4} (\gamma^{ab})_\alpha{}^\beta \cR_{c \bd \, a b}~.
\end{alignat}
The last equation above implies a constraint on $\cR_{c\bd\, a b}$ due to the $\gamma$-traceless
left-hand side, but this kills the remaining representation, leading to
\begin{align}
\cR_{a \bb\, c d} = 0~, \qquad
\cR_{\alpha \bb \,c}{}^\gamma = 0~.
\end{align}
We are nearly finished. The remaining Bianchi identities at dimension two are
\begin{alignat}{3}
\cB_{a \bb} &= 0 
&\quad &\implies &\quad
% \cR_{a \bd\, \bb}{}^\bd &= - \cR_{d \bb\, a}{}^d ~, \eol
\cR_{a \bd\, \bb}{}^\bd &= -\cR_{\bb d\, a}{}^d ~, \eol
\cB_{a b} &= 0
&\quad &\implies &\quad
\cR_{a b} &= 0~, \eol
\cB_{\overline{a b}} &= 0
&\quad &\implies &\quad
\cR_{\overline{a b}} &= 0~, \eol
\cB &=0
&\quad &\implies &\quad
\cR + \bar \cR &= 0~, \eol
\cB_{\alpha}{}^\beta &= 0
&\quad &\implies &\quad
% \cR_\alpha{}^\gamma{}_\gamma{}^\beta &= 0 \quad \implies \quad \cR = 0~.
\cR_{\alpha\gamma}{}^{\gamma\beta} &= 0 \quad \implies \quad \cR = 0~.
\end{alignat}
The first four equations are analogous to bosonic ones.
The first tells us that there is a single ``Einstein tensor''
given by
% $\cR_{a \bb} = \cR_{a c \,\bb}{}^{c} = -\cR_{a \bc \,\bb}{}^{\bc}$,
$\cR_{a \bb} = \cR_{a \bd \,\bb}{}^{\bd} = -\cR_{\bb d \,a}{}^{d}$,
but this vanishes due to the $\cB_{\alpha}{}^\beta{}_{c\bd}$ Bianchi discussed
above. The fourth equation tells us that there is a single scalar curvature,
given by
$\cR = \cR_{ab}{}^{ab} = -\cR_{\overline{ab}}{}^{\overline{ab}}$.
The final Bianchi identity tells us this scalar curvature vanishes as well.

Remarkably, all of the torsions and curvatures have been eliminated through dimension two.
This was to be expected because there are no \emph{on-shell} component curvatures
or covariant fields at this level.

%%%%%%%%%%%%%%%%%%%%%%%%%%%%%%%%%%%%%%%%%%%%%%%%%%%%%%%%%%%%%%%%%%%%%%%%%%%%%%%%%
\section{Derivation of component supersymmetric DFT}\label{S:Components}
%%%%%%%%%%%%%%%%%%%%%%%%%%%%%%%%%%%%%%%%%%%%%%%%%%%%%%%%%%%%%%%%%%%%%%%%%%%%%%%%%
The analysis by Siegel \cite{Siegel:1993th} did not directly address the component structure 
of double field theory. Since then, the component supersymmetric formulation has been
independently derived by Hohm and Kwak \cite{Hohm:2011nu} to second order in fermions and 
by Jeon, Lee, and Park to all orders \cite{Jeon:2011sq}.
Already one can see from the linearized transformations 
\eqref{E:SDFT.comptrafo1} -- \eqref{E:SDFT.comptrafo2}
that the results of
\cite{Hohm:2011nu, Jeon:2011sq} are recovered to this order if we solve
the fermionic part of the section condition by setting $\pa^\mu = 0$,
leaving the doubled bosonic space untouched. Our goal in the remainder of this paper
will be to flesh out this connection to all orders.

It may be surprising that this is not as easy as one might expect. There are standard
techniques developed for relating component fields to superfields, and in the case where
the superfield description is geometric -- that is, involving connections, curvatures, gauge
transformations, diffeomorphisms, and so on -- the path from superspace to components
is straightforward. Much of that intuition implicitly assumes that
diffeomorphisms and the supervielbein have a simple $\g{GL}(D|s)$ structure. 
Since that no longer is the case here, we will find that greater care must be taken.

\subsection{Conventional $\cN=1$ superspace: a review and a redo}
As a brief review and a warm-up, let's recall how conventional superspace leads
to a component theory. We take a conventional $\cN=1$ superspace for 
some $\g{GL}(D|s)$ structure group. The supervielbein is
\begin{align}
E_\tM{}^\tA &=
\begin{pmatrix}
e_\rmm{}^\ra & \psi_\rmm{}^\alpha \\
E_\mu{}^\ra & E_\mu{}^\alpha
\end{pmatrix}~.
\end{align}
The fields denoted $e_\rmm{}^\ra$ and $\psi_\rmm{}^\alpha$ correspond to
the physical vielbein and gravitino; the other two fields are unphysical.
To see that $e$ and $\psi$ transform appropriately, recall that
superdiffeomorphisms act as
\begin{align}
\delta_\xi E_\tM{}^\tA = \xi^\tN \pa_\tN E_\tM{}^\tA + \pa_\tM \xi^\tN E_\tN{}^\tA~.
\end{align}
Specializing to a purely bosonic superdiffeomorphism, $\xi^\tM = (\xi^\rmm, 0)$,
it is easy to see that $e$ and $\psi$ transform as conventional 1-forms
with parameter $\xi^\rmm$. Considering now \emph{covariant} superdiffeomorphisms
\begin{align}
\delta^{\rm cov}_\xi E_\tM{}^\tA = \cD_\tM \xi^\tA + E_\tM{}^\tB \xi^\tC T_{\tC\tB}{}^\tA
\end{align}
one can see that specializing to a fermionic parameter $\xi^\tA = (0, \eps^\alpha)$
gives\footnote{The two classes of transformations,
$\xi^\tM = (\xi^\rmm,0)$ and $\xi^\tA = (0, \eps^\alpha)$,
span the full space of transformations, so we lose nothing by focusing on these.}
\begin{align}\label{E:ConventionalSUSY}
\delta e_\rmm{}^\ra &= e_\rmm{}^\rb \eps^\gamma \, T_{\gamma \rb}{}^\ra
    + \psi_\rmm{}^\beta \eps^\gamma T_{\gamma \beta}{}^\ra~, \eol
\delta \psi_\rmm{}^\alpha &= \cD_\rmm \eps^\alpha
    + e_\rmm{}^\rb \eps^\gamma \, T_{\gamma \rb}{}^\alpha
    + \psi_\rmm{}^\beta \eps^\gamma T_{\gamma \beta}{}^\alpha~.
\end{align}
For the vielbein, the first torsion term usually vanishes and the second
is a $\gamma$-matrix. For the gravitino, its torsion terms usually
involve additional fields present in the supergravity multiplet. These transformations
involve no explicit $\theta$ derivatives and can consistently be applied at
$\theta=0$.
The additional fields $E_\mu{}^\ra$ and $E_\mu{}^\alpha$ can be set to zero at
lowest order in $\theta$ as they transform with leading terms 
$\cD_\mu \xi^\ra$ and $\cD_\mu \xi^\alpha$. By a judicious choice of the higher $\theta$-dependence
of $\xi^\ra$ and $\xi^\alpha$ (which do not affect the bottom components of
$e_\rmm{}^\ra$ and $\psi_\rmm{}^\alpha$), one can eliminate the bottom components of
$E_\mu{}^\ra$ and $E_\mu{}^\alpha$.\footnote{This is not quite the 
whole story because one should inquire about the higher
$\theta$-components of these superfields. One can show using a
normal coordinate expansion in $\theta$, that the $\theta$-expansion of $E_\tM{}^\tA$ is
determined purely from the bottom components of $E_\tM{}^\tA$, the torsion tensor,
the Riemann tensor, and their various covariant derivatives.}

A parallel analysis as above is not really available in double field theory. The main
hangup is that the supervielbein in DFT is a constrained object: its entries are not 
independent. Let us redo the above analysis in a way that will generalize more
directly to the DFT situation. Admittedly, this will look rather pointlessly complicated
for $\g{GL}(D|s)$.

We begin by decomposing the supervielbein as
\begin{align}
E_\tM{}^\tA &=
\begin{pmatrix}
e_\rmm{}^\ra & \psi_\rmm{}^\alpha \\
\Xi_\mu{}^\rn e_\rn{}^\ra & \phi_\mu{}^\alpha + \Xi_\mu{}^\rn \psi_\rn{}^\alpha
\end{pmatrix}
\end{align}
where $\Xi$ and $\phi$ are new parametrizations for the second row.
Denoting the tangent-space valued gravitino with a tilde for compactness,
$\tilde \psi_\ra{}^\alpha = e_\ra{}^\rmm \psi_\rmm{}^\alpha$,
this expression can be written as the product of four factors:
\begin{align}\label{E:ConvE.Decomposed}
E_\tM{}^\tA &=
\begin{pmatrix}
1 & 0 \\
\Xi & 1
\end{pmatrix}
\times
\begin{pmatrix}
1 & 0 \\
0 & \phi
\end{pmatrix}
\times
\begin{pmatrix}
e & 0 \\
0 & 1
\end{pmatrix}
\times
\begin{pmatrix}
1 & \tilde\psi \\
0 & 1
\end{pmatrix} \eol
    &= 
\exp( \Xi_\mu{}^\rmm X_\rmm{}^\mu) \times
\exp( a_\mu{}^\nu X_\nu{}^\mu) \times
\exp( a_\rmm{}^\rn X_\rn{}^\rmm ) \times
\exp( \tilde\psi_\rmm{}^\mu X_\mu{}^\rmm) \eol
    &= \cV_{-1} \times \cV_{0'} \times \cV_0 \times \cV_{+1}
\end{align}
We have decomposed the $\g{GL}(D|s)$ generators $X_\tM{}^\tN$ as 
$X_\rmm{}^\nu$, $X_\mu{}^\nu$, $X_\rmm{}^\rn$, and $X_\mu{}^\rn$. An external automorphism
of the algebra permits us to assign level $-1$ to the first generator, $0$ to the second
and third, and $+1$ to the last. We can label the three factors as $\cV_{\ell}$ with respect
to this level.
In the above parametrization, $e = \exp(a_\rmm{}^\rn)$ and $\phi = \exp(a_\mu{}^\nu)$,
and we will work with $e$ and $\phi$ directly.

Let's check that this makes sense. A Lorentz transformation acts on the right as 
$\delta E = -E \Lambda$ and is purely level 0 (and $0'$). It follows that
\begin{align}
\delta_\Lambda \cV_{+1} = -[\cV_{+1}, \Lambda]~, \qquad
\delta_\Lambda (\cV_{0'} \cV_0) = \cV_{0'} \cV_0 \Lambda~, \qquad
\delta_\Lambda \cV_{-1} = 0~.
\end{align}
In other words, the gravitino $\tilde\psi$ is covariant on both indices, $e$ and $\phi$
transform only from the right, and $\Xi$ is invariant.

Next, we check the action of diffeomorphisms. For simplicity, we ignore the transport term,
considering only general coordinate transformations, where
$\delta_{\rm g.c.} E = \cK E$ for $\cK = \pa_M \xi^N$.
This has an obvious level decomposition:
\begin{align}
\cK_{-1} = \pa_\mu \xi^n~, \qquad 
\cK_0 = \pa_m \xi^n~, \qquad 
\cK_{0'} = \pa_\mu \xi^\nu~, \qquad
\cK_{+1} = \pa_m \xi^\nu~.
\end{align}
We focus on bosonic transformations with $\xi^\tM = (\xi^\rmm,0)$. Then only $\cK_{-1}$ and
$\cK_0$ are nonzero, leading to
\begin{align}
\delta_{\rm g.c.} \cV_{-1} = \cK_{-1} + [\cK_0, \cV_{-1}] ~, \qquad
\delta_{\rm g.c.} \cV_{0} = \cK_0 \cV_0 ~, \qquad
\delta_{\rm g.c.} \cV_{0'} = 0~, \qquad
\delta_{\rm g.c.} \cV_{+1} = 0~.
\end{align}
Evidently, $\Xi$ shifts by $\pa_\mu \xi^\rn$ (so its lowest component can be set to zero),
$e$ transforms as a 1-form, and $\phi$ and $\tilde\psi$ are scalars.

Finally, we look at covariant diffeomorphisms, which are given by
\begin{align}\label{E:ConventionalK}
\delta E = E \cK~, \qquad
\cK_\tB{}^\tA = \cD_\tB \xi^\tA + \xi^\tC T_{\tC \tB}{}^\tA~.
\end{align}
We are interested in a fermionic transformation with $\xi^\tA = (0, \eps^\alpha)$.
The first term of $\cK$ contributes only to $\cK_\rb{}^\alpha$ and $\cK_\beta{}^\alpha$,
but the torsion term contributes to all elements of $\cK$, at least in principle.
Now we need to equate this to an arbitrary variation of $E$. This can be written
as $\delta E = E \cJ$, but it's actually going to be more useful to parametrize
the current a little differently. We choose to define $\mathring \cJ$ via
\begin{align}
\delta E &= \cV_{-1} \cV_{0'} \cV_0 \times \mathring \cJ \times \cV_{+1}~.
\end{align}
This is related to $\cJ = E^{-1} \delta E $ by composition by $\cV_{+1}$,
$\mathring \cJ = \cV_{+1} \cJ (\cV_{+1})^{-1}$.

At this point, decomposing the vielbein with respect to the level $\ell$ starts
to become notationally cluttered. Let's denote
$\cV_\Xi = \cV_{-1} \cV_{0'}$ and $\cV_\Psi = \cV_{+1}$, so that
\begin{align}
E = \cV_\Xi \times \cV_0 \times \cV_\Psi~.
\end{align}
The pieces $\cV_0$ and $\cV_\Psi$ involve the component graviton and gravitino, while
$\cV_\Xi$ involves pieces that can be set to zero at lowest level in $\theta$.
Then we have
\begin{align}
\delta E &= \cV_\Xi \,\cV_0 \times \mathring \cJ \times \cV_\Psi~, \qquad
\mathring \cJ = \cV_\Psi \,\cJ \,\cV_\Psi^{-1}~.
\end{align}
The current $\mathring \cJ$ decomposes as follows:
\begin{alignat}{2}
\mathring \cJ_\ra{}^\rb &\equiv \Delta e_\ra{}^\rb = e_\ra{}^\rmm \delta e_\rmm{}^\ra~, 
& \qquad
\mathring \cJ_\ra{}^\beta &= \delta \tilde\psi_\ra{}^\beta = \delta (e_\ra{}^\rmm \psi_\rmm{}^\beta)~,  \eol
\mathring \cJ_\alpha{}^\beta &= \phi_\alpha{}^\mu \delta \phi_\mu{}^\beta
&\qquad
\mathring \cJ_\alpha{}^\rb &= \phi_\alpha{}^\mu \delta \Xi_\mu{}^\rn e_{\rn}{}^\ra
~.
\end{alignat}
We want to equate this with $\mathring \cK = \cV_\Psi \cK (\cV_\Psi)^{-1}$.
Writing $\cV_\Psi = \exp \psi$ in terms of the level +1 generator $\psi$,
we can compute easily that
\begin{align}
\mathring \cK_{+1} &=  \cK_{+1} 
    + [\psi,  \cK_0 + \cK_{0'}]
    + \tfrac{1}{2} [\psi,[\psi,  \cK_{-1}]]~, \eol
\mathring \cK_{0} + \mathring \cK_{0'}
    &=  \cK_0 +  \cK_{0'} + [\psi,  \cK_{-1}]~, \eol
\mathring \cK_{-1} &=  \cK_{-1}~.
\end{align}
From this, one can read off the graviton variation,
\begin{align}
\Delta e_\ra{}^\rb = e_\ra{}^\rmm \delta e_\rmm{}^\rb = 
    \eps^\gamma T_{\gamma \ra}{}^\rb
    + \psi_\ra{}^\alpha \eps^\gamma T_{\gamma \alpha}{}^\rb
\end{align}
which coincides with \eqref{E:ConventionalSUSY}.
For the gravitino, the answer is more involved:
\begin{align}
\delta (e_\ra{}^\rmm \psi_\rmm{}^\beta)
    &= 
    \cD_\ra \eps^\beta + \eps^\gamma T_{\gamma \ra}{}^\beta
    + \psi_\ra{}^\alpha (\cD_\alpha \eps^\beta + \eps^\gamma T_{\gamma \alpha}{}^\beta)
    - \eps^\gamma T_{\gamma \ra}{}^\rb \psi_\rb{}^\beta
    - \psi_\ra{}^\alpha \eps^\gamma T_{\gamma \alpha}{}^\rb \psi_\rb{}^\beta \eol
    &= e_\ra{}^\rmm \cD_\rmm \eps^\beta
    + \eps^\gamma T_{\gamma \ra}{}^\beta
    + \psi_\ra{}^\alpha \eps^\gamma T_{\gamma \alpha}{}^\beta
    - \Delta e_\ra{}^\rb \,\psi_\rb{}^\beta~.
\end{align}
The two covariant derivative terms have combined into $e_\ra{}^\rmm \cD_\rmm$, which is independent
of $\theta$ derivatives. The first three terms in the final expression
coincide with the expected result
\eqref{E:ConventionalSUSY}, and the last arises just from the transformation of the
vielbein. Finally, one can also show that the leading contribution to $\Delta \phi_{0'}$
is as before $\pa_\mu \xi^\alpha$, so this term can be eliminated to lowest order in $\theta$.

\subsection{Level decomposition of the doubled supervielbein}

In parallel with the analysis of the conventional $\cN=1$ supervielbein, we will
first perform a level decomposition of the generators
$X^{\cM \cN}$ of $\g{OSp}(10,10|32)$:
\begin{align}
\underbrace{X^{\mu \nu}}_{\text{level -2}}, \quad 
\underbrace{X^{\mu \hn}}_{\text{level -1}}, \quad 
\underbrace{X^{\hm \hn}~, X_\mu{}^\nu}_{\text{level 0}}~, \quad 
\underbrace{X_{\mu \hm}}_{\text{level +1}}~, \quad 
\underbrace{X_{\mu \nu}}_{\text{level +2}}~,
\end{align}
where the level denotes the difference between the number of lower and upper fermionic indices.
As in the previous example, we are going to place the positive level generators to the right
in $\cV_\cM{}^\cA$, so they are naturally going to lead to fields possessing tangent space
indices. With that in mind, we introduce fields associated with these generators in
Table \ref{T:OSpV.Pieces}.

\begin{table}[t]
\centering
\renewcommand{\arraystretch}{1.3}
\begin{tabular}{ccc}
\toprule
field & generator & level \\ \midrule
$C^{\alpha \beta}$ & $X_{\alpha \beta}$ & $+2$ \\
$\chi_a{}^{\beta}$ & $X_\beta{}^a$&  $+1$ \\ 
$\Psi_{\bar a}{}^{\beta}$ & $X_\beta{}^{\bar a}$ & $+1$ \\
$V_{\hat m}{}^{\hat a}$ & $X_{\hat m}{}^{\hat n}$ & $0$ \\
$\phi_{\mu}{}^{\alpha}$ & $X_{\mu}{}^\nu$ & $0$ \\
$\Xi_{\mu}{}^\hm$ & $X_\hm{}^{\mu}$ & $-1$ \\
$B_{\mu \nu}$ & $X^{\mu \nu}$ & $-2$ \\
\bottomrule
\end{tabular}
\captionsetup{width=0.6\textwidth}
\caption{Constituent fields of the $\g{OSp}$ supervielbein.
Positive level fields are written with Lorentz indices.}
\label{T:OSpV.Pieces}
\end{table}

The fields associated to the positive level generators we have denoted $C^{\alpha \beta}$,
$\chi_a{}^\alpha$, and $\Psi_\ba{}^\alpha$. Using the additional tangent space symmetry,
we are going to gauge away $C^{\alpha \beta}$ and the spin-3/2 part of
$\chi_a{}^\alpha$, fixing the latter to 
$\chi_a{}^\alpha = \tfrac{1}{10} (\gamma_a)^{\alpha \beta} \chi_\beta$.
The field $V_\hm{}^\ha$ will become the bosonic doubled vielbein.
The remaining fields have analogues with fields in conventional $\cN=1$ superspace.
$\phi_\mu{}^\alpha$ is the identically named field there, and corresponds to the inverse
of the $\cN=1$ superspace vielbein component $E_\alpha{}^\mu$.
$\Xi_\mu{}^\rn$ corresponds to the similarly named field there.
The remaining fields $\Xi_{\mu \rn}$ and $B_{\mu \nu}$ correspond to 
the fermionic legs of the super 2-form $B_{\tM\tN}$.
As in conventional $\cN=1$ superspace,
the bottom components of $\phi_\mu{}^\alpha$, $\Xi_\mu{}^\rn$, and $B_{\mu\nu}$ 
can be eliminated by using the higher $\theta$-components of superdiffeomorphisms.

We want to decompose the supervielbein analogously to how we did in conventional superspace \eqref{E:ConvE.Decomposed}. 
To fix our conventions, we normalize
the fields associated with the \emph{nonzero} generators so that they fill out a graded symmetric element
$\cA_{\cA\cB}$ of $\g{OSp}(10,10|32)$ as
\begin{equation}\label{E:AGen}
\begin{aligned}
\cA_{\cA\cB} &=
\begin{pmatrix}
0 & 0 & \Xi & \chi \\
0 & 0 & \bar \Xi & \bar\Psi \\
-\Xi & -\bar \Xi & B & 0 \\
-\chi & -\bar\Psi & 0 & C \\
\end{pmatrix}
\end{aligned} \qquad 
\begin{gathered}
\Xi = \Xi_{m \nu}~, \qquad
\bar \Xi = \Xi_{\bar m \nu}~, \qquad
B = B_{\mu \nu}~, \\
\chi = \chi_a{}^\beta~,\qquad
\bar\Psi = \Psi_{\bar a}{}^\beta~, \qquad
C = C^{\alpha \beta}~.
\end{gathered}
\end{equation}
We will be using the left/right basis for the generators, so that
$\Xi_{\hm \nu}$ has been decomposed into $\Xi_{m \nu}$ and $\Xi_{\bar m \nu}$.
We have also distinguished between world and tangent space indices
based on how we are going to place these generators into the
coset, although this is perhaps sloppy notation.

We have denoted the gravitino by $\bar\Psi$ above because later on it will be
convenient to also use $\Psi = \Psi_a{}^\beta$ as an alias for $\chi_a{}^\beta$.
Then $\Psi$ and $\bar\Psi$ appear in an analogous way as $\Xi$ and $\bar\Xi$
in some of the formulae.
We won't do this just yet because we want to keep in mind $\chi$ is the dilatino
and not a second gravitino.

Exponentiating the above generators using $\cV = \exp (\cA_{\cA}{}^\cB)$ for each level
gives
\begin{align}
\cV_{+2} &= 
\begin{pmatrix}
1 & 0 & 0 & 0 \\
0 & 1 & 0 & 0 \\
0 & 0 & 1 & 0 \\
0 & 0 & C & 1
\end{pmatrix}~, \qquad
\cV_{-2} = 
\begin{pmatrix}
1 & 0 & 0 & 0 \\
0 & 1 & 0 & 0 \\
0 & 0 & 1 & -B \\
0 & 0 & 0 & 1
\end{pmatrix}~, \qquad
B = B_{\mu \nu}~, \qquad
C = C^{\alpha \beta}~,
\eol
\cV_{+1} &=
\begin{pmatrix}
1 & 0 & \chi & 0 \\
0 & 1 & \bar\Psi & 0 \\
0 & 0 & 1 & 0 \\
-\chi^T & \bar\Psi^T & -\tfrac{1}{2} (\chi^T\chi- \bar\Psi^T \bar\Psi) & 1
\end{pmatrix}~, \qquad
\begin{gathered}
\chi = \chi_a{}^\beta~,\qquad
\bar\Psi = \Psi_{\bar a}{}^\beta~, \\
(\chi^T)^{\alpha b} =  \chi^{b \alpha}~, \quad
(\bar\Psi^T)^{\alpha \bb} = \bar\Psi^{\bb \alpha}~,
\end{gathered}~, \eol
\cV_{-1} &= 
\begin{pmatrix}
1 & 0 & 0 & -\Xi \\
0 & 1 & 0 & -\bar\Xi \\
-\Xi^T & \bar\Xi^T & 1 & \tfrac{1}{2} (\Xi^T \Xi - \bar\Xi^T \bar \Xi) \\
0 & 0 & 0 & 1
\end{pmatrix}~, \qquad
\begin{gathered}
\Xi = \Xi_{m \nu}~, \qquad
\bar \Xi = \Xi_{\bar m \nu}~, \\
(\Xi^T)_\mu{}^{n} =  \Xi^{n}{}_{\mu}~, \quad
(\bar\Xi^T)_\mu{}^{\bar n} = \bar\Xi^{\bar n}{}_{\mu}~,
\end{gathered}
\end{align}
At level 0, there are two commuting generators $X_\hm{}^\hn$ and $X_\mu{}^\nu$.
Their exponentiated elements are
\begin{align}
\cV_0  &=
\left(\begin{array}{c|c}
V_{\hat m}{}^{\hat a}  &
\begin{array}{cc}
0 & 0 \\
0 & 0
\end{array} \\\hline
\begin{array}{cc}
0 & 0 \\
0 & 0
\end{array} &
\begin{array}{cc}
1 & 0 \\
0 & 1
\end{array}
\end{array}\right)~, \qquad
\cV_\phi =
\begin{pmatrix}
1 & 0 & 0 & 0 \\
0 & 1 & 0 & 0 \\
0 & 0 & \phi_\mu{}^\alpha & 0 \\
0 & 0 & 0 & \phi_\alpha{}^\mu
\end{pmatrix}~.
\end{align}
Here $V_\hm{}^\ha$ will become the component DFT vielbein.

Using these building blocks, we can construct a generic orthosymplectic element as
\begin{align}
\cV = \underbrace{\cV_{-2} \cV_{-1} \cV_\phi}_{\cV_\Xi} \times \cV_0 \times 
\underbrace{\cV_{+1} \cV_{+2}}_{\cV_\Psi}
    = \cV_\Xi \times \cV_0 \times\cV_\Psi
\end{align}
The interpretation of these three factors will be similar to conventional $\cN=1$
superspace. The factor $\cV_0$ contains the bosonic double vielbein. $\cV_\Psi$
involves the gravitino, dilatino, and other pieces that will be set to zero by
a tangent space transformation. $\cV_\Xi$ involves fields that live purely in
superspace and have no component analogues, as their $\theta=0$ parts can be
eliminated by a $\theta$-dependent diffeomorphism.

Let's see now if the decomposition we have chosen is actually sensible from the perspective
of the component theory. Specifically, does it lead to a sensible parametrization of the derivatives $D_\cA$? It is helpful to first split off the gravitino (and dilatino) piece and write
$\zcV = \cV_\Xi \cV_0$. This is given by
\begin{equation}
\begin{aligned}
\zcV^{-1} &=
\begin{pmatrix}
V_a{}^m & V_a{}^{\bar m} & 0 & {\rm x} \\
V_{\bar a}{}^a & V_{\bar a}{}^{\bar m} & 0 & {\rm x} \\
{\rm x} & {\rm x} & \phi_\alpha{}^\mu & {\rm x} \\
0 & 0 & 0 & \phi_\mu{}^\alpha
\end{pmatrix}
\end{aligned}
~, \qquad
\begin{aligned}
\zcV_\alpha{}^{\hat n} &= -\phi_{\alpha}{}^\nu \Xi^{\hat n}{}_\nu~, \qquad
\zcV_{\hat a}{}_\nu = V_{\hat a}{}^{\hat m} \Xi_{\hat m \nu}~, \\
\zcV_{\alpha}{}_\nu &= \phi_\alpha{}^\mu (B_{\mu \nu} 
    + \tfrac{1}{2} \Xi_{\hat m \mu} \Xi^{\hat m}{}_\nu)~.
\end{aligned}
\end{equation}
The entries marked x in the matrix are given above.
As we will solve the fermionic part of the section condition 
by taking $\pa^\mu = 0$, it follows that
\begin{align}
\zD_\cA = 
\zcV_\cA{}^\cM \pa_\cM &=
\begin{pmatrix}
V_{\hat a}{}^{\hat m} \pa_{\hat m} \\ 
\zcV_\alpha{}^{\hat m} \pa_{\hat m} + \phi_\alpha{}^\mu \pa_\mu \\
0
\end{pmatrix}
\equiv
\begin{pmatrix}
\zD_{\hat a} \\ 
\zD_{\alpha} \\
0
\end{pmatrix}
\end{align}
where we define $\zD_{\ha}$ and $\zD_\alpha$ by these expressions.
Now we can build $D_\cA = \cV_\cA{}^\cM \pa_\cM = (\cV_\Psi^{-1})_\cA{}^\cB \zD_\cB$ and this leads to
\begin{align}\label{E:DtoComp}
D_\cA  = (\cV_\Psi^{-1})_\cA{}^\cB \zD_\cB =
\begin{pmatrix}
\zD_\ha - \Psi_\ha{}^\alpha \zD_\alpha \\
\zD_\alpha \\
\Psi^{\hb \alpha} \zD_{\hb}
- \Big(C^{\alpha \beta} + \frac{1}{2} \Psi^{\hc \alpha} \Psi_{\hc}{}^\beta \Big) \zD_\beta
\end{pmatrix}
\end{align}
where we have used the condensed notation
$\Psi_\ha{}^\beta = (\chi_a{}^\beta, \Psi_\ba{}^\beta)$
and $\Psi^\ha{}^\beta = (\chi^{a \beta}, -\Psi^{\ba \beta})$.
It is extremely useful that $D_\alpha = \zD_\alpha$. We can then rewrite 
easily the other derivatives as
\begin{align}\label{E:CompDs}
D_\ha = \zD_\ha - \Psi_\ha{}^\beta D_\beta~, \qquad
D^\alpha = 
\Psi^{\hb \alpha} D_{\hb}
- \Big(C^{\alpha \beta} - \tfrac{1}{2} \Psi^{\ha \alpha} \Psi_{\ha}{}^\beta \Big) D_\beta~.
\end{align}
The bosonic derivative $D_\ha$ is now directly analogous to the conventional superspace
derivative, where it is given by the component flat derivative $\zD_\ha$ modified by
a gravitino connection. The new operator in DFT is the dual fermionic derivative
$D^\alpha$. Its explicit form will be crucial in subsequent computations.
In neither expression above do the pieces of $\cV_\Xi$ appear
explicitly; they appear implicitly only via $D_\alpha$.

\emph{It is crucial that the above expressions
hold only upon imposing $\pa^\mu=0$}. Without this condition, the expressions are
significantly more complicated.

\subsection{Generalized diffeomorphisms}
The next task is to figure out how the elements of the supervielbein transform,
in order to verify that they have been correctly identified.

We begin with generalized diffeomorphisms. As in conventional superspace, it is best
to split these up between standard diffeomorphisms $\xi^\cM$ and covariant ones 
$\xi^\cA$. It is not immediately obvious how to treat the dual fermionic direction,
but it will turn out that we should keep this as a standard diffeomorphism. In other words,
we are going to examine $\xi^\cM = (\xi^\hm, 0, -\tilde\xi_\mu)$ and $\xi^\cA = (0, \eps^\alpha, 0)$.
We begin with the former. Ignoring the transport term,
\begin{align}
\delta_{\rm g.c.} \cV_{\cM}{}^\cA = 
    \cK_\cM{}^\cN \cV_\cN{}^\cA~, \qquad
\cK_\cM{}^\cN = \pa_\cM \xi^\cN - \pa^\cN \xi_\cM (-)^{nm}
\end{align}
Since we restrict to $\pa^\mu=0$, it turns out that we only have non-positive
levels for $\cK$. Using the fact that the $-1$ generators commute with the $-2$ generators,
one can show that
\begin{gather}
\delta_{\rm g.c.} \cV_{\Xi} = \cK_{-2} \cV_{\Xi} + \cK_{-1} \cV_{\Xi}  + [\cK_0, \cV_{\Xi}]~, \qquad
\delta_{\rm g.c.} \cV_0 = \cK_0 \cV_0~, \qquad
\delta_{\rm g.c.} \cV_{\Psi} = 0~.
\end{gather}
It is useful here that the level zero element $\cK_0$ only involves
$\pa_\hm \xi^\hn$ and not $\pa_\mu \xi^\nu$ or $\pa^\mu \tilde \xi_\nu$, as this
simplifies the transformation of $\cV_\Xi$.
Because the $\cK_0$ piece is just the bosonic $\g{O}(10,10)$ general 
coordinate transformation, $\cV_0$ transforms as a DFT vielbein
should, while the gravitino and dilatino are scalar fields.
The negative level transformations are
completely soaked up by the fields in $\cV_\Xi$, which transform as
\begin{align}
\delta B_{\mu \nu} = \pa_\mu \tilde\xi_\nu + \pa_\nu \tilde\xi_\mu ~, \qquad
\delta \Xi_{\hat m \nu} = \pa_{\hat m} \tilde\xi_\nu -\pa_\nu \xi_{\hat m}~.
\end{align}
As in conventional superspace, these fields can be set to zero at lowest level in $\theta$.

Because the fields in $\cV_\Psi$ are inert under $\tilde\xi_\mu$ and transform as scalars
under $\xi^\hm$, their gauge-fixing is not disturbed. That is, 
if $C^{\alpha \beta} = 0$, it stays zero. 
If $\chi_a{}^\alpha$ is $\gamma$-traceless, it remains so.
This means component diffeomorphisms do not need to be defined with
a compensating tangent space transformation. This is important: it ensures that
their algebra remains unchanged in going from 
superspace to components.

\subsection{Supersymmetry transformations}
\label{S:SUSYtrafo}
Supersymmetry transformations are encoded in covariant diffeomorphisms.
Let's analyze these next. Including a compensating tangent
space transformation $\lambda_\cA{}^\cB$ (which will be necessary in this case),
the supervielbein transforms as $\delta \cV = \cV \,(\cK - \lambda)$
where
\begin{align}
\cK_{\cA}{}^{\cB} = \nabla_\cA \xi^\cB - \nabla^\cB \xi_\cA (-1)^{ab}
    + \xi^\cC \cT_{\cC \cA}{}^{\cB}~.
\end{align}
What are the nonzero pieces?
While our focus is on $\xi^\cA = (0, \eps^\alpha, 0)$, we must be careful
about expressions like $\nabla_\cA \xi^\cB = D_{\cA} \xi^{\cB} - \Omega_{\cA}{}^{\cB \cC} \xi_\cC$,
which might lead to an unexpected contribution when the $\Omega$ connection
is not purely double Lorentz.
Luckily, the $\Omega$ piece in our case involves just $\Omega_{\cA}{}^{\beta}{}_\gamma \eps^\gamma$;
there is no pollution from $\Omega_{\cA}{}^{\beta \gamma}$ and $\Omega_{\cA}{}_{b}{}^\gamma$
because we have set $\xi_\alpha$ and $\xi_a$ to zero. We replace
$\nabla$ with $\cD$ here to emphasize $\cD$ carries only the double Lorentz connection.
For the torsion term, only $\cT_{\gamma \beta a}$ is non-vanishing. 
This means the only nonzero elements of $\cK_\cA{}^\cB$ are 
\begin{alignat}{5}
\cK^{\alpha \beta} &= 2\, \cD^{(\alpha} \eps^{\beta)}~, &\quad
\cK_a{}^\beta &= \cD_a \eps^\beta~, &\quad
\cK_\ba{}^\beta &= \cD_\ba \eps^\beta~, &\quad
\cK_\alpha{}^\beta &= \cD_\alpha \eps^\beta~, &\quad
\cK_{\alpha}{}^b &= \kappa (\gamma^{b})_{\alpha \gamma} \eps^\gamma~, \eol
&& 
\cK^\alpha{}^b &= -\cD^b \eps^\alpha~, &\quad
\cK^\alpha{}^\bb &= \cD^\bb \eps^\alpha~, &\quad
\cK^\alpha{}_\beta &= -\cD_\beta \eps^\alpha~, &\quad
\cK_{a \beta} &= \kappa (\gamma_{a})_{\beta \gamma} \eps^\gamma~,
\end{alignat}
corresponding to levels $+2$, $+1$, $+1$, $0$, and $-1$. The possible
compensating $\lambda$ transformations are at levels $+2$, $+1$ and $0$.

As in the conventional $\cN=1$ superspace example, we can parametrize an
arbitrary variation as
\begin{align}
\delta \cV = \cV \times \cJ = \cV_{\Xi} \cV_0 \times \zJ \times \cV_\Psi \quad \implies \quad
\zJ = \cV_\Psi \cJ \cV_\Psi^{-1}~.
\end{align}
For the case of a supersymmetry transformation, $\cJ = \cK - \lambda$.
Now work out $\zJ$ level-by-level, using the fact that $\cV_\Psi \equiv \exp \Psi$
for a generator $\Psi$ at level $+1$ only. (We now work in the gauge where $C=0$.)
\begin{align}
\zJ_{+2} &= \cK_{+2} + [\Psi, \cK_{+1}] 
    + \tfrac{1}{2!} [\Psi, [\Psi, \cK_{0}]] 
    + \tfrac{1}{3!} [\Psi, [\Psi, [\Psi, \cK_{-1}]]] 
    \eol & \quad
    - \lambda_{+2} - [\Psi, \lambda_{+1}]
    - \tfrac{1}{2!} [\Psi, [\Psi, \lambda_{0}]] ~, \eol
\zJ_{+1} &= \cK_{+1} 
    + [\Psi, \cK_0]
    + \tfrac{1}{2} [\Psi, [\Psi, \cK_{-1}]] 
%     \eol & \quad
    - \lambda_{+1}
    - [\Psi, \lambda_{0}]~, \eol
\zJ_0 &= \cK_0 + [\Psi, \cK_{-1}] - \lambda_0~, \eol
\zJ_{-1} &= \cK_{-1}, \eol
\zJ_{-2} &= 0~.
\end{align}
From the explicit expressions for $\cV_\ell$, we find for the non-negative levels,
\begin{alignat}{2}
(\zJ_2)^{\alpha \beta} &= -\Psi^\ha{}^{(\alpha} \delta \Psi_\ha{}^{\beta)}~, &\qquad
(\zJ_1)_\ha{}^\beta &= \delta \Psi_\ha{}^{\beta} ~, \eol
(\zJ_0)_\ha{}^\hb &= V_\ha{}^\hm \delta V_\hm{}^\hb \equiv J_\ha{}^\hb~, &\qquad
(\zJ_0)_\alpha{}^\beta &= \phi_\alpha{}^\mu \delta \phi_\mu{}^\beta~.
\end{alignat}
The expression for $\zJ_2$ is just going to tell is what the
compensating $\lambda_{+2} = \lambda^{\alpha\beta}$ transformation needs to be, 
but its explicit form is irrelevant.
The transformation of $\phi_\mu{}^\alpha$ isn't really relevant here either -- 
it's going to involve a leading term $\pa_\mu \eps^\alpha$ indicating that at lowest
order in $\theta$ we can set $\phi_\mu{}^\alpha = \delta_\mu{}^\alpha$.
For the bosonic DFT vielbein, we find that
\begin{align}\label{E:SUSY.vielbein}
J_{a \bb} = \kappa \, (\eps \gamma_a \Psi_\bb)~, \qquad
J_{a b}
    = \tfrac{1}{5} \kappa \,(\eps \gamma_{ab} \chi) - \lambda_{a b}~, \qquad
J_{\overline{ab}} = - \lambda_{\overline{ab}}~.
\end{align}
The expression for $J_{a \bb}$ is exactly as expected. The nonzero
expression for $J_{ab}$ is interesting, but not really illuminating. It
tells us that in order to make contact with supersymmetric DFT, where $J_{ab}$ is
usually taken to vanish, one should make a compensating $\lambda_{ab}$ transformation
to cancel the dilatino term. Naturally such a choice of $\lambda$ will in turn 
affect the gravitino and dilatino transformations as well as the algebra of
supersymmetry transformations.

The gravitino transformation is a bit more complicated. A direct computation leads to
\begin{align}
\delta \Psi_\ba{}^\beta
    &= \cD_\ba \eps^\beta + \Psi_\ba{}^\gamma \cD_\gamma \eps^\beta
    + \tfrac{1}{10} \kappa\, (\eps \gamma^b \Psi_\ba)\, (\gamma_b \chi)^\beta
    - \tfrac{1}{4} \Psi_\ba{}^\gamma (\gamma^{c d})_\gamma{}^\beta \, \lambda_{cd}
    - \lambda_\ba{}^\bb \Psi_{\bb}{}^\beta ~, \eol
    &= \zcD_\ba \eps^\beta 
    + \tfrac{1}{10} \kappa\, (\eps \gamma^b \Psi_\ba)\, (\gamma_b \chi)^\beta
    - \tfrac{1}{4} \Psi_\ba{}^\gamma (\gamma^{c d})_\gamma{}^\beta \, \lambda_{cd}
    - \lambda_\ba{}^\bb \Psi_{\bb}{}^\beta ~.
\end{align}
In going from the first to the second line, we have combined 
the two derivative terms to give $\zcD_\ba = V_\ba{}^\hm \cD_\hm$ using \eqref{E:DtoComp}.
This is purely a bosonic derivative of $\eps$, with the $\theta$ derivative cancelling out.
This is exactly what needed to happen.
Actually, this is a little bit subtle because $\cD$ carries the spin connection,
which we have not yet defined. We will come back to this in the next section.

For the dilatino, we find
\begin{align}
\delta \chi_a{}^\beta
    &= \zcD_a \eps^\beta 
    + \kappa (\eps \gamma^b \chi_a) \chi_b{}^\beta
    - \tfrac{1}{2} \kappa \,(\eps \gamma_a \chi^b) \chi_b{}^\beta
    + \tfrac{1}{2} \kappa \,(\eps \gamma_a \Psi^\bb) \Psi_\bb{}^\beta
    \eol & \quad
    + \lambda_a{}^b \chi_b{}^\beta
    - \tfrac{1}{4} \chi_a{}^\gamma (\gamma^{c d})_\gamma{}^\beta \, \lambda_{cd}
    - \lambda_a{}^\beta ~.
\end{align}
The last term is the compensating tangent space transformation that we will use to
maintain the gauge $\chi_a{}^\alpha = \tfrac{1}{10} (\gamma_a)^{\alpha \beta} \chi_\beta$.
Contracting with a $\gamma$ matrix collapses this expression to
\begin{align}
\delta \chi_\alpha
    &= (\gamma^a)_{\alpha \beta} \zcD_a \eps^\beta 
    - \tfrac{3}{200} \kappa (\eps \gamma^{ab} \chi) (\gamma_{ab} \chi)_\alpha
    + \tfrac{1}{20} \kappa\, (\eps \chi) \chi_\alpha
    + \tfrac{1}{4} \lambda^{ab} (\gamma_{ab} \chi)_\alpha
    + \tfrac{1}{2} \kappa \,(\eps \gamma^a \Psi^\bb) (\gamma_a \Psi_\bb)_\alpha~.
\end{align}

We have not yet addressed how to define the component dilation $\phi$.
Recall the superdilaton $\Phi$ transforms as \eqref{E:SDFT.VPhi.trafos.b}.
Decomposing indices and neglecting the transport term, this reads
\begin{align}
\delta_{\rm g.c.} \log \Phi = \pa_\hm \xi^\hm - \pa_\mu \xi^\mu + \pa^\mu \tilde \xi_\mu~.
\end{align}
At the component level, the last term can be dropped because we take $\pa^\mu=0$, but 
the second term remains problematic as it will obstruct the construction of a
sensible supersymmetry transformation. The solution is to shift the definition of
the component dilaton relative to the superdilaton:
\begin{align}\label{E:CompDilaton}
\phi = \Phi \times \det \phi_\mu{}^\alpha \quad \implies \quad
\delta \log \phi = \pa_\hm \xi^\hm + 2 \,\pa^\mu \tilde \xi_\mu~.
\end{align}
Now the second term drops out when $\pa^\mu=0$
and the component dilaton transforms as a scalar field.\footnote{Another
way of arriving at this same conclusion is to recall that
the component DFT dilaton is related to the supergravity
dilaton by a factor of $e = \det e_\rmm{}^\ra$, that is,
$\phi = e\, e^{-2 \varphi}$.
The superdilaton in Siegel's superspace DFT is similarly related to
the conventional (non-density) superspace dilaton by a factor of $E = \sdet E_\tM{}^\tA$, i.e.
$\Phi = E\, e^{-2 \varphi}$. In conventional superspace, the component
and superspace dilatons coincide (hence both are $\varphi$ above).
This implies that $\phi = \Phi \times e / E = \Phi \times \det \phi_\mu{}^\alpha$.}

We can easily work out the supersymmetry transformation of the 
proposed component dilaton. Observe that
\begin{align}
\delta \log \det \phi_\mu{}^\alpha = 
\phi_\alpha{}^\mu \delta \phi_\mu{}^\alpha & = (\zJ_0)_\alpha{}^\alpha 
    = \cD_\alpha \eps^\alpha - \kappa\, \eps^\alpha \chi_\alpha~.
\end{align}
Combining with the transformation of the superdilaton,
\begin{align}
\delta \log \Phi = \xi^\cA \cT_\cA + \nabla_\cA \xi^\cA \,(-1)^a
    = - \cD_\alpha \eps^\alpha
\end{align}
we recover the expected component transformation,
\begin{align}
\delta \log \phi = -\kappa\, \eps^\alpha \chi_\alpha~.
\end{align}

\subsection{The component spin connection}
The expressions for the gravitino and dilatino supersymmetry transformations involve the
component spin connection, which we will now specify how to compute.
The form of these transformations suggest we should define\footnote{This definition can equivalently be written as
$\omega_\hm = (\cV_\Xi^{-1})_\hm{}^\cN \Omega_\cN$.
It is interesting that this is \emph{not} the simple $\omega_m = \Omega_m$ relation from
conventional superspace connections, and is another example of how DFT differs.}
\begin{align}\label{E:ComponentOmega}
\omega_{\ha \, \hb \hc} := \Omega_{\ha \, \hb \hc} + \Psi_\ha{}^\alpha \Omega_{\alpha\, \hb \hc}
    = (\cV_\Psi)_\ha{}^\cD \Omega_{\cD\, \hb \hc}~.
\end{align}
These are precisely the corrections needed so that under Lorentz transformations,
only a bosonic derivative of the Lorentz parameter appears.
The key feature of \eqref{E:ComponentOmega} is that it allows us to extend the simple
relation $\zD_\ha = (\cV_\Psi)_\ha{}^\cB D_\cB$ to the covariant derivative as well.
This is crucial when we consider how to translate the superspace torsion tensor to components.

Let's work this out now.
Arbitrary variations $\cJ$ and $J$ of the supervielbein $\cV$ and component vielbein 
$V$ are given by
\begin{align}
\cJ_{\cA \cB} = (\cV^{-1} \delta \cV)_\cA{}^\cC \eta_{\cC \cB}~, \qquad 
J_{\ha \hb} = (V^{-1} \delta V)_\ha{}^\hc \eta_{\hc \hb}~.
\end{align}
Using the explicit form of $\cV$, these are related by
\begin{align}
J_{\ha \hb} = (\cV_\Psi)_\hb{}^\cB (\cV_\Psi)_\ha{}^\cA \cJ_{\cA \cB}~.
\end{align}
This is for an arbitrary variation, but applies equally well if we take a derivative.
Choosing the variation $\delta$ to be action of the flattened component derivative
$\mathring D_\ha = (\cV_\Psi)_\ha{}^\cB D_\cB$, we find that
\begin{align}\label{E:CompJfromSuperJ}
(J_{\hc}){}_{\hb \ha} = (\cV_\Psi)_\ha{}^\cA (\cV_\Psi)_\hb{}^\cB (\cV_\Psi)_\hc{}^\cC 
    (\cJ_{\cC})_{\cB \cA }
\end{align}
where $\cJ_\cC$ is built from $D_\cC$.
This means that the component flux tensor $F_{\hc \hb \ha}$ is related to the
superspace one $\cF_{\cC \cB \cA}$ just by contracting with $\cV_\Psi$ factors.
The same turns out to be true for the component Lorentz connection: here it is crucial
that $(\cV_\Psi)_\cA{}_\beta$ vanishes so that only the doubled Lorentz connection
from superspace contributes. In other words, the expression \eqref{E:CompJfromSuperJ}
holds just as well for the covariant derivative.
Then the component torsion tensor is related to 
the superspace one via the extremely simple result
\begin{align}\label{E:CompTorsion}
T_{\hc \hb \ha } = (\cV_\Psi)_\ha{}^\cA (\cV_\Psi)_\hb{}^\cB (\cV_\Psi)_\hc{}^\cC \cT_{\cC \cB \cA}
\end{align}

This expression can be read in one of two ways. The most obvious
is as a constraint equation on the component torsion $T_{\hc \hb \ha}$
in terms of the constrained superspace torsion. The other way is to
forget briefly that $\cT_{\hc \hb \ha}$ is constrained to vanish
and instead read this as its \emph{definition} in terms of
the component torsion $T_{\hc \hb \ha}$ and the subleading gravitino
corrections involving $\cT_{\gamma \hb \ha}$, $\cT_{\gamma \beta \ha}$,
and so forth. In this way, the resulting object $\cT_{\hc \hb \ha}$ is
called the \emph{supercovariant torsion tensor}; at the component level,
it contains the necessary gravitino (and dilatino) additions to
$T_{\hc \hb \ha}$ to render it supercovariant, meaning that it transforms
under supersymmetry without a derivative of the parameter $\eps$. 
Setting $\cT_{\hc \hb \ha}$ to zero is then understood as the supercovariantized
version of the bosonic torsion constraint.
Since we're mainly interested in figuring out how $\omega$ is constrained here,
the first point of view suffices, but the second will be important to keep
in mind because we will need to return to the concept of supercovariant
curvatures soon.

Expanding out \eqref{E:CompTorsion} using the explicit expression for $\cV_\Psi$ gives the 
relation between the component torsion tensor $T_{\hc \ha \hb}$ and its
supercovariantized version $\cT_{\hc \ha \hb}$.
Using the constraints imposed on the superspace torsion, we find
\begin{align}
T_{a b c} = \tfrac{3}{100} \kappa\, (\chi \gamma_{abc} \chi)~, \qquad
T_{\ba b c} = \tfrac{1}{5} \kappa\, (\Psi_\ba \gamma_{bc} \chi)~, \qquad
T_{a \overline{bc}} = -\kappa\, (\Psi_{\bb} \gamma_a \Psi_\bc)~, \qquad
T_{\overline{abc}} = 0~.
\end{align}
These give fermion bilinear corrections to the component spin connections.
The first and the last constrain only the antisymmetric parts of the
respective spin connections; the trace constraints are found in
the dilaton torsion, and the hook irrep is undetermined as usual.

Deriving the constraints on the dilaton torsion is a good deal more involved.
At the component level, the dilaton torsion is
\begin{align}
T_\ha 
    = \zD_\ha \log \phi + J^\hb{}_{\hb \ha}
\end{align}
where here the currents are understood to be built using the covariant
derivative.
Using the definition for the superdilaton torsion tensor \eqref{E:DilTorsion} and the relation
\eqref{E:CompDilaton} between
the component $\phi$ and the superdilaton $\Phi$, this can be rewritten
\begin{align}
T_a 
    &= (\cV_{\Psi})_a{}^\cA \Big(
        \cT_\cA + \cT_{\cA \beta}{}^\beta
        + \nabla_\cA \phi_\mu{}^\beta \phi_\beta{}^\mu
        + 2 \cJ^{\beta}{}_{\beta \cA} - \cJ_{\cA \beta}{}^\beta
        - \cJ^\hb{}_{\hb \cA}
    \Big) 
    + J^\hb{}_{\hb \ha}
\end{align}
A number of simplifications are now needed. First, we will use \eqref{E:CompJfromSuperJ} to rewrite
$J^{\hb}{}_{\hb \ha}$ in terms of $\cJ_{\cC \cB \cA}$. Second, we will use
$
\phi_\alpha{}^\mu \delta \phi_\mu{}^\alpha = \zJ_\alpha{}^\alpha = \cJ_\alpha{}^\cB (\cV_\Psi^{-1})_\cB{}^\alpha
$.
Finally, we will need to explicitly use the expression \eqref{E:CompDs} for $D^\beta$
to rewrite
$\cJ^\beta{}_{\beta \cA}$ in terms of $\cJ_{\hb \beta \cA}$ and $\cJ_{\gamma \beta \cA}$.
(For this to work, it's crucial that the $\Omega^\beta$ connections end up dropping out of these
particular expressions.) Making all these substitutions, one finds the currents nontrivially
recombine to form torsion tensors. The end result is
\begin{align}
T_\ha
    &= \cT_\ha + \cT_{\ha \beta}{}^\beta + \Psi_\ha{}^\alpha (\cT_\alpha + \cT_{\alpha \beta}{}^\beta)
        + \Psi^{\hb\alpha} \cT_{\alpha \hb \ha}
        - \Psi_\ha{}^\alpha \Psi^{\hb \beta} \cT_{\beta \alpha\, \hb}~.
\end{align}
This provides the dictionary between the component $T_\ha$ and the superfield
$\cT_\ha$. In accord with the usual supergravity terminology, we can call the expression for
$\cT_\ha$ the \emph{supercovariant dilaton torsion}, in comparison with the
usual dilaton torsion $T_\ha$.
Using the explicit expressions for the torsion tensors,
\begin{align}
T_a = - \kappa \,\chi_a{}^\alpha \chi_b{}^\beta (\gamma^b)_{\alpha \beta}
    = 0~, \qquad
T_\ba = - \kappa \,\Psi_\ba{}^\alpha \chi_b{}^\beta (\gamma^b)_{\alpha \beta}
    = -\kappa \,(\Psi_\ba \chi)~.
\end{align}
From these expressions, one can derive the trace contributions to
$\omega^b{}_{ba}$ and $\omega^\bb{}_{\overline{ba}}$.

\subsection{Supercovariant gravitino curvatures and fermionic equations of motion}
We briefly have mentioned that $\cT_\ha$ and $\cT_{\ha \hb \hc}$ are supercovariant
torsion tensors and we found the relations between them and the component analogues.
We're going to need to repeat that for the gravitino curvature $\cT_{\ha \hb}{}^\gamma$
and the Lorentz curvature $\cR_{\ha \hb \,\hc \hd}$ in order to understand how the
superspace geometry reproduces equations of motion in a supercovariant form.

Let's start with the fermionic equations of motion. We will need to build supercovariant gravitino
curvatures that generalize $\zD_{[\bc} \Psi_{\bb]}{}^\alpha$ and
$\zD^\bb \Psi_\bb{}^\alpha$. These are going to be related to
$\cT_{\overline{cb}}{}^\alpha$ and $\cT^\alpha$.
Let's start with the component gravitino (and dilatino) flux. The
current $\cJ$ we want, built with a flat component derivative $\zD_{\hc}$, is
\begin{align}
(\zJ_{\hc})_{\hb}{}^\alpha := (\zD_\hc \cV_\Psi \cV_\Psi^{-1})_\hb{}^\alpha
    = \zD_\hc \Psi_\hb{}^\alpha ~.
\end{align}
Here we leave the $\Omega$ connections out to begin with, as these will
turn out to be non-trivial. The above current is also given by
\begin{align}
(\zJ_{\hc})_{\hb}{}^\alpha &= (\cV_\Psi)_\hc{}^\cC \times \Big(\cV_\Psi \cJ_\cC \cV_\Psi^{-1}\Big){}_\hb{}^\alpha
\end{align}
where $(\cJ_\cC)_\cB{}^\cA$ is a superspace current built by acting with $D_\hC$.
A straightforward but laborious computation shows that
\begin{align}
2 \mathring D_{[\hc} \Psi_{\hb]}{}^\alpha 
    &= \cF_{\hc \hb}{}^\alpha
    + 2 \,\Psi_{[\hc}{}^\gamma \cF_{\gamma \hb]}{}^\alpha - \cF_{\hc \hb}{}^\ha  \Psi_\ha{}^\alpha
    \eol & \quad
    - \Psi_\hc{}^\gamma \Psi_\hb{}^\beta \,\cF_{\beta \gamma}{}^\alpha
    - 2 \Psi_{[\hc}{}^\gamma \,\cF_{\gamma \hb]}{}^\ha \Psi_\ha{}^\alpha
    - \tfrac{1}{2} (\Psi^T\Psi)^{\alpha \beta} \cF_{\beta \hc \hb}
    \eol & \quad
    + \Psi_\hc{}^\gamma \Psi_\hb{}^\beta \cF_{\beta \gamma}{}^\ha \, \Psi_\ha{}^\alpha
    + (\Psi^T\Psi)^{\alpha \beta} \Psi_\hc{}^\gamma 
        \cF_{\gamma \beta \hb}
    \eol & \quad
    + \tfrac{1}{2} (\Psi^T\Psi)^{\alpha\delta}
        \Psi_\hc{}^\gamma \Psi_\hb{}^\beta   \cF_{\beta \gamma \delta }~.
\end{align}
Restoring the $\Omega$ connections in the superspace flux tensors on the right-hand
side above leads to
\begin{align}\label{E:GravFlux}
2 \,\zcD_{[\hc} \Psi_{\hb]}{}^\alpha 
    + 2 \,\omega_{[\hc \hb]}{}^\alpha
    + \omega^\alpha{}_{\hc\hb}
    &= \cT_{\hc \hb}{}^\alpha
    + 2 \,\Psi_{[\hc}{}^\gamma \cT_{\gamma \hb]}{}^\alpha - \cT_{\hc \hb}{}^\ha  \Psi_\ha{}^\alpha
    \eol & \quad
    - \Psi_\hc{}^\gamma \Psi_\hb{}^\beta \,\cT_{\beta \gamma}{}^\alpha
    - 2 \Psi_{[\hc}{}^\gamma \,\cT_{\gamma \hb]}{}^\ha \Psi_\ha{}^\alpha
    - \tfrac{1}{2} (\Psi^T\Psi)^{\alpha \beta} \cT_{\beta \hc \hb}
    \eol & \quad
    + \Psi_\hc{}^\gamma \Psi_\hb{}^\beta \cT_{\beta \gamma}{}^\ha \, \Psi_\ha{}^\alpha
    + (\Psi^T\Psi)^{\alpha \beta} \Psi_\hc{}^\gamma 
        \cT_{\gamma \beta \hb}
    \eol & \quad
    + \tfrac{1}{2} (\Psi^T\Psi)^{\alpha\delta}
        \Psi_\hc{}^\gamma \Psi_\hb{}^\beta   \cT_{\beta \gamma \delta }~.
\end{align}
On the left-hand side we have collected the non-Lorentz tangent space
component connection $\omega_{\hc \hb}{}^\alpha$ as well as a component 
of $\omega_{\cA \hc \hb}$ in the dual fermionic direction.
These are necessary for complete covariance under tangent space
transformations. The last term on the left is notable. Inspired by \eqref{E:ComponentOmega},
we have extended the definition of $\omega_\ha$ to $\omega_\cA$ via
\begin{align}
\omega_{\cA\, \cB \cC} := (\cV_\Psi)_\cA{}^\cD \Omega_{\cD\, \cB \cC}~.
\end{align}
The object $\omega^\alpha{}_{\hc\hb}$ appearing above turns out to be gauge covariant
under double Lorentz transformations, because its transformation involves only
$\pa^\mu \lambda_{\hc \hb}$, which vanishes. So the above expression is completely
covariant under double Lorentz transformations, although not under the full
$\cH$ tangent space group, as we have gauge fixed the superspace vielbein.

After imposing subleading torsion constraints, \eqref{E:GravFlux} collapses to
\begin{align}
2 \,\zcD_{[\hc} \Psi_{\hb]}{}^\alpha 
    + 2 \,\omega_{[\hc \hb]}{}^\alpha
    + \omega^\alpha{}_{\hc\hb}
&= \cT_{\hc \hb}{}^\alpha
    + \Psi_\hc{}^\gamma \Psi_\hb{}^\beta \cT_{\beta \gamma}{}^\ha \, \Psi_\ha{}^\alpha
    + (\Psi^T\Psi)^{\alpha \beta} \Psi_\hc{}^\gamma 
        \cT_{\gamma \beta \hb}~.
\end{align}
We can view this as a \emph{definition} of the supercovariant gravitino
curvature $\cT_{\hc \hb}{}^\alpha$, which is subsequently set to zero as
a superspace constraint. In the component theory, some of this is a conventional
constraint, defining what we mean by the fermionic component connections 
$\omega_{\hc \hb}{}^\alpha$ and $\omega^\alpha{}_{\hc\hb}$, and
some of it leads to the fermionic equations of motion.
We will see how this works in just a moment.

The above computation can be repeated for $D^\hb \Psi_\hb{}^\alpha$. The result is
\begin{align}
\zcD^\hb \Psi_\hb{}^\alpha + \omega^\hb{}_{\hb}{}^\alpha + \tfrac{1}{2} \omega^{\beta}{}_{\hb \hc} (\gamma^{\hb \hc})_\beta{}^\alpha
    &= \cT^\alpha + \cT^\beta{}_\beta{}^\alpha 
    - \Psi^{\hb \alpha} \cT_\hb - \Psi^{\hb \alpha} \cT^\gamma{}_{\gamma \hb}
    + \Psi^{\hb \gamma} \cT_{\gamma \hb}{}^\alpha
    \eol & \quad
    - \tfrac{1}{2} (\Psi^T\Psi)^{\alpha \beta} (\cT_\beta + \cT^\gamma{}_{\gamma \beta})
    + \Psi^{\hb \alpha} \Psi^{\hc \gamma} \cT_{\gamma \hb \hc}
    + \tfrac{1}{2} (\Psi^T\Psi)^{\alpha \beta} \Psi^{\hb \gamma} \cT_{\gamma \beta \hb}~.
\end{align}
Imposing the subleading torsion constraints collapses this to
\begin{align}\label{E:DilatinoFlux}
\zcD^\hb \Psi_\hb{}^\alpha + \omega^\hb{}_{\hb}{}^\alpha + \tfrac{1}{2} \omega^{\beta}{}_{\hb \hc} (\gamma^{\hb \hc})_\beta{}^\alpha
    &= \cT^\alpha 
    + \Psi^{\hb \alpha} \Psi^{\hc \gamma} \cT_{\gamma \hb \hc}
    + \tfrac{1}{2} (\Psi^T\Psi)^{\alpha \beta} \Psi^{\hb \gamma} \cT_{\gamma \beta \hb}~.
\end{align}
This defines the supercovariant tensor $\cT^\alpha$.

Let's see how these various tensors decompose. Starting with
\eqref{E:GravFlux}, we see that the expression $\cT_{\overline{cb}}{}^\alpha$
defines what we might call the supercovariant gravitino curvature. However,
this expression is actually constrained to vanish from superspace.
This does not mean that the gravitino is pure gauge, of course; it just
acts as a \emph{definition} of $\omega^\alpha{}_{\overline{cb}}$
in terms of $\cD_{[\bc} \Psi_{\bb]}{}^\alpha$, 
which is the closest thing to a gravitino curvature, see \eqref{E:FermOmegas} below.
Why is it not an actual curvature? From the transformation law of $\Omega$, one can show that
$\delta \omega^\alpha{}_{\overline {bc}} = \Psi^{\ba \alpha} \, \Lambda_{\ba | \overline{bc}}$.
From the explicit form of $\omega^\alpha{}_{\overline {bc}}$, this transformation
comes from the shift symmetry of the undetermined piece of the spin connection.

Next, let's consider $\cT_{\bc b}{}^\alpha$. It is given by\footnote{Here
and below $\Psi^T \Psi$ still involve both $\Psi$ and $\chi$.}
\begin{align}
\zcD_{\bc} \chi_{b}{}^\alpha - \zcD_{b} \Psi_{\bc}{}^\alpha  
    + \omega_{\bc b}{}^\alpha
    &= \cT_{\bc b}{}^\alpha
    + \Psi_\bc{}^\gamma \chi_b{}^\beta \cT_{\beta \gamma}{}^a \, \chi_a{}^\alpha
    + \tfrac{1}{2} (\Psi^T\Psi)^{\alpha \beta} \Psi_\bc{}^\gamma \cT_{\gamma \beta b}~.
\end{align}
Again, $\cT_{\bc b}{}^\alpha$ is constrained to vanish in superspace.
The part of this that is $\gamma$-traceless in $b\alpha$ just defines
$\omega_{\bc b}{}^\alpha$. To find the residual part, we contract with
a $\gamma$ matrix, giving
\begin{empheq}[box=\fbox]{align}
\label{E:EOM.gravitino}
\zcD_{\bc} \chi_\alpha - (\gamma^b)_{\alpha\beta}\, \zcD_{b} \Psi_{\bc}{}^\beta 
    &= 
    - \tfrac{1}{2} \kappa\, (\gamma^b \Psi^\bd)_\alpha\, (\Psi_\bd \gamma_b \Psi_{\bc})
    - \tfrac{3}{200} \kappa\,
        (\Psi_\bc \gamma_{ab} \chi) \, (\gamma^{ab} \chi)_\alpha
    + \tfrac{1}{20} \kappa\,
        (\Psi_\bc \chi) \, \chi_\alpha~.
\end{empheq}
This is the supercovariantized gravitino equation of motion.

Finally we take $\cT_{cb}{}^\alpha$ and $\cT^\alpha$ in tandem.
Multiplying $\cT_{cb}{}^\alpha$ by $\tfrac{1}{2} \gamma^{cb}$ gives
the same combination of $\omega$'s as in $\cT^\alpha$, which lets us remove them.
Fixing $\cT_{cb}{}^\alpha$ and $\cT^\alpha$ to zero leads to
\begin{empheq}[box=\fbox]{align}
\label{E:EOM.dilatino}
(\gamma^b)^{\alpha \beta} \zcD_b \chi_\beta
    -\zcD^\bb \Psi_\bb{}^\alpha 
    &= 
    - \tfrac{1}{2} \kappa \,\Psi^{\bb \alpha}  \, (\Psi_\bb \chi)
    + \tfrac{1}{20}  \kappa \, (\gamma_{c b} \Psi^\bd)^\alpha (\Psi_\bd \gamma_{cb} \chi) ~.
\end{empheq}
This is the supercovariantized dilatino equation of motion.

For completeness, we give the fermionic $\omega$'s that we have uncovered:
\begin{align}\label{E:FermOmegas}
2 \,\omega_{[c b]}{}^\alpha
    + \omega^\alpha{}_{cb}
    &= 
        \tfrac{1}{5} (\gamma_{[c})^{\alpha \beta} \zcD_{b]} \chi_\beta
        - \tfrac{1}{500} \kappa\, (\chi \gamma_{cba} \chi)\, (\gamma^a \chi)^{\alpha}
        + \tfrac{1}{10} \kappa \,\Psi^{\bd \alpha} \, (\Psi_\bd \gamma_{cb} \chi)~, \eol
\omega^\alpha{}_{\overline{cb}}
    &= - 2 \,\cD_{[\bc} \Psi_{\bb]}{}^\alpha
    + \kappa (\Psi_\bc \gamma^a \Psi_\bb) (\gamma_a \chi)^\alpha~, \eol
\omega_{\bc b}{}^\alpha
    &= \Big[\cD_{b}{\Psi_{\bc}{}^{\alpha}} 
    -\tfrac{1}{2} \kappa \,\Psi^{\bd}{}^{\alpha} (\Psi_{\bd} \gamma_{b} \Psi_{\bc})
    + \tfrac{3}{800} \kappa \,
        (\chi \gamma_{b c d} \chi) 
        \Psi_{\bc}{}^{\delta} (\gamma^{c d})_{\delta}{}^{\alpha}
        \Big] \proj
\end{align}
The fact the two $\omega$'s in the first equation are determined only in
this combination is a consequence of the $\Lambda_{c,b}{}^\alpha$ shift symmetry since
only this combination is invariant under this shift.

\subsection{Supercovariant Lorentz curvatures and bosonic equations of motion}

Our next task is to compute the supercovariant Lorentz curvatures. Using the dictionary that
a component connection $h$ is related to a superspace connection $H$ via
$h_\cA{}^\ub := (\cV_\Psi)_\cA{}^\cB H_\cB{}^\ub$, one can show that
\begin{align}\label{E:defR.supercovariant}
(\cV_\Psi)_\hb{}^\cB (\cV_\Psi)_\hc{}^\cC \cR_{\cC \cB}{}^\ua
    &= \Big[2 \mathring D_\hc h_\hb{}^\ua + F_{\hc \hb}{}^\hd h_\hd{}^\ua
    - \Big(
    \cT_{\hc \hb \alpha} + 2 \,\Psi_\hc{}^\gamma \cT_{\gamma \hb \alpha}
    - \Psi_\hc{}^\gamma \Psi_\hb{}^\beta \cT_{\beta \gamma \alpha}
    \Big) h^{\alpha \ua}
    \eol & \quad
    - h_\hb{}^\ub h_\hc{}^\uc f_{\uc \ub}{}^\ua
    + 2 \,h_\hc{}^\uc f_{\hb \uc}{}^\ua
    + 2 \,h_\hc{}^\uc \Psi_\hb{}^\beta f_{\beta \uc}{}^\ua
    + \Big(
        p^{\ua \ud} - \tfrac{1}{2} h^{\hc \ua} h_\hc{}^\ud
    \Big) f_{\ud \hc \hb}\Big]_{[\hc \hb]}
\end{align}
where we have defined the component $p^{\ua \ub}$ via
\begin{align}
p^{\ua \ub} := P^{\ua \ub} + h^\gamma{}^{[\ua} h_\gamma{}^{\ub]}~.
\end{align}
This expression, like that of $h_\ha{}^\ub$ can be justified on the grounds
that it eliminates fermionic derivatives from the gauge transformation of $p$.
The term on the right-hand side of \eqref{E:defR.supercovariant} that is 
independent of the fermionic pieces like $\Psi$ and $h^\alpha$ we define as
$R_{\hc \hb}{}^\ua$, i.e.
\begin{align}
(\cV_\Psi)_\hb{}^\cB (\cV_\Psi)_\hc{}^\cC \cR_{\cC \cB}{}^\ua
    &= R_{\hc \hb}{}^\ua 
    - \Big(
    \cT_{\hc \hb \alpha} + 2 \,\Psi_\hc{}^\gamma \cT_{\gamma \hb \alpha}
    - \Psi_\hc{}^\gamma \Psi_\hb{}^\beta \cT_{\beta \gamma \alpha}
    \Big) h^{\alpha \ua}
    + 2 \,h_\hc{}^\uc \Psi_\hb{}^\beta f_{\beta \uc}{}^\ua
\end{align}
In the above expressions, we have not yet imposed the torsion constraints, so that
the reader may check the results in general.

Now let's specialize to the Lorentz curvature. We won't give all of the full
expressions, as only a few are relevant for constructing the equations of motion.
First, we observe that the Lorentz curvature with all barred indices requires no
fermionic corrections, i.e.
\begin{align}
\cR_{\overline{dc}\,\overline{ba}}
    &= R(\omega,h,p)_{\overline{dc}\, \overline{ba}}
\end{align}
where by $h$ we mean specifically the hook irrep.
Recall we constrained the left-hand side of the above to vanish in superspace.
At the component level, this works very similarly. The larger representations determine
the component $p$ and (parts of) $h$, but these fields drop out of the complete trace.
That gives the analogue of the dilaton equation of motion,
written purely in terms of the $\g{SO}(9,1)_R$ connection:
\begin{empheq}[box=\fbox]{align}
R(\omega,h,p)_{\overline{ba}}{}^{\overline{ba}} = R(\omega)_{\overline{ba}}{}^{\overline{ba}} = 0~.
\end{empheq}
Next we consider a mixed Riemann tensor with one unbarred index:
\begin{align}
\cR_{d \bc \,\overline{ba}}
    &= R(\omega,h)_{d \bc \, \overline{ba}}
        - \kappa\, \Psi_\bc{}^\alpha (\gamma_d)_{\alpha \beta}\, \omega^{\beta}{}_{\overline{ba}}~.
\end{align}
Again the constraint from superspace is that this vanishes. Part of this determines
$h$. Taking one trace projects out the $h$ terms and gives
\begin{empheq}[box=\fbox]{align}
0 &= R(\omega)_{d \bc\, \bb}{}^{\bc}
    - \kappa\, \Psi^{\bc \alpha} (\gamma_d)_{\alpha \beta}\, \omega^{\beta}{}_{\overline{bc}}~.
\end{empheq}
One must plug in the expression for $\omega^{\beta}{}_{\overline{bc}}$ from \eqref{E:FermOmegas}
to give the analogue of the Einstein equation.

While these are perfectly valid equations, they are chiral, being built purely from
the right-handed spin connections. 
To compare e.g. with \cite{Jeon:2011sq} it would be helpful to have the left-handed
versions. These are more involved because the tangent space is quite
elaborate in the left-handed sector. To keep things simple, we just look at the two
specific representations we want. The first is the complete trace of $\cR_{dc \, ba}$,
\begin{align}
\cR_{ba}{}^{ba}
    &= R(\omega)_{ba}{}^{ba}
    + \tfrac{1}{5} \kappa \, \chi_\alpha (\gamma^{ab})^\alpha{}_\beta \, \omega^\beta{}_{ab}
    - \tfrac{8}{5} \kappa \, \chi_\alpha \omega^b{}_b{}^\alpha
    - 2 \kappa\, \omega_b{}^{\alpha \beta} (\gamma^b)_{\alpha \beta}~.
\end{align}
There is a wrinkle here that $\omega_b{}^{\alpha \beta}$ hasn't been determined yet.
It arises from analyzing the supercovariant expression for $\cT_{\hc}{}^{\beta \alpha}$.
This can be computed from
\begin{align}
(\cV_\Psi)_\hc{}^D \cT_D{}^{\beta \alpha}
    &= \Big[\omega_\hc{}^{\beta \alpha}
    + \cD_\hc \Psi_\hd{}^\alpha \Psi^{\hd \beta}
    - 2 \omega^\alpha{}_\hc{}^\beta
    - 2 \Psi^\hd{}^\alpha \omega_{\hc \hd}{}^\beta
    - 2 \Psi^\hd{}^\alpha \omega^\beta{}_{\hc\hd}
    \eol & \quad
    + \tfrac{1}{4} \Psi_\hc{}^\gamma \omega^\alpha{}_{a b} (\gamma^{a b})_{\gamma}{}^\beta
    + \tfrac{1}{4} \Psi_\hc{}^\gamma \omega^\beta{}_{a b} (\gamma^{a b})_{\gamma}{}^\alpha
    + 2 \Psi^\hd{}^\alpha \cT_{\hc\hd}{}^\beta
    \eol & \quad
    + (\Psi^T \Psi)^{\alpha \gamma} \cT_{\hc \gamma}{}^\beta
    + 2 \Psi_\hc{}^\gamma \Psi^\hd{}^\alpha \cT_{\hd \gamma}{}^\beta
    - \Psi^\hb{}^\alpha \Psi^\hd{}^\beta \cT_{\hc \hb \hd}
    \eol & \quad
    + (\Psi^T \Psi)^{\alpha \delta} \Psi_\hc{}^\gamma \cT_{\gamma \delta}{}^\beta
    - (\Psi^T \Psi)^{\alpha\delta} \Psi^\hd{}^\beta \cT_{\delta \hc \hd}
    - \Psi^\hd{}^\alpha \Psi^\he{}^\beta \Psi_\hc{}^\gamma \cT_{\gamma \hd \he}
    \eol & \quad
    - (\Psi^T \Psi)^{\alpha \delta} \Psi_\hc{}^\gamma \Psi^\hd{}^\beta \cT_{\gamma \delta \hd}
    - \tfrac{1}{4} (\Psi^T \Psi)^{\alpha \gamma} (\Psi^T \Psi)^{\beta \delta} \cT_{\hc \gamma \delta}
    \eol & \quad
    - \tfrac{1}{4} \Psi_\hc{}^\gamma (\Psi^T \Psi)^{\alpha \delta} (\Psi^T \Psi)^{\beta \epsilon} \cT_{\gamma \delta \epsilon}
    \Big]^{(\alpha\beta)}
\end{align}
This looks fiendishly complicated, but it collapses upon imposing the torsion constraints to
\begin{align}
0 = \cT_{\hc}{}^{\beta\alpha}
    &= \Big[\omega_\hc{}^{\beta \alpha}
    + \cD_\hc \Psi_\hd{}^\alpha \Psi^{\hd \beta}
    - 2 \,\omega^\alpha{}_c{}^\beta
    - 2 \Psi^d{}^\alpha \omega_{\hc d}{}^\beta
    - 2 \Psi^\hd{}^\alpha \omega^\beta{}_{\hc\hd}
    \eol & \quad
    + \tfrac{1}{4} \Psi_\hc{}^\gamma \omega^\alpha{}_{ab} (\gamma^{ab})_{\gamma}{}^\beta
    + \tfrac{1}{4} \Psi_\hc{}^\gamma \omega^\beta{}_{ab} (\gamma^{ab})_{\gamma}{}^\alpha
    \eol & \quad
    - \kappa\, (\Psi^T \Psi)^{\alpha \delta} \Psi_\hc{}^\gamma \Psi^d{}^\beta 
        (\gamma_d)_{\gamma \delta}
    - \tfrac{1}{4} \kappa\, (\Psi^T \Psi)^{\alpha \gamma} (\Psi^T \Psi)^{\beta \delta} 
        (\gamma_c)_{\gamma \delta}\Big]^{(\alpha \beta)}~.
\end{align}
This leads to
\begin{align}
0 &= R(\omega)_{ba}{}^{ba}
    - \tfrac{12}{5} \kappa\, \Big(
        \chi_\alpha \omega^b{}_b{}^\alpha
        - \tfrac{1}{2} (\chi \gamma^{ab})_\alpha \omega^\alpha{}_{ab}
    \Big)
    + \tfrac{4}{25} \kappa\, (\chi \gamma^c \cD_c \chi)
    + 2 \kappa\, (\Psi_\bd \gamma^c \cD_c \Psi_\bd)
    \eol & \quad
    + \tfrac{1}{50} \kappa^2 (\Psi^\ba \gamma^c \gamma^d \chi) (\Psi_\ba \gamma_d \gamma_c \chi)
    - \tfrac{1}{100} \kappa^2 (\Psi^\ba \gamma^c \gamma^d \chi) (\Psi_\ba \gamma_c \gamma_d \chi)
    + \tfrac{1}{2} \kappa^2 (\Psi^\ba \gamma^c \Psi^\bb) (\Psi_\ba \gamma_c \Psi_\bb)
\end{align}
and it helps to use
\begin{align}
\omega^b{}_b{}^\alpha - \tfrac{1}{2} (\gamma^{ab})^\alpha{}_\beta \omega^\beta{}_{ab}
    &= \tfrac{9}{10} (\gamma^b \cD_b \chi)^\alpha
    - \tfrac{1}{20} \kappa\, (\gamma^{cb} \Psi^\ba)^\alpha  (\Psi_\ba \gamma_{cb} \chi)~,
\end{align}
which follows from \eqref{E:FermOmegas} to give
\begin{align}
0 &= R(\omega)_{ba}{}^{ba}
    -2 \kappa\, (\chi \gamma^c \cD_c \chi)
    + 2 \kappa\, (\Psi_\bd \gamma^c \cD_c \Psi_\bd)
    + \tfrac{16}{25} (\Psi^\ba \chi)\, (\Psi_\ba \chi)
    + \tfrac{1}{2} (\Psi^\ba \Psi^\bb)\, (\Psi_\ba \Psi_\bb)~.
\end{align}
Combining this with $R(\omega)_{\overline{ba}}{}^{\overline{ba}} = 0$ gives
\begin{empheq}[box=\fbox]{align}
\label{E:EOM.dilaton}
0 &= R(\omega)_{ba}{}^{ba} - R(\omega)_{\overline{ba}}{}^{\overline{ba}}
    -2 \kappa\, (\chi \gamma^c \cD_c \chi)
    + 2 \kappa\, (\Psi^\bd \gamma^c \cD_c \Psi_\bd)
    + \tfrac{16}{25} (\Psi^\ba \chi)\, (\Psi_\ba \chi)
    + \tfrac{1}{2} (\Psi^\ba \Psi^\bb)\, (\Psi_\ba \Psi_\bb)~.
\end{empheq}
It is tempting to call this the dilaton field equation, but this is not
guaranteed. Superspace identifies only the on-shell locus, so this is actually
a combination of the dilaton field equation and the fermionic field equations.

Repeating for the generalized Ricci tensor and skipping the intermediate steps, we find
\begin{empheq}[box=\fbox]{align}
\label{E:EOM.graviton}
0 &=  - \tfrac{1}{2} R(\omega)_{a}{}^\bc{}_{\overline{bc}} + \tfrac{1}{2} R(\omega)_{\bb}{}^c{}_{a c} 
- \tfrac{1}{2} (\chi \,\cD_{a}{\Psi_{\bb}})
+ \tfrac{3}{2} (\Psi^{\bc} \gamma_a \cD_{\bc}{\Psi_{\bb}})
- (\Psi^\bc \gamma_a  \cD_{\bb}{\Psi_{\bc}})
- \tfrac{2}{5} (\Psi_{\bb} \gamma_{a b} \cD^{b}{\chi})
\eol & \quad
+ \tfrac{1}{10} (\chi \gamma_{a b} \cD^{b}{\Psi_{\bb}})
% \eol & \quad 
% %  + 1/2000 \chi_{\alpha} \chi_{\beta} \chi_{\gamma} \gamma_{b c \delta}^{\alpha} \gamma_{a b c}^{\beta \gamma} \Psi_{\bb}^{\delta} 
+ \tfrac{7}{320} (\Psi_{\bb} \gamma^{b c} \chi)
     (\Psi^{\bd} \gamma_{a b c} \Psi_{\bd})
- \tfrac{1}{960} (\Psi_{\bb} \gamma_{a b c d} \chi)
     (\Psi^{\bd} \gamma^{b c d} \Psi_{\bd})~.
\end{empheq}
Again, this is only the double vielbein field equation modulo the fermionic
field equations.

\subsection{Comparing with component DFT and $\cN=1$ supergravity}
Our results for the field equations and supersymmetry transformations look
rather different from those found by Jeon, Lee, and Park \cite{Jeon:2011sq},
and it is a crucial check that they actually agree. 
For the field equations, this is straightforward to verify after making
numerous convention swaps. Most of these are obvious, but a few we will
highlight. For the fermions, the charge conjugation tensor in \cite{Jeon:2011sq}
is imaginary, leading to
\begin{gather}
\psi_{\bar p} \rightarrow \Psi_{\ba}{}^\alpha~, \quad 
\bar\psi_{\bar p} \rightarrow i \Psi_{\ba}{}^\alpha~.
\end{gather}
There is also an additional sign in the definition of the dilatino, so
(remembering the charge conjugation has an opposite sign for the opposite chirality)
\begin{gather}
\rho \rightarrow -\chi_\alpha~, \qquad
\bar\rho \rightarrow i \chi_\alpha~.
\end{gather}
The $\g{SO}(9,1)_R$ metric has the opposite sign,
$\eta_{\bar p \bar q} \rightarrow - \eta_{\overline{ab}}$, and the
spin connection uniformly differs by a sign \emph{when the indices are lowered}
\begin{align}
\Phi_{A\, p q} \rightarrow - \omega_{\hm\, a b}~, \qquad
\Phi_{A\, \bar p \bar q} \rightarrow - \omega_{\hm\, \overline{ab}}~.
\end{align}
We also must fix $\kappa=1$.
With these modifications, the fermionic field equations
\eqref{E:EOM.gravitino} and \eqref{E:EOM.dilatino} coincide exactly with those
given in \cite{Jeon:2011sq},
while the bosonic field equations \eqref{E:EOM.dilaton} and \eqref{E:EOM.graviton}
hold modulo the former. This is an intricate comparison involving
numerous Fierz identities, and we were saved countless hours  of computation by using
\texttt{Cadabra} \cite{Peeters.Cadabra1, Peeters.Cadabra2}.
Similarly, the supersymmetry transformations discussed in
section \ref{S:SUSYtrafo} match, provided we choose the compensating
Lorentz parameters in \eqref{E:SUSY.vielbein} so that $J_{ab} = J_{\overline{ab}}=0$.

We may also compare with 10D $\cN=1$ supergravity in the string frame, see e.g.
\cite{Bergshoeff:1988nn}. To keep things simple, we'll just focus on the
supersymmetry transformations.
Taking the explicit parametrization \eqref{E:CompDFT.eb} for the double vielbein, we find that
\begin{align}
e_{\ra}{}^\rmm \delta e_{\rmm \rb} 
    &= \tfrac{1}{10} \,\kappa \,(\eps \gamma_{ab} \chi)
    - \tfrac{1}{2} (\lambda_{ab} - \lambda_{\overline{ab}})
        - \kappa \, (\eps \gamma_{(a} \Psi_{\bb)})~.
\end{align}
To identify the conventional supersymmetry transformation requires that we choose
\begin{align}
\lambda_{ab} = \tfrac{1}{5} \kappa \, (\eps \gamma_{ab} \chi )
    - 2 \kappa \,(\eps \gamma_{[a} \Psi_{\bb]})~, \qquad
\lambda_{\overline{ab}} = 0~.
\end{align}
Moreover, one must keep in mind the factor of $1/\sqrt{2}$ in the covariant
derivative, which requires that we rescale the component gravitino,
$\psi_\rmm{}^\alpha = \sqrt{2} e_\rmm{}^\ra \, \Psi_\ba{}^\alpha$
to have a canonical supersymmetry transformation. All told, this leads to
\begin{align}
\delta e_\rmm{}^\ra 
    = -\frac{1}{\sqrt 2} \, \kappa\, (\eps \gamma^\ra \psi_{\rmm})~.
\end{align}
Matching conventions requires $\kappa = -1/\sqrt{2}$, which we fix for the remainder
of this discussion. The supergravity  dilaton $\varphi$ is related to $\phi = e^{-2d}$ by
$\phi = e\, e^{-2 \varphi}$.
This allows us to identify the supergravity dilatino via
\begin{align}
\delta \varphi = - \frac{1}{\sqrt 2} \eps\lambda~, \qquad
\lambda_\alpha = \frac{1}{2} \chi_\alpha - \frac{1}{2\sqrt 2} (\gamma^\ra \psi_\ra)_\alpha~.
\end{align}
We normalize the dilatino $\lambda$ the same way as \cite{Bergshoeff:1988nn},
but we keep the standard $\varphi$, which differs from the
dilaton used in \cite{Bergshoeff:1988nn}.

Now let's analyze the fermion variations.
The gravitino transformation involves the connection $\omega_{\ba bc}$.
This is found from the torsion component
\begin{align}
T_{\bc a b}
    &= \frac{1}{2 \sqrt 2} H_{\rm abc}
    - \frac{1}{2\sqrt 2} (C_{\rm cba} + C_{\rm abc} + C_{\rm acb})
    + \omega_{\bc a b}
\end{align}
Solving for $\omega$ and translating to $\hat\omega$ gives
\begin{align}
\omega_{\bc a b}
    &= \frac{1}{\sqrt 2} \hat \omega_{\rm cab}
    - \frac{1}{2 \sqrt 2} H_{\rm abc}
    + T_{\bc a b}
    + \frac{1}{4\sqrt 2} \Big(
        \psi_\rc \gamma_\ra \psi_\rb
    - \psi_\rc \gamma_\rb \psi_\ra
    + \psi_\ra \gamma_\rc \psi_\rb
    \Big)
\end{align}
where we have used the standard supercovariant torsion tensor from supergravity,
\begin{align}
\hat \omega_{\rm c b a} = -\tfrac{1}{2} (C_{\rm cba} + C_{\rm abc} + C_{\rm acb})
    + \tfrac{1}{4} (\psi_\rc \gamma_\ra \psi_\rb)
    - \tfrac{1}{4} (\psi_\rc \gamma_\rb \psi_\ra)
    + \tfrac{1}{4} (\psi_\ra \gamma_\rc \psi_\rb)~.
\end{align}
The field strength $H$ should be supercovariantized to
$\hat H_{\rm abc} = H_{\rm abc} 
    + \tfrac{3}{2} (\psi_{[\ra} \gamma_\rb \psi_{\rc]})
$ which gives
\begin{align}
\omega_{\bc a b} &= T_{\bc a b} + \frac{1}{\sqrt 2} \Big(
    \hat \omega_{\rm c a b} - \frac{1}{2} \hat H_{\rm c a b}
    + \psi_\rc \gamma_{[\ra} \psi_{\rb]}
    \Big) = T_{\bc a b} + 
    \frac{1}{\sqrt 2} \Big(
    \hat \omega^-_{\rc \ra \rb}
    + (\psi_\rc \gamma_{[\ra} \psi_{\rb]})
    \Big)
\end{align}
where we have defined $\hat\omega^-$ by the above.
The component gravitino transformation then becomes
\begin{align}
\delta \psi_\rmm{}^\alpha
    = \pa_\rmm \eps^\alpha - \tfrac{1}{4} \hat\omega^-_{\rmm\,\ra \rb} (\gamma^{\ra \rb} \eps)^\alpha
    + \frac{1}{\sqrt 2} \Big(
    \eps^\alpha \, (\psi_\rmm \lambda)
    - \psi_\rmm{}^\alpha \, (\eps \lambda)
    + (\gamma^\ra \lambda)^\alpha \,(\psi_\rmm \gamma_\ra \eps)
    \Big)~.
\end{align}
This matches \cite{Bergshoeff:1988nn}.

To identify the dilatino transformation, we need
\begin{alignat}{3}
T_{abc} &= \frac{1}{2\sqrt 2} (H_{\rm abc} + 3 \,C_{[\rm abc]}) + 3 \,\omega_{[abc]}
&\,\, &\implies &\,\,
\omega_{[abc]}
    &= \frac{1}{3} T_{[abc]} 
    + \frac{1}{\sqrt 2} \Big(
    \hat \omega^-_{[\rm abc]} + \frac{1}{3} \hat H_{\rm abc}
    + \frac{1}{2} (\psi_{[\ra} \gamma_\rb \psi_{\rc]})
    \Big)~, \eol
T_a &= \frac{1}{\sqrt 2} (D_\ra \log \phi + C_{\ra \rb}{}^\rb)
    + \omega^b{}_{b a} 
    &\,\,&\implies &\quad
\omega^b{}_{ba} &= T_a + \frac{1}{\sqrt 2} \Big(
    \hat\omega^\rb{}_{\rb\ra} - D_\ra \log \phi - \tfrac{1}{2} (\psi_\ra \gamma^\rb \psi_\rb)
    \Big)~.
\end{alignat}
Putting this together, one finds
\begin{align}
\delta \lambda_\alpha
    &= - \frac{\sqrt{2}}{4} (\gamma^\rmm \eps)_\alpha \,\Big(\pa_\rmm \varphi 
        - \tfrac{1}{4} (\psi_\rmm \lambda)\Big)
    - \frac{\sqrt{2}}{48} (\gamma^{\rm abc} \eps)_\alpha \, \Big(
    \hat H_{\rm abc} - \tfrac{1}{4} (\lambda \gamma_{\rm abc} \lambda)
    \Big) 
\end{align}
This also matches \cite{Bergshoeff:1988nn}, keeping in mind that 
$(\bar\lambda \gamma_{abc} \lambda)= -(\lambda \gamma_{abc} \lambda)$
using our convention for the charge conjugation matrix.

\section{Open questions}
\label{S:Discussion}

Our initial goal in this work was to reproduce the component results of $\cN=1$
DFT using a manifestly supersymmetric starting point. While we have achieved that,
at the same time we have stumbled upon an intriguing expansion of the tangent space
group. This enlarged tangent space seems necessary in order to explicitly gauge away unphysical components of the supervielbein and suggests the introduction of an
ever higher set of connections. These in turn require curvature constraints in order to render
them composite, with the consequence that only physical components of curvature
tensors appear. As we have emphasized in section \ref{S:BosonicDFT}, this is not
dependent upon supersymmetry per se, but can be imposed even in the bosonic theory,
where it is required if we demand that the undetermined part of the spin connection
be gauged.

Several issues remain to be addressed, and we pose these here as questions:
\begin{itemize}
\item \textbf{Is the enlarged tangent space description correct?}

We have proposed a simple Lie superalgebra
as the basis for the tangent space group $\ext\cH_L \times \ext{SO(9,1)}_R$.
Here $\ext\cH_L$ is understood as the dual of the super-Maxwell$_\infty$ algebra,
and $\ext{SO(9,1)}_R$ is the dual of the on-shell Maxwell$_\infty$ algebra.
For the lowest lying levels, this can explicitly be verified,
providing the connections necessary for building
torsions and curvatures along with the required gauge transformations.
What remains to be confirmed is that the higher connections $H_\cA{}^\ub$ and the 
Pol\'a\v{c}ek-Siegel fields $P^{\ua \ub}$ are completely determined 
(modulo gauge symmetries) by imposing constraints on the curvatures.
We have checked this explicitly through dimension two, which
is sufficient to understand the two derivative equations of motion and supersymmetry
transformations.
Possibly the proposal we make requires modifications beyond dimension two,
or perhaps it is indeed complete to all orders. This will require further
investigation.

\item \textbf{Does the $\cN=2$ formulation work in a similar way?}

An extension to $\cN=2$ double field theory should be possible, in particular
to reproduce the component results of $\cN=2$ supersymmetric DFT \cite{Jeon:2012hp}.
Superspaces relevant for these
cases have been considered by Cederwall \cite{Cederwall:2016ukd} and by
Hatsuda, Kamimura, and Siegel \cite{Hatsuda:2014qqa,Hatsuda:2014aza}.
Cederwall's proposal is the more conventional of the two, relying upon
an $\g{OSp}(D,D|2s)$ vielbein and a separate $\Omega$ connection valued just in
$\g{Spin}(D-1,1)_L \times \g{Spin}(D-1,1)_R$.
The formulation of Hatsuda et al. employed an enlarged Pol\'a\v{c}ek-Siegel
megavielbein directly, but similarly restricted the local tangent space.
(In both approaches, there is an elegant interpretation of the RR field strengths
lying in the supervielbein.) Both differ from the $\cN=1$ approach
we have advocated (and its natural $\cN=2$ extension), which requires 
an enhanced tangent space beyond $\g{Spin}(D-1,1)_L \times \g{Spin}(D-1,1)_R$
to explicitly eliminate the unphysical components of the supervielbein.

Presumably the relationship between our approach (which builds on Siegel's 1993
formulation \cite{Siegel:1993th}) and these ones involves a process of
``degauging'' where one fixes the additional gauge symmetries and reinterprets
the gauge-fixed higher connections as curvatures involving undetermined components
of the spin connection. This would be very similar to the relationship between
conformal gravity (where additional dilation and special conformal connections
are introduced) and conventional Poincar\'e gravity.\footnotemark\,
In the latter case, when
describing conformal gravity, one must require by hand that the local Weyl
transformation $\delta e_{\rmm}{}^\ra = -\sigma\, e_\rmm{}^\ra$ be a symmetry
of the action. Only the traceless projection of the Riemann tensor -- the
Weyl tensor -- is physical in this scheme, and other pieces are unphysical.
\footnotetext{See e.g. \cite{Freedman:2012zz}
for a pedagogical discussion of conformal (super)gravity,
and \cite{Butter:2009cp} for its superspace analogue.}

\item \textbf{Can the $\cN=2$ formulation tell us about super-$E_{11}$?}

A major motivation to study the $\cN=2$ formulation is to learn something about
super-$E_{11}$, whose component formulation has
recently been explored \cite{Bossard:2019ksx}. One might expect that in
super-DFT, the RR forms should appear only via their field strengths, but in extending
the formalism to include their $p$-form potentials, the link to super-$E_{11}$
would be uncovered. Some work along these lines has already been explored by
Cederwall, who proposed to identify RR potentials and field strengths using the
language of OSp spinors \cite{Cederwall:2016ukd}. It would be interesting to
explore this further.

\item \textbf{Can higher derivative or off-shell formulations be encoded in super-DFT?}

The formulation of superspace DFT we have discussed is on-shell, much like the conventional
10D $\cN=1$ superspace it reduces to when the section condition is imposed. Naturally,
this is because it is describing at the component level the supersymmetric DFT of \cite{Hohm:2011nu, Jeon:2011sq}, which is a two-derivative theory whose supersymmetry transformations
close only up to the equations of motion. Two natural questions suggest themselves.
First, can we extend superspace DFT to encode higher derivative $\alpha'$ corrections?
This certainly seems plausible, and relatively recently, the first-order corrections
to $\cN=1$ supersymmetric DFT were constructed at the component level \cite{Lescano:2021guc}.
It would be quite interesting to understand how this construction can be geometrized in
superspace, and if that can aid in its extension to higher orders. An even more ambitious
proposal would be to try to understand whether any connection can be drawn to the off-shell pure
spinor superspace \cite{Cederwall:2009ez,Cederwall:2010tn, Berkovits:2018gbq}, which might allow the direct construction of higher order invariants.

\end{itemize}

%%%%%%%%%%%%%%%%%%%%%%%%%%%%%%%%%%%%%%%%%%%%%%%%%%%%%%%%%%%%%%%%%%%%%%%%%%%%%%%%%
\section*{Acknowledgements}
It is a pleasure to thank Falk Hassler, Gianluca Inverso, William Linch, 
Jakob Palmkvist, Axel Kleinschmidt, and Warren Siegel for discussions and comments.
This work is partially supported by the NSF under grants NSF-1820921 and
PHY-1606531, and the Mitchell Institute for Fundamental Physics and Astronomy 
at Texas A\&M University.

%%%%%%%%%%%%%%%%%%%%%%%%%%%%%%%%%%%%%%%%%%%%%%%%%%%%%%%%%%%%%%%%%%%%%%%%%%%%%%%%%

\appendix

%%%%%%%%%%%%%%%%%%%%%%%%%%%%%%%%%%%%%%%%%%%%%%%%%%%%%%%%%%%%%%%%%%%%%%%%%%%%%%%%%
\section{Notations and conventions}
\label{A:Notation}
%%%%%%%%%%%%%%%%%%%%%%%%%%%%%%%%%%%%%%%%%%%%%%%%%%%%%%%%%%%%%%%%%%%%%%%%%%%%%%%%%
\subsection{$\g{Spin}(9,1)$ conventions}
Our conventions for $\g{Spin}(9,1)$ are as follows. The metric $\eta_{ab}$ is
mostly positive signature, with $\gamma$-matrices
\begin{gather}
\{\gamma^a, \gamma^b\} = 2 \,\eta^{ab}~, \qquad 
\gamma^{a_1 \cdots a_n} := \gamma^{[a_1} \cdots \gamma^{a_n]}~.
\end{gather}
(Anti)-symmetrization always involves factors of $1/n!$. We use a 
Weyl basis for the $\gamma$-matrices so that
\begin{align}
\gamma^a =
\begin{pmatrix}
0 & (\gamma^a)^{\alpha \beta}  \\
(\gamma^a)_{\alpha \beta} & 0
\end{pmatrix}~, \qquad
\gamma_{11} = 
\begin{pmatrix}
1 & 0 \\
0 & -1
\end{pmatrix}~, \qquad
C = 
\begin{pmatrix}
0 & 1 \\
-1 & 0
\end{pmatrix}
\end{align}
The chiral (Pauli) $\gamma$-matrices $(\gamma^a)_{\alpha\beta}$ obey the 10D identities
\begin{align}
(\gamma^a)_{(\alpha \beta} (\gamma_a)_{\gamma) \delta} = 0~, \qquad
(\gamma^{abc})_{[\alpha \beta} (\gamma_{ab})_{\gamma]}{}^\delta = 0~.
\end{align}
The gravitino and dilatino decompose as
\begin{align}
\psi =
\begin{pmatrix}
\psi^\alpha \\
0
\end{pmatrix}~, \qquad
\chi = 
\begin{pmatrix}
0 \\
\chi_\alpha
\end{pmatrix}~.
\end{align}
We use explicit Weyl-component notation throughout and suppress indices in obvious ways
so that, for example,
\begin{align}
(\psi \chi) = \psi^\alpha \chi_\alpha = - (\chi \psi)~, \qquad
(\psi \gamma_{abc} \psi) = \psi^\alpha (\gamma_{abc})_{\alpha\beta} \psi^\beta~, \qquad
(\chi \gamma_{abc} \chi) = \chi_\alpha (\gamma_{abc})^{\alpha \beta} \chi_\beta~, \qquad
\end{align}
This last point is important because in Dirac notation with Majorana fermions, we would take
$\bar \Psi = \Psi^T C$, so $\bar\chi = (-\chi_\alpha, 0)$ involves a minus sign while 
$\bar\psi = (0, \psi^\alpha)$ does not.

\subsection{$\g{OSp}(p,q|2s)$ conventions}
The supergroup $\g{OSp}(p,q|2s)$ is the group of linear transformations that preserve the
canonical invariant $\eta_{AB}$, graded with $p+q$ bosonic and $2s$ fermionic indices,
\begin{align}
\eta_{AB} =
\begin{pmatrix}
\eta_{\ha\hb} & 0 \\
0 & \veps_{\hat\alpha\hat\beta}
\end{pmatrix}
\end{align}
where $\eta_{\ha\hb}$ is the $\g{SO}(p,q)$ metric and $\veps_{\hat\alpha\hat\beta}$ is the canonical
symplectic element
$\begin{pmatrix}
0 & 1 \\
-1 & 0
\end{pmatrix}$.
We will be interested in the two cases
$\g{OSp}(10,10|32)$ and $\g{OSp}(9,1|32)$ where $\hat\alpha$ is a 32-component
Dirac index. The generators $M_{AB}$ of this algebra obey
\begin{align}
[M_{AB}, M_{CD}] = (-)^{a(b+c)} \eta_{BC} M_{A D} - (-)^{bc} \eta_{AC} M_{B D}
    - (-)^{bc} M_{A C} \eta_{BD} + (-)^{d (b+c)} \eta_{A D} M_{B C}
\end{align}
On the fundamental vector representation $V^A$,
\begin{align}
\delta_\lambda V^A \equiv 
    \tfrac{1}{2} \lambda^{BC} M_{CB} V^A = \lambda^{A B} V_B \qquad \implies \quad
M_{CB} V^A = 2\, V_{[C} \delta_{B]}{}^A 
\end{align}
where we use a NW-SE convention for contracting super-indices, so that
$V_A = V^B \eta_{B A}$.

%%%%%%%%%%%%%%%%%%%%%%%%%%%%%%%%%%%%%%%%%%%%%%%%%%%%%%%%%%%%%%%%%%%%%%%%%%%%%%%%%
\section{Pol\'a\v{c}ek-Siegel formulation of DFT with connections}
\label{A:PS}
%%%%%%%%%%%%%%%%%%%%%%%%%%%%%%%%%%%%%%%%%%%%%%%%%%%%%%%%%%%%%%%%%%%%%%%%%%%%%%%%%
In this appendix, we review the Pol\'a\v{c}ek-Siegel approach to connections in
DFT \cite{Polacek:2013nla}. We will take a somewhat more streamlined
approach, eliminating the discussion of background derivatives $D_M$, while
simultaneously extending the allowed gauging beyond the Lorentz group to a
generic gauge group. We restrict purely to the field theoretic description
of their results and avoid discussing the motivations from the form of
current algebras on the string worldsheet.

To simplify notation somewhat, we will denote the doubled coordinates of
DFT by $x^m$, without any further adornment, and the corresponding tangent
space indices by $a,b,\cdots$. Now let the coordinates be enhanced to
$z^M = (y^\um, x^m, \tilde y_\um)$ subject to a section condition
$\pa^M \otimes \pa_M = 0$ where $\eta^{MN}$ is given by
\begin{align}
\eta^{MN} =
\begin{pmatrix}
0 & 0 & \delta^\um{}_\un \\
0 & \eta^{mn} & 0 \\
\delta_\um{}^\un & 0 & 0
\end{pmatrix}~.
\end{align}
We will assume the section condition for the additional coordinates
is already solved by $\pa^\um = 0$, so that no fields depend on $\tilde y_\um$.
The dependence on $y^\um$ will be chosen in a particular way to
parametrize how fields transform under double Lorentz
transformations as well as any other local gauge symmetries present.
We refer to this local gauge group that extends the double Lorentz group as $\ext\cH$.
We assume that all coordinates are bosonic to avoid introducing gradings, 
but it is trivial to extend to supercoordinates.

Let the \emph{megavielbein} on this space be denoted $\cV_M{}^A$. It transforms under
diffeomorphisms in the usual manner as
\begin{align}\label{E:PS.Diff}
\delta \cV_M{}^A &= \xi^N \pa_N \cV_M{}^A + (\pa_M \xi^N - \pa^N \xi_M) \cV_M{}^A~.
\end{align}
There are no tangent space transformations, as these will be encoded in $\xi$ itself.
Rewriting the above transformation in terms of $\xi^A = \xi^M \cV_M{}^A$ leads to
the covariant form of generalized diffeomorphisms,
\begin{align}\label{E:PS.CovDiff}
\delta \cV_M{}^A &= \cV_M{}^B (
    \nabla_B \xi^A - \nabla^A \xi_B + \xi^C \cT_{C B}{}^A
    )
\end{align}
where $\nabla_A := \cV_A{}^M \pa_M$. The totally antisymmetric torsion tensor
$\cT_{CBA}$ is given by
\begin{align}
\cT_{CBA} = - 3 \nabla_{[C} \cV_B{}^M \cV_{M A]}~.
\end{align}
These obey Bianchi identities
\begin{align}\label{E:PS.Bianchi}
4 \nabla_{[A} \cT_{BCD]} + 3 \cT_{[A B}{}^F \cT_{CD] F} = 0~.
\end{align}
Of course, the torsion tensor also appears in the algebra of covariant derivatives
\begin{align}
[\nabla_A, \nabla_B] &= - \cT_{A B}{}^C \nabla_C~,
\end{align}
where we have made use of the section condition.
The standard form of the Bianchi identity follows from the above,
\begin{align}
\Big(\nabla_{[A} \cT_{BC]}{}^D + \cT_{[AB}{}^F \cT_{F| C]}{}^D \Big) \nabla_D = 0~,
\end{align}
and coincides with \eqref{E:PS.Bianchi} upon using the section condition.

Now let us decompose our derivatives as $\nabla_A = (\nabla_\ua, \nabla_a, \nabla^\ua)$. 
The derivative $\nabla_\ua$ will be identified soon with the generator of local $\ext\cH$
transformations, and $\nabla_a$ will coincide with the usual covariant derivative.
The additional derivative $\nabla^\ua$ will turn out to be a composite
operator, which we will explore in due course. We will be interested in the following
algebra, where some of the torsion tensors $\cT_{CBA}$ have been set to 
constants $f_{CBA}$:\footnote{In their original paper \cite{Polacek:2013nla},
Pol\'a\v{c}ek and Siegel employed just the doubled Lorentz group, and so a number
of the $f$'s we have here actually vanish, in particular $f_{\ua b}{}^\uc$.}
\begin{subequations}\label{E:PS.algebra}
\begin{align}
[\nabla_a, \nabla_b] &= - \cT_{a b}{}^\uc \nabla_\uc - \cT_{a b}{}^c \nabla_c
    - f_{a b \uc} \nabla^\uc~, \\
[\nabla_\ua, \nabla_b] &= -f_{\ua b}{}^c \nabla_c - f_{\ua b}{}^\uc \nabla_\uc~, \\
[\nabla_\ua, \nabla_\ub] &= - f_{\ua \ub}{}^\uc \nabla_\uc~, \\[2ex]
[\nabla^\ua, \nabla^\ub] &= -\cT_c{}^{\ua \ub} \nabla^c - \cT^{\ua \ub \uc} \nabla_\uc
    - f_\uc{}^{\ua \ub} \nabla^\uc~, \\
[\nabla^\ua, \nabla_b] &= - \cT_{b c}{}^\ua  \nabla^c - f_{b \uc}{}^\ua \nabla^\uc
    + \cT_{b}{}^{\ua \uc} \nabla_\uc~, \\
[\nabla^\ua, \nabla_\ub] &= -f_{\ub \uc}{}^\ua \nabla^\uc
    - f_{\ub c}{}^\ua \nabla^c
    + f_{\ub}{}^{\ua\uc} \nabla_\uc~.
\end{align}
\end{subequations}
In addition to setting some of the torsion components to constants, we have also chosen to
fix $\cT_{\ul{cba}}$ and $\cT_{\ul{cb} a}$ to zero. The reason for this choice is that
we want $\nabla_\ua$ to furnish a closed subalgebra: this is the $\ext\cH$ algebra that will
be gauged. The torsion tensors that have been set to constants will turn out to correspond
to choices of how various components of the megavielbein transform under local $\ext\cH$ transformations.
The remaining non-constant tensors $\cT_{abc}$ and $\cT_{ab}{}^\uc$ will correspond to 
the standard torsion tensor and $\ext\cH$-curvature, while $\cT_c{}^{\ua \ub}$ and $\cT^{\ul{abc}}$
will be new tensors, whose role will be explained in due course.

We are not going to explicitly solve the above constraints, but give a
megavielbein that does the job. Following Pol\'a\v{c}ek and Siegel, let us first presume
the megavielbein can be gauged to a triangular form,
\begin{align}\label{E:PS.Triangular}
\cV_M{}^A =
\begin{pmatrix}
{\rm x} & 0 & 0 \\
{\rm x} & {\rm x} & 0 \\
{\rm x} & {\rm x} & \rm x
\end{pmatrix}~, \qquad
\cV_A{}^M =
\begin{pmatrix}
{\rm x} & 0 & 0 \\
{\rm x} & {\rm x} & 0 \\
{\rm x} & {\rm x} & {\rm x}
\end{pmatrix}~,
\end{align}
where the x's denote non-vanishing entries.
We emphasize that the entries are ordered so that $\nabla_A = (\nabla_\ua, \nabla_a, \nabla^\ua)$.
Defining the antisymmetric current $\cJ_{B A}$ via $\delta \cV_M{}^A = \cV_M{}^B \cJ_{B C} \eta^{C A}$, a consistency condition for the above gauge is that
$\cJ_{\ub \ua} = \cJ_{\ub a} = 0$.
This immediately implies that covariant diffeomorphisms \eqref{E:PS.CovDiff} must be constrained as
\begin{align}
0 &= \cJ_{\ub \ua} = \nabla_\ub \xi_\ua - \nabla_\ua \xi_\ub + \xi_\uc f^\uc{}_{\ub \ua}~, \eol
0 &= \cJ_{\ub a} = \nabla_\ub \xi_a - \nabla_a \xi_\ub 
    + \xi_\uc f^\uc{}_{\ub a}
    + \xi^c f_{\ub a c}~.
\end{align}
These conditions are easily solved by taking
\begin{align}\label{E:PS.CC}
\xi_\ua = 0~, \qquad \nabla_\ub \xi_a &= - f_{\ub a}{}^c \xi_c~.
\end{align}
The first condition has the interpretation that once we have imposed 
the gauge \eqref{E:PS.Triangular}, the dual $\ext\cH$ transformations associated with $\nabla^\ua$
no longer play any role. This is good: we are looking for a minimal way to incorporate
the gauging of a group $\ext\cH$, and having to accommodate another dual group would
be excessive. In the triangular gauge, imposing $\xi_\ua=0$ is equivalent to
imposing $\xi_\um = 0$.
The second condition in \eqref{E:PS.CC} tells us how $\xi_a$ transforms
under the $\ext\cH$ group, which will amount to a condition on its $y$-dependence.

Let's check that the above conditions are consistent.
The closure of two diffeomorphisms leads to
\begin{align}
\xi_{12}{}^A := (\Lie_{\xi_1} \xi_2)^A = \xi_1{}^B \nabla_B \xi_2{}^A - \xi_2{}^B \nabla_B \xi_1{}^A
    + \xi_2{}^B \nabla^A \xi_{1 B}
    + \xi_1{}^B \xi_2{}^C \cT_{C B}{}^A~.
\end{align}
We must take $\xi_{12 \ua}$ to vanish. As required,
this holds given the conditions we have already imposed:
\begin{align}
\xi_{12 \ua} = \xi_2{}^b \nabla_\ua \xi_{1 b} 
    + \xi_1{}^b \xi_2{}^c \cT_{c b \ua} = - \xi_2{}^b f_{\ua b}{}^c \xi_{1 c}
    + \xi_1{}^b \xi_2{}^c f_{\ua c b} = 0~.
\end{align}
We also should check that the second condition \eqref{E:PS.CC} satisfies integrability:
\begin{align}
[\nabla_\ua, \nabla_\ub] \xi^c
    &= -\nabla_\ua \xi^d f_{d \ub}{}^c + \nabla_\ub \xi^d f_{d \ua}{}^c 
    = \xi^e f_{e \ua}{}^d f_{d \ub}{}^c  - \xi^e f_{e \ub}{}^d f_{d \ua}{}^c 
    = - f_{\ua \ub}{}^\ud \nabla_\ud \xi^c~,
\end{align}
with the last equality following from the Jacobi identity.

Now let us assign names to the non-vanishing entries of the megavielbein \eqref{E:PS.Triangular}.
Using orthogonality, these can be parametrized as
\begin{align}\label{E:megaV}
\cV_M{}^B &=
\begin{pmatrix}
K_\um{}^\ua & 0 & 0 \\
0 & \cV_m{}^a & 0 \\
0 & 0 & K_\ua{}^\um
\end{pmatrix}
\times
\begin{pmatrix}
\delta_\ua{}^\ub & 0 & 0 \\
h_a{}^\ub & \delta_a{}^b & 0 \\
-p^{\ua \ub} - \tfrac{1}{2} h^{c \ua} h_c{}^\ub & - h^{b \ua} & \delta^\ua{}_\ub
\end{pmatrix}~, \eol
\cV_B{}^M &=
\begin{pmatrix}
\delta_\ub{}^\ua & 0 & 0 \\
-h_b{}^\ua & \delta_b{}^a & 0 \\
p^{\ub \ua} - \tfrac{1}{2} h^{c \ub} h_c{}^\ua & h^{a \ub} & \delta^\ub{}_\ua
\end{pmatrix} \times
\begin{pmatrix}
K_\ua{}^\um & 0 & 0 \\
0 & \cV_a{}^m & 0 \\
0 & 0 & K_\um{}^\ua
\end{pmatrix}~,
\end{align}
where we give both the megavielbein and its inverse for convenience.
The element $\cV_m{}^a$ will be the usual doubled vielbein and $h_a{}^\ub$ will be
the $\ext\cH$-connection. The field $p^{\ua \ub}$ is an antisymmetric tensor,
which we call the Pol\'a\v{c}ek-Siegel (PS) field; it transforms non-linearly under 
the $\ext\cH$ gauge group and will allow us to build $\ext\cH$-curvatures.

The element $K_\um{}^\ua$ (with inverse $K_\ua{}^\um$)
will turn out to not play a physical role and will drop
out of all relevant formulae. In \cite{Polacek:2013nla}, this field (along with a piece
of $\cV_m{}^a$) dresses $\pa_M$ to become their background $D_M$ derivatives. Here we
have found it simpler to exhibit it explicitly and just ensure that it drops out.
We should emphasize that it is perfectly possible to impose the gauge where $K_\um{}^\ua$
is set to some fixed \emph{coordinate-dependent} value. This would involve imposing
$\nabla_\ub \xi^\ua = -\xi^\uc f_{\uc \ub}{}^\ua$. However, this is not strictly
necessary, so we will avoid doing that here.

In the triangular gauge, the various derivatives decompose as follows:
\begin{subequations}
\begin{align}
\nabla_\ua &= K_\ua{}^\um \pa_\um~, \\
\nabla_a &= \cV_a{}^m \pa_m - h_a{}^\ub \nabla_\ub~, \\
\nabla^\ua &= (p^{\ua \ub} - \tfrac{1}{2} h^{m \ua} h_m{}^{\ub} ) \nabla_\ub 
    + h^{b \ua} \cV_b{}^m \pa_m + K_\um{}^\ua \pa^\um~.
\end{align}
\end{subequations}
In the expression for $\nabla^\ua$, the last term can be dropped because
we will always assume that no field depends on $\tilde y_\um$. Each of
these operators will turn out to be covariant. This is more or less obvious
for the first two, assuming that they will behave as the $\ext\cH$ generator
and the $\ext\cH$-covariant derivative, respectively. This is less obvious
for the third operator, but this will turn out to also be true by virtue
of the section condition.

Now let us show how the megavielbein \eqref{E:megaV} solves the torsion constraints.
It is easy to see that the conditions
$\cT_{\ul{cba}} = \cT_{\ul{cb} a} = 0$
hold automatically in the triangular gauge. The torsions that are fixed
to constants turn out to constrain the $y^\um$ dependence of the various fields:
\begin{subequations}\label{E:PS.TC}
\begin{align}
\label{E:PS.TC1}
\cT_{\uc \ub}{}^\ua &= f_{\uc \ub}{}^\ua \quad \implies \quad
    2 \nabla_{[\uc} K_{\ub]}{}^\um K_\um{}^\ua = -f_{\uc \ub}{}^\ua~, \\
\label{E:PS.TC2}
\cT_{\uc b a} &= f_{\uc b a} \quad \implies \quad
\nabla_\uc \cV_b{}^m = -f_{\uc b}{}^a \cV_a{}^m~, \\
\label{E:PS.TC3}
\cT_{c \ub}{}^\ua &= f_{c\ub}{}^\ua \quad \implies \quad
\nabla_\ub h_m{}^\ua = 
    -h_m{}^\uc f_{\uc \ub}{}^\ua - \cV_m{}^c f_{c \ub}{}^\ua
    + K_\ub{}^\un \pa_m K_\un{}^\ua ~, \\
\label{E:PS.TC4}
\cT_\uc{}^{\ub \ua} &= f_\uc{}^{\ub \ua} \quad \implies\quad
\nabla_\uc p^{\ub \ua} = -f_\uc{}^{\ub \ua}
    - h^{d [\ub} f_{d \uc}{}^{\ua]}
    + 2 p^{\ud [\ub} f_{\ud \uc}{}^{\ua]}
    - h^{m [\ub} K_\uc{}^\un \pa_m K_\un{}^{\ua]}~.
\end{align}
\end{subequations}
The remaining torsion tensors are field-dependent and lead to curvatures.
The standard $\ext\cH$-covariant torsion tensor of DFT is
\begin{align}
\cT_{cba} &= -3 \nabla_{[c} \cV_b{}^m \cV_{m a]}~, \qquad
\nabla_c \cV_b{}^m := \cV_c{}^n \pa_n \cV_b{}^m + h_c{}^\uc f_{\uc b}{}^a \cV_a{}^m~,
\end{align}
where we have used \eqref{E:PS.TC2} in the definition of $\nabla_c \cV_b{}^m$.
The $\ext\cH$-curvature is $\cT_{cb}{}^\ua$, which is given by
\begin{align}
\cT_{c b}{}^\ua &= 
    D_c h_b{}^\ua
    - D_b h_c{}^\ua
    + \cF_{cb}{}^d h_d{}^\ua
    - h_b{}^\ub h_c{}^\uc f_{\uc \ub}{}^\ua
    + h_c{}^{\uc} f_{b \uc}{}^\ua
    - h_b{}^{\uc} f_{c \uc}{}^\ua
    \eol & \quad
    + \Big(p^{\ua \ud} 
    - \tfrac{1}{2} h^{d \ua} h_d{}^\ud 
    \Big) f_{\ud cb}~,
\end{align}
where $D_a := \cV_a{}^m \pa_m$ and $\cF_{ab}{}^c = -3 D_{[a} \cV_b{}^m \cV_{m c]}$ is the
generalized flux of the DFT vielbein.
Here we see that the PS field $p^{\ua \ub}$ permits the construction of $\cT_{cb}{}^\ua$
explicitly. An important consistency condition is that there remains a combination
\begin{align}
\cT_{a b}{}^\uc f_{\uc c d} + \cT_{c d}{}^\uc f_{\uc a b}
\end{align}
where the PS field drops out. For the case where $\ext\cH$ is just the doubled Lorentz group,
this is the symmetrized Riemann tensor that one can construct. That the PS field permits
the direct construction of $T_{a b}{}^\uc$ was the key observation of \cite{Polacek:2013nla}.

Finally, we give the remaining invariant tensors. The tensor $\cT_c{}^{\ub \ua}$ is the
covariantized gradient of the PS field,
\begin{align}
\cT_c{}^{\ub \ua}
    &= \bigg[- D_c p^{\ub \ua}
    - 2 \,h_c{}^{\uc} p^{\ub \ud} f_{\ud \uc}{}^\ua
    + 2 \,p^{\ub \ud} f_{c \ud}{}^\ua
    \eol & \quad
    + h^{d \ub} D_c h_d{}^\ua
    - 2 \,h^{d \ub} D_d h_c{}^\ua
    + \cF_c{}^{ba} h_b{}^\ub h_a{}^\ua
    \eol & \quad
    - h_c{}^\uc f_\uc{}^{\ub \ua}
    - h^{d \,\ub} h_d{}^\ud \Big(
        f_{c \ud}{}^\ua+ h_c{}^{\uc} f_{\uc\ud}{}^\ua
    \Big) \bigg]_{[\ub \ua]}
\end{align}
where $\ub \ua$ are antisymmetrized. Similarly, $\cT^{\uc \ub \ua}$ is the
antisymmetric $\nabla^\uc$ derivative of $p^{\ub \ua}$. Expanding out all terms we find
\begin{align}
\cT^{\ul{cba}}
    &= 3 \times \bigg[- h^{c \uc} D_c p^{\ub \ua}
%     - h^{m \uc} h^{n \ub} \pa_n h_{m}{}^{\ua}
    + h^{c \uc} h^{b \ub} D_{c} h_{b}{}^{\ua}
    + \frac{1}{3} h^{c \uc} h^{b \ub} h^{a \ua} \cF_{c b a} 
    + p^{\uc \ud} f_\ud{}^{\ub \ua}    
    + p^{\uc \ud} p^{\ub \ue} f_{\ue \ud}{}^\ua
    \eol & \qquad\qquad
    - h^{d \uc} h_d{}^{\ud} \Big(
        \frac{1}{2} f_\ud{}^{\ub \ua}
        + p^{\ub \ue} f_{\ue \ud}{}^\ua
        + \frac{1}{4} h^{e \ub} h_e{}^{\ue} f_{\ud \ue}{}^\ua
    \Big)
    \bigg]_{[\uc \ub \ua]}~.
\end{align}
This condition actually follows from the simpler result that $\cT^{\ul{pnm}} = 0$
from the section condition.

We have not yet specified how the various components of the megavielbein transform.
For that, we return to the conventional form of doubled diffeomorphisms \eqref{E:PS.Diff}
on the megavielbein. First, we check that the zero elements remain zero,
\begin{align}
0 &= \delta \cV_\um{}^a = (\pa_\um \xi^n - \pa^n \xi_\um) \cV_n{}^a
    + (\pa_\um \xi_\un - \pa_\un \xi_\um) \cV^{\un a}~, \eol
0 &= \delta \cV_\um{}_\ua = (\pa_\um \xi_\un - \pa_\un \xi_\um) \cV^{\un}{}_{\ua}~.
\end{align}
These are consistent if we choose $\xi_\um=0$ and $\pa_\um \xi^n = 0$, which
follow from \eqref{E:PS.CC} and \eqref{E:PS.TC2}.
The DFT vielbein transforms, using \eqref{E:PS.TC2}, as
\begin{align}
\delta \cV_m{}^a = \Lie_\xi \cV_m{}^a + \cV_m{}^b \Lambda^\uc f_{\uc b}{}^a~, \qquad
\Lambda^\ua := \xi^\um K_\um{}^\ua~.
\end{align}
The $\ext\cH$ connection transforms, using \eqref{E:PS.TC3}, as
\begin{align}
\delta h_m{}^\ua
    &= \Lie_\xi h_m{}^\ua
    + \pa_m \Lambda^\ua
    + h_m{}^\ub \Lambda^\uc f_{\uc \ub}{}^\ua
    + \cV_m{}^b \Lambda^\uc f_{\uc b}{}^\ua~.
\end{align}
Finally, the PS field transforms as
\begin{align}
\delta p^{\ua \ub}
    = \xi^m \pa_m p^{\ua \ub}
    - \Lambda^\uc f_{\uc}{}^{\ua \ub}
    - 2 \,\Lambda^\uc p^{\ud [\ua} f_{\uc \ud}{}^{\ub]}
    - h^{m [\ua} \pa_m \Lambda^{\ub]}
    - \Lambda^\uc h^{c [\ua} f_{c \uc}{}^{\ub]}~.
\end{align}

In writing the transformations this way, we have eliminated all of the $y$ derivatives,
so the $y$-dependence can be thought of as merely a trick to arrive at these transformation
rules. One could in principle eliminate the $y$ dependence explicitly by
introducing $y$-dependent twist matrices for the various connections in order to
solve the conditions \eqref{E:PS.TC} directly. However, it is simpler instead merely to
verify that the algebra of $\ext\cH$ transformations closes on all of the fields
above.

There are a few more issues we have not yet explicitly addressed. First, we mention
how the torsions and curvatures each transform. From the extended Bianchi identity,
one can show that
\begin{align}
\delta_\Lambda \cT_{bcd} := \Lambda^\ua \nabla_\ua \cT_{bcd}
    = - 3 \Lambda^\ua \Big(f_{\ua [b|}{}^e \cT_{e| cd]} + f_{\ua [b|}{}^\ue f_{\ue |c d]}\Big)~.
\end{align}
This follows as well from the explicit definition of the torsion tensor.
If we assume that there is a background constant value for the torsion,
$\mathring \cT_{abc} = f_{abc}$, for which \eqref{E:PS.algebra} corresponds to a closed Lie algebra,
then the above transformation can be written
\begin{align}
\delta_\Lambda \cT_{bcd}
    = - 3 \Lambda^\ua f_{\ua [b|}{}^e \Big(\cT_{e| cd]} - \mathring \cT_{e|cd]}\Big)
    \equiv - 3 \Lambda^\ua f_{\ua [b|}{}^e \Delta \cT_{e| cd]}
\end{align}
with $\Delta \cT$ denoting the deviation from the background.

The Bianchi identity for the torsion tensor becomes
\begin{align}
0 &= 4 \,\nabla_{[a} \cT_{b c d]}
    + 3 \,\cT_{[a b}{}^e \cT_{c d] e}
    + 6 \,\cR_{[a b}{}^\ue f_{\ue c d]}
\end{align}
where we have written the $\ext\cH$-curvature as
$\cR_{a b}{}^\uc := \cT_{ab}{}^\uc$ to improve readability.
Observe that the PS field $p^{\ua \ub}$ drops out of the combination of curvatures
appearing above. In addition, due to the projection with $f_{\ua cd}$,
only some of the $\ext\cH$-curvatures play a role in the Bianchi identity. Naturally
these are those with nonzero $f_{\ua cd}$ as only these $\ext\cH$ connections appear
in the torsion tensor.

The $\ext\cH$-curvature itself transforms as
\begin{align}
\delta_\Lambda \cR_{b c}{}^\ud
    &= - \Lambda^\ua \Big(2 \cT_{\ua [b}{}^E \cT_{c]}{}^\ud{}_E
    + \cT_{b c}{}^E \cT_{\ua}{}^\ud{}_E \Big) \eol
    &= 
    - \Lambda^\ua \Big(
    f_{\ua b}{}^e \Delta \cR_{e c}{}^\ud 
    + f_{\ua c}{}^e \Delta \cR_{b e}{}^\ud 
    + \Delta \cR_{b c}{}^\ue f_{\ue \ua}{}^\ud
    + \Delta \cT_{b c}{}^e f_{e \ua}{}^\ud\Big)~,
\end{align}
where $\Delta \cR$ denotes the deviation of $\cR$ from its
background value. (We allow for this possibility, but will only
be concerned with cases where the background $\mathring \cR$ vanishes.)
Its Bianchi identity is a bit more complicated:
\begin{align}
3 \nabla_a \cR_{b c}{}^\ud
    - \nabla^\ud \cT_{abc}
    + 3 \cT_{ab}{}^e \cR_{e c}{}^\ud
    + 3 \cR_{ab}{}^\ue f_{\ue c}{}^\ud
    + 3 \cR_{c}{}^{\ud \ue} f_{\ue a b} \Big\vert_{[abc]}= 0~.
\end{align}
This is quite an awkward identity as $\nabla^\ua$ involves a naked connection.

The additional gauge transformations are
\begin{align}
\delta_\Lambda \cR_b{}^{\uc \ud}
    &= \Lambda^\ua \Big(- f_{\ua b}{}^e \Delta \cR_{e}{}^{\uc \ud}
    + 2 \Delta \cR_b{}^{\ue \ud} f_{\ue \ua}{}^\uc
    + 2 f_{\ua}{}^{\uc e} \Delta \cR_{e b}{}^\ud \Big)\Big\vert_{[\ul{cd}]}~, \\
\delta_\Lambda \cR^{\ul{bcd}}
    &= -3 \Lambda^\ua \Big( f_{e \ua}{}^\ub \cR^{e \uc \ud}
    + f_\ua{}^{\ub \ue} f_\ue{}^{\uc \ud}
    + \cR^{\uc \ud \ue} f_{\ue \ua}{}^\ub \Big)\Big\vert_{[\ul{bcd}]} ~.
\end{align}
and the remaining Bianchi identities are
\begin{align}
0 &= 2 \nabla_a \cR_b{}^{\uc \ud}
    + 2 \nabla^\uc \cR_{a b}{}^\ud
    + T_{ab}{}^e \cR_e{}^{\uc \ud}
    + \cR_{ab}{}^\ue f_\ue{}^{\uc \ud}
    + \cR^{\uc \ud \ue} f_{\ue a b}
    \eol & \quad
    - 2 \,\cR^e{}_a{}^\uc \cR_{e b}{}^{\ud}
    - 2 \,\cR_a{}^{\uc \ue} f_{\ue b}{}^\ud
    - 2 \,\cR_b{}^{\ud \ue} f_{\ue a}{}^\uc \Big\vert_{[\ul{cd}]}~, \\[2ex]
0 &= \nabla_a \cR^{\ul{bcd}} - 3 \nabla^\ub \cR_a{}^{\ul{cd}}
    + 3 \cR^e{}_a{}^\ub \cR_e{}^{\ul{cd}}
    + 3 \cR_a{}^{\ub \ue} f_\ue{}^{\ul{cd}}
    + 3 f_{\ue a}{}^{\ub} \cR^{\ue \uc \ud} \Big\vert_{[\ul{bcd}]}~.
\end{align}

There is actually one more feature of DFT that we have not discussed: the dilaton.
In the enlarged Pol\'a\v{c}ek-Siegel theory, the dilaton is a scalar $\widehat \Phi$
that transforms as
\begin{align}
\delta \widehat \Phi  = \pa_M (\xi^M \widehat \Phi)
    = \pa_m (\xi^m \widehat \Phi) + \pa_\um (\xi^\um \widehat \Phi)~.
\end{align}
Its curvature $\widehat \cT_A$ and Bianchi identity are given by
\begin{align}
\widehat \cT_A = \nabla_A \log \widehat\Phi + \nabla_M \cV_A{}^M~, \qquad
2 \nabla_{[A} \widehat \cT_{B]} = -\cT_{AB}{}^C \widehat \cT_C - \nabla^C \cT_{CAB}~.
\end{align}
The actual dilaton we want differs from this dilaton as
\begin{align}
\log \Phi = \log \widehat \Phi - \log \det K_\um{}^\ua~.
\end{align}
It follows that
\begin{subequations}
\begin{align}
\widehat \cT_a &= \cT_a + f_{a \ub}{}^\ub~, \\[2ex]
\widehat \cT_\ua &= \nabla_\ua \log\Phi + f_{\ua \ub}{}^\ub~, \\[2ex]
\widehat \cT^\ua &= 
    D^b h_b{}^\ua + \cF^b h_b{}^\ua
    + h^{d \ub} f_{d \ub}{}^\ua
    + p^{\ub \uc} f_{\uc \ub}{}^{\ua}
%     \eol & \quad
    + f_\ub{}^{\ub \ua}
    + (p^{\ua \ub} - \frac{1}{2} h^{d \ua} h_d{}^\ub) \, \nabla_\ub \log\Phi
\end{align}
\end{subequations}
where we have defined
\begin{align}
\cF_a := D_a \log\Phi + \pa_m \cV_a{}^m~, \qquad
\cT_a := \nabla_a \log\Phi + \nabla_m \cV_a{}^m~.
\end{align}
Evidently, $\widehat \cT_a$ is merely a constant shift from $\cT_a$.
Assuming the dilaton transforms homogeneously, so that $\nabla_\ua \log\Phi$ is a
constant (or zero), we find that $\widehat \cT_\ua$ is also a constant.
Finally, the last curvature $\widehat \cT^\ua$ is partly related to the
tensor we previously called $\widehat \cR_{\ha \hb}$ in \eqref{E:BDFT.hatR}
for the case where the $\ext\cH$ group involved just the doubled Lorentz group.

The Bianchi identity now decomposes to
\begin{align}
2 \nabla_{[a} \cT_{b]} 
    &= - \cT_{a b}{}^c \cT_c
    - \nabla^c \cT_{c a b}
    - \cR_{a b}{}^\uc \nabla_\uc \log \Phi
    - \cR_{a}{}^{d\uc} f_{\uc d b}
    + \cR_{b}{}^{d\uc} f_{\uc d a}
    - (\widehat \cT^\uc - f_\ud{}^{\ud \uc}) f_{\uc a b}~.
\end{align}
Presuming that $\nabla_\ua \log \Phi$ is a constant, it follows that $\cT_a$ 
transforms under $\ext\cH$ transformations as
\begin{align}
\delta_\Lambda \cT_b &= - \Lambda^\ua \Big(f_{\ua b}{}^{c} \widehat \cT_c + f_{\ua b}{}^{\uc} \widehat \cT_\uc\Big)
    = - \Lambda^\ua \Big(
    f_{\ua b}{}^{c} \cT_c 
    + f_{\ua b}{}^{\uc} \nabla_\uc \log\Phi
    + f_{\ua b}{}^C f_{C \ud}{}^{\ud} \Big)~.
\end{align}

\section{Construction of the extended super-Maxwell$_\infty$ algebra}
\label{A:SMaxwell}

In this appendix, we detail the construction of a superalgebra which appears
to coincide with the $\ext\cH_L$ tangent space algebra required for $\cN=1$ DFT,
introduced in section \ref{S:SGeo.Detour}. As mentioned in the main body,
it arises as the dual of what we call the super-Maxwell$_\infty$ algebra.
We first discuss what precisely we mean by the super-Maxwell$_\infty$ algebra,
and then review a construction, based on the notion of a local Lie 
superalgebra \cite{Kac:1977em}, that leads to both the full superalgebra
involving both it and its dual, that we need to construct the connections
and curvatures of $\cN=1$ DFT.

\subsection{The super-Maxwell$_\infty$ algebra}

Here we detail the construction of the free differential 
algebra, up through dimension 4, that extends the 10D $\cN=1$ super-Poincar\'e algebra
\begin{align}\label{E:FDA.QQP}
\{Q_\alpha, Q_\beta\} = (\gamma^a)_{\alpha \beta} P_a~.
\end{align}
$Q_\alpha$ and $P_a$ are assigned their usual engineering dimensions 
of $+\tfrac{1}{2}$ and $+1$. Relative to the main body of the text, we have
eliminated the constant $-\kappa$ to simplify formulae, but will restore it later.

Let us now construct the free Lie algebra that extends the super-Poincar\'e algebra.
Following \cite{Gomis:2017cmt,Gomis:2018xmo}, we denote this the super-Maxwell$_\infty$ algebra.
Denote the generators encountered in this algebra
by $Y$ with various subscripts and superscripts. We use a shorthand notation 
where, for example, $Y_{A,B}$ denotes $[P_A, P_B]$ and $Y_{A,B,C}$ denotes
$[P_A, [P_B, P_C]]$ where $P_A = (Q_\alpha, P^a)$.
In this way, \eqref{E:FDA.QQP} could be denoted
$Y_{\alpha,\beta} = (\gamma^a)_{\alpha \beta} Y_a$.

Let's begin by analyzing the generator at dimension $\tfrac{3}{2}$. The Bianchi
identity for three $Q$'s is
\begin{align}
Y_{(\alpha,\beta,\gamma)} = (\gamma^b)_{(\beta \gamma} Y_{\alpha), b} = 0~.
\end{align}
This is solved by taking
\begin{align}
[Q_\alpha, P_b] \equiv Y_{\alpha,b} = (\gamma_b)_{\alpha \beta} \,\tQ^\beta
\end{align}
for some fermionic operator $\tQ^{\alpha}$ at dimension $\tfrac{3}{2}$.
In keeping with our DFT motivation, we will consider $\tQ$ together with $Q$ and $P$ 
as our super $P_A$ operators.

Here we pause and make an important observation. In analyzing the algebra, it is 
extremely helpful to remember that the algebra \eqref{E:FDA.QQP} is obeyed by the 
covariant superspace derivatives that describe the super-Yang-Mills algebra in superspace.
In that case, the operator $\tQ^\alpha$ encountered above is just $\bm\lambda^\alpha$,
the gaugino superfield. We will find that other generators we encounter at higher
dimension admit a similar simple interpretation.

At dimension 2, a new generator can be introduced as
\begin{align}
Y_{ab} := Y_{a,b} \equiv [P_a, P_b]~.
\end{align}
In the super-Yang-Mills case, this is just the non-abelian field strength $\bm F_{ab}$.
The Jacobi identity at this dimension reads
\begin{align}
0 &= \{Q_\alpha, [Q_\beta, P_b]\} + \{Q_\beta, [Q_\alpha, P_b]\} + [P_b, \{Q_\alpha, Q_\beta\}] \eol
    &= Y_{\alpha, \beta, b} + Y_{\beta, \alpha, b} + Y_{b, \alpha, \beta} \eol
    &= (\gamma_b)_{\beta \gamma} Y_{\alpha,}{}^\gamma
    + (\gamma_b)_{\alpha \gamma} Y_{\beta,}{}^\gamma
    + (\gamma^c)_{\alpha \beta} Y_{b c}
\end{align}
This is solved by taking
\begin{align}
Y_{\alpha,}{}^\beta = \tfrac{1}{4} (\gamma^{ab})_\alpha{}^\beta Y_{ab}~.
\end{align}
In the case of the super-Yang-Mills algebra, this is the condition that
the spinor derivative of the gaugino superfield is the field strength, i.e.
$\cD_\alpha \bm\lambda^\beta \propto (\gamma^{ab})_\alpha{}^\beta\, \bm F_{ab}$.

At dimension $\tfrac{5}{2}$, a new generator may be denoted
\begin{align}
 Y_b{}^\alpha := Y_{b,}{}^\alpha \equiv [P_b, \tilde Q^\alpha]~.
\end{align}
This will turn out to be the only generator at this dimension.
The commutator of $Q_\alpha$ with $Y_{bc}$ just reproduces this generator:
\begin{align}
[Q_\alpha, Y_{bc}] \equiv Y_{\alpha, bc}
    = (\gamma_b)_{\alpha \gamma} Y^\gamma{}_c + (\gamma_c)_{\alpha \gamma} Y_b{}^\gamma
    = -2 (\gamma_{[b})_{\alpha \gamma} Y_{c]}{}^\gamma~.
\end{align}
Actually, this generator turns out to carry a constraint: its spin-1/2 part,
corresponding to the $\gamma$ trace, vanishes. This follows from the Jacobi identity:
\begin{align}
0 &= [Q_\alpha, \{Q_\beta, Q^{\gamma}\}]
    + [Q_\beta, \{Q_\alpha, Q^{\gamma}\}]
    + [Q^\gamma, \{Q_\alpha, Q_\beta\}] \eol
    &= 2 Y_{(\alpha, \beta),}{}^\gamma + Y^{\gamma}{}_{, \alpha,\beta}
    = \tfrac{1}{2} (\gamma^{ab})_{(\alpha}{}^\gamma Y_{\beta), ab}
    - (\gamma^c)_{\alpha \beta} Y_c{}^{\gamma} \eol
    &= -\tfrac{1}{2} (\gamma_a)_{\alpha \beta} \,Y_b{}^\delta \,(\gamma^b \gamma^a)_\delta{}^\gamma 
    + \delta_\alpha{}^\gamma Y_b{}^\delta\, (\gamma_b)_{\beta \delta}~.
\end{align}
This vanishes only if $Y_b{}^\beta$ is $\gamma$-traceless. In the super-Yang-Mills algebra, this
condition can be interpreted as the gaugino field equation.

At dimension 3, we encounter
\begin{align}
Y_{a,bc} \equiv [P_a, Y_{bc}] = [P_a, [P_b, P_c]]~.
\end{align}
The antisymmetric part of this vanishes, as a consequence of the Jacobi identity.
In the super-Yang-Mills algebra, this is $\cD_a \bm F_{bc}$, and the vanishing
of its totally antisymmetric part is the Bianchi identity.
Other possible generators we encounter are
\begin{align}
Y_{\alpha, b}{}^\beta \equiv \{Q_\alpha, Y_b{}^\beta\}~, \qquad
Y^{\alpha \beta} := \{Q^{\alpha}, Q^{\beta}\}~.
\end{align}
From the Jacobi identity involving $Q$, $P$, and $\tilde Q$,
it is easy to show that the first of these is given in
terms of the other two:
\begin{align}
Y_{\alpha, b}{}^\beta
    = (\gamma_b)_{\alpha \gamma} Y^{\gamma \beta} 
        + \frac{1}{4} (\gamma^{cd})_{\alpha}{}^\beta Y_{b,cd}
\end{align}
Actually, because the left-hand side is $\gamma$-traceless in its last two indices, a constraint
must be satisfied:
\begin{align}\label{E:FDA.dim3C}
Y^b{}_{,ba} = - (\gamma_a)_{\alpha \beta} Y^{\alpha \beta}~.
\end{align}
In the super Yang-Mills algebra, this is the equation of motion for the gauge connection,
setting the divergence of the field strength tensor equal to the gaugino bilinear
$\{\bm\lambda^\alpha, \bm\lambda^\beta\}$.
The Jacobi identity involving two $Q$'s and $Y_{ab}$ tells us no new information.
For later use, it will be helpful to decompose $Y_{a,bc}$ into its irreducible (traceless)
hook representation $Y_{a|bc}$ and the trace:
\begin{align}\label{E:FDA.dim3.Ydecompose}
Y_{a,bc} &= Y_{a|bc} + \tfrac{1}{9} \eta_{a b} Y^d{}_{,dc} - \tfrac{1}{9} \eta_{a c} Y^d{}_{,db} \eol
    &= Y_{a|bc} - \tfrac{2}{9} \eta_{a [b} (\gamma_{c]})_{\alpha \beta} Y^{\alpha \beta}~.
\end{align}
At this level, $Y_{a|bc}$ and $Y^{\alpha\beta}$ may be considered the independent operators.

At dimension $\tfrac{7}{2}$, we identify
\begin{align}
Y_{a,b}{}^\beta \equiv [P_a, Y_b{}^\beta]~.
\end{align}
By construction it is $\gamma$-traceless on $b\beta$.
This turns out to be the only operator present at this dimension as
other commutators just lead ultimately to this one:
\begin{subequations}
\begin{align}
Y^{\alpha}{}_{,ab} \equiv [\tilde Q^\alpha, Y_{ab}]
    &= - 2 \,Y_{[a,b]}{}^\alpha~, \\
Y_{\alpha,}{}^{\beta \gamma} \equiv [Q_\alpha, Y^{\beta \gamma}]
    &= (\gamma^{a b})_\alpha{}^{(\beta} Y_{a,b}{}^{\gamma)}~, \\
Y_{\alpha,b,cd} \equiv [Q_\alpha, Y_{b,cd}]
    &= -2 (\gamma_b)_{\alpha \gamma} Y_{[c,d]}{}^\gamma
    - (\gamma_{c})_{\alpha \gamma} Y_{b,d}{}^\gamma
    + (\gamma_{d})_{\alpha \gamma} Y_{b,c}{}^\gamma~.
\end{align}
\end{subequations}
As a check, one can verify that the last two equations are consistent with
\eqref{E:FDA.dim3C}, with no assumption on $Y_{a,b}{}^\beta$ aside from 
it being $\gamma$-traceless in its last two indices.
In the super Yang-Mills case, this object is two covariant derivatives of
the gaugino superfield $\cD_a \cD_b \bm\lambda^\beta$.
One might expect that instead we should deal with
the traceless and $\gamma$-traceless projection of $\cD_{(a} \cD_{b)} \bm\lambda^\beta$
and separately the operator $[\bm F_{ab}, \bm \lambda^\beta]$. However, these
two representations precisely match the former, i.e.
$\drep{10000} \times \drep{10010} = \drep{20010} + \drep{01000} \times \drep{00010}$,
so we will remain with the former.

At dimension 4, we identify the following two operators:
\begin{align}
Y^\alpha{}_{,b}{}^\beta &\equiv [\tilde Q^\alpha, Y_b{}^\beta]~, \qquad
Y_{a,b,cd} \equiv [P_a, Y_{b,cd}]~.
\end{align}
In the super Yang-Mills algebra, these are $\{\bm\lambda^\alpha, \cD_b \bm\lambda^\beta\}$
and $\cD_a \cD_b \bm F_{cd}$, respectively.
Other potential operators are related to these. For example,
\begin{subequations}
\begin{align}
Y_{a,}{}^{\beta \gamma} \equiv [P_a, Y^{\beta \gamma}]
    = Y^\beta{}_{,a}{}^\gamma + Y^\gamma{}_{,a}{}^\beta~, \\
Y_{ab,cd} \equiv [Y_{ab}, Y_{cd}]
    = Y_{a,b,cd} - Y_{b,a,cd}~.
\end{align}
\end{subequations}
The operator generated by $Q_\alpha$ on $Y_{a,b}{}^\beta$ is a bit
more involved:
\begin{align}\label{E:QYab.beta.v1}
\{Q_\alpha, Y_{a, b}{}^\beta\}
    = (\gamma_a)_{\alpha \gamma} Y^\gamma{}_{,b}{}^\beta
    + (\gamma_b)_{\alpha \gamma} Y^\gamma{}_{,a}{}^\beta
    + (\gamma_b)_{\alpha \gamma} Y^\beta{}_{,a}{}^\gamma
    + \tfrac{1}{4} (\gamma^{cd})_\alpha{}^\beta Y_{a,b,cd}~.
\end{align}
Checking the $\gamma$-traceless condition implies that
\begin{align}
Y_{a,}{}^b{}_{,bc} = -2 (\gamma_c)_{\alpha \beta} Y^\alpha{}_{,a}{}^\beta~.
\end{align}
This condition is actually just a direct consequence of \eqref{E:FDA.dim3.Ydecompose}, and
amounts to
\begin{align}\label{E:FDA.dim4.Ydecompose}
Y_{a,b,cd} = Y_{a,b|cd} 
    - \tfrac{2}{9} \eta_{b c} (\gamma_d)_{\alpha \beta} Y^\alpha{}_{,a}{}^\beta
    + \tfrac{2}{9} \eta_{b d} (\gamma_c)_{\alpha \beta} Y^\alpha{}_{,a}{}^\beta~.
\end{align}

Analyzing the remaining Jacobi identities at this dimension leads to only one
additional constraint. This is best formulated in terms of the reducible operator
$Y_{a,b,cd}$, where it reads
\begin{align}\label{E:FDA.Ydim4C}
Y_{[a,b],cd} + Y_{[c,d],ab} = 0~.
\end{align}
In terms of (reducible) Young tableaux, this eliminates the
$\tiny\yng(2,2)$ representation, leaving the 
$\tiny\yng(3,1)$ and $\tiny\yng(2,1,1)$ representations.
This identity is obvious for the super Yang-Mills case: $Y_{[a,b],cd}$ is analogous to
$\cD_{[a} \cD_{b]} \bm F_{cd} \propto [\bm F_{ab},\bm F_{cd}]$.
We solve this constraint by writing
\begin{align}
Y_{[a,b],cd} = \frac{1}{2} Y_{ab,cd}~.
\end{align}
There is no constraint on $Y_{ab,cd}$ aside from pairwise antisymmetry,
$Y_{ab,cd} = - Y_{cd,ab}$. Again, this is obvious from considering the super Yang-Mills case,
where it is the commutator of two field strengths.

We would like to identify whatever residual piece of $Y_{a,b|cd}$ is independent
of either $Y_{ab,cd}$ or $Y^\alpha{}_{,b}{}^\beta$.
The $\g{SO}(9,1)$ reps found in $Y_{a,b|cd}$, $Y_{ab,cd}$ and $Y^\alpha{}_{,b}{}^\beta$ are
\begin{align}
Y_{a,b|cd} &: \drep{01000} + \drep{10100} + \drep{20000} + \drep{02000} + \drep{21000}~, \eol
Y_{ab,cd} &: \drep{01000} + \drep{10100}~, \eol
Y^\alpha{}_{,b}{}^\beta &: \underline{\drep{01000}} + \drep{10100} + \underline{\drep{20000}} + \drep{00011} + \drep{10020} ~.
\end{align}
The underlined representations for $Y^\alpha{}_{,b}{}^\beta$
correspond to those that occur in \eqref{E:FDA.dim4.Ydecompose}, 
so these are the ones that in principle could be related to the corresponding
reps in $Y_{a,b|cd}$ by the constraint \eqref{E:FDA.Ydim4C}. Meanwhile,
this constraint itself decomposes as
$\drep{00000} + \drep{00011} + \drep{02000} + \drep{20000}$.
The upshot of this is that of the reps in $Y_{a,b|cd}$, only the
$\drep{21000}$ is an independent generator. This corresponds to
a traceless tensor $Y_{a|b|cd}$, 
which is symmetric in $ab$, antisymmetric in $cd$, and obeying
$Y_{a|\,\small[b|cd\small]} = 0$.
The antisymmetric tensor $\drep{01000}$ and the irreducible hook $\drep{10100}$ correspond
to $Y_{ab,cd}$; the $\drep{20000}$ is identified with that 
piece of $Y^\alpha{}_{,b}{}^\beta$; and the $\drep{02000}$ is constrained to vanish.
The resulting decomposition can be written
\begin{align}
Y_{a,b,cd}
    &= \Bigg[Y_{a|b|cd}
    + \tfrac{1}{2} Y_{ab,cd}
    + \tfrac{1}{2} Y_{a c, b d}
    \eol & \quad
    + \eta_{a c} \Big(
        \tfrac{1}{8} Y_{b}{}^e{}_{,de}
        - \tfrac{1}{30} (\gamma_{b})_{\alpha \beta} Y^\alpha{}_{,d}{}^\beta
        - \tfrac{11}{30} (\gamma_{d})_{\alpha \beta} Y^\alpha{}_{,b}{}^\beta
    \Big)
    \eol & \quad
    + \eta_{b c} \Big(
        \tfrac{1}{8} Y_{a}{}^e{}_{,de}
        - \tfrac{1}{30} (\gamma_{a})_{\alpha \beta} Y^\alpha{}_{,d}{}^\beta
        - \tfrac{11}{30} (\gamma_{d})_{\alpha \beta} Y^\alpha{}_{,a}{}^\beta
    \Big)
    \eol & \quad
    + \eta_{a b} \Big(
        \tfrac{1}{8} Y_{c}{}^e{}_{,de}
        + \tfrac{1}{3} (\gamma_{c})_{\alpha \beta} Y^\alpha{}_{,d}{}^\beta
    \Big)  \Bigg]_{[cd]}
\end{align}
Of course, what we really need is the commutator of $P_a$ with the independent
generator $Y_{b|cd}$, which we denote
\begin{align}
[P_a, Y_{b|cd}] = Y_{a,b|cd}
    &= \Bigg[Y_{a|b|cd}
    + \tfrac{1}{2} Y_{ab,cd}
    + \tfrac{1}{2} Y_{a c, b d}
    \eol & \quad
    + \eta_{a c} \Big(
        \tfrac{1}{8} Y_{b}{}^e{}_{,de}
        - \tfrac{1}{30} (\gamma_{b})_{\alpha \beta} Y^\alpha{}_{,d}{}^\beta
        - \tfrac{11}{30} (\gamma_{d})_{\alpha \beta} Y^\alpha{}_{,b}{}^\beta
    \Big)
    \eol & \quad
    + \eta_{b c} \Big(
        \tfrac{1}{8} Y_{a}{}^e{}_{,de}
        - \tfrac{1}{30} (\gamma_{a})_{\alpha \beta} Y^\alpha{}_{,d}{}^\beta
        + \tfrac{7}{90} (\gamma_{d})_{\alpha \beta} Y^\alpha{}_{,a}{}^\beta
    \Big)
    \eol & \quad
    + \eta_{a b} \Big(
        \tfrac{1}{8} Y_{c}{}^e{}_{,de}
        + \tfrac{1}{3} (\gamma_{c})_{\alpha \beta} Y^\alpha{}_{,d}{}^\beta
    \Big)  \Bigg]_{[cd]}
\end{align}
This can also be written compactly as
\begin{align}
Y_{a,b|cd} &= Y_{a|b|cd} + \Big[
    \tfrac{3}{4} Y_{a b,c d} 
    + \tfrac{3}{8} \eta_{a c} Y_{b}{}^e{}_{,de}
    + \tfrac{3}{10} \eta_{a c} (\gamma_{b})_{\alpha \beta} Y^{\alpha}{}_{,d}{}^\beta
    - \tfrac{7}{10} \eta_{a c} (\gamma_{d})_{\alpha \beta} Y^{\alpha}{}_{,b}{}^\beta
    \Big]_{b|cd}
\end{align}
where the projection onto the irreducible hook is implied by the brackets.

This was rather involved, so let's pause to emphasize that at dimension 4,
there are three independent generators:
\begin{gather}
Y^\alpha{}_{,b}{}^\beta~, \qquad
Y_{a b,cd}~, \qquad
Y_{a|b|cd}~.
\end{gather}
The first is $\gamma$-traceless in its last two indices. The second is pairwise
antisymmetric. Both of these are reducible. The third is irreducible, corresponding to
the $\drep{21000}$ representation. In the super-Yang-Mills case, the $\drep{21000}$
corresponds to the irreducible representation involving two symmetrized covariant derivatives
of the field strength, where the divergences and Bianchi identity terms have been
projected out.

We will not exhaustively analyze the generators at dimension $\tfrac{9}{2}$, except
to say that by comparing with the super-Yang-Mills case, we expect to encounter four.
The first, $Y^{\alpha|\beta\gamma}$ corresponds to the product of three gauginos,
$[\bm\lambda^\alpha, \{\bm\lambda^\beta, \bm\lambda^\gamma\}]$;
it lies in the hook representation because the totally symmetric part vanishes due to
the Jacobi identity for the non-abelian gauge group in which the gauginos are valued.
The second, $Y_{ab,c}{}^\gamma$ corresponds to $[\bm F_{ab}, \cD_c \bm\lambda^\gamma]$.
The third, $Y^\alpha{}_{,b|cd}$ corresponds to $[\bm\lambda^\alpha, \cD_b \bm F_{cd}]$.
The fourth is $Y_{(abc)}{}^\gamma$,
which is $\gamma$-traceless and $\eta$-traceless, corresponds to 
$\cD_{(a} \cD_b \cD_{c)} \bm\lambda^\gamma$ with the traces removed (as they correspond
to the previously mentioned operators).

It is clear from the super-Yang-Mills analogy that this algebra is infinite in extent,
although it can be truncated at any level. This ceases to be true when one includes the
negative dimension generators of $\ext{\cH}$.

Let us reintroduce factors of $\kappa$ to more easily match our conventions in the main
body. This involves replacing
\begin{align}
P_a \rightarrow -\kappa\, P_a~, \qquad
\tilde Q^\alpha \rightarrow \kappa^2 \tilde Q^{\alpha}~.
\end{align}
We keep all defining relations for $Y$'s formally the same. This means we take
\begin{gather*}
Y_{a b} \rightarrow \kappa^2 Y_{ab}~, \qquad
Y_{\alpha,}{}^\beta \rightarrow \kappa^2 Y_{\alpha,}{}^\beta ~, \\
Y_b{}^\alpha \rightarrow -\kappa^3 Y_b{}^\alpha~, \qquad
Y_{\alpha, bc} \rightarrow \kappa^2 Y_{\alpha, bc}~, \\
Y_{a,bc} \rightarrow -\kappa^3 Y_{a,bc}~, \qquad
Y_\alpha{}_{,b}{}^\beta \rightarrow -\kappa^3 Y_\alpha{}_{,b}{}^\beta ~, \qquad
Y^{\alpha \beta} \rightarrow \kappa^4 Y^{\alpha \beta}~, \\
Y_{a,b}{}^\beta \rightarrow \kappa^4 Y_{a,b}{}^\beta~, \qquad
Y^\alpha{}_{,ab} \rightarrow \kappa^4 Y^\alpha{}_{,ab}~, \qquad
Y_{\alpha,}{}^{\beta \gamma} \rightarrow \kappa^4 Y_{\alpha,}{}^{\beta \gamma}~, \qquad
Y_{\alpha,b,cd} \rightarrow -\kappa^3 Y_{\alpha,b,cd}~, \\
Y^\alpha{}_{,b}{}^\beta \rightarrow -\kappa^5 Y^\alpha{}_{,b}{}^\beta~, \qquad
Y_{a,b,cd} \rightarrow \kappa^4 Y_{a,b,cd}~, \qquad
Y_{a,}{}^{\beta \gamma} \rightarrow -\kappa^5 Y^{\beta\gamma}~, \qquad
Y_{ab,cd} \rightarrow \kappa^4 Y_{ab,cd}~.
\end{gather*}
For example, we find at dimension 2,
\begin{align}
Y_{\alpha,}{}^\beta = \frac{1}{4} (\gamma^{ab})_\alpha{}^\beta Y_{ab}~,
\end{align}
but at dimension 2.5,
\begin{align}
Y_{\alpha, bc}
    = 2\kappa\, (\gamma_{[b})_{\alpha \gamma} Y_{c]}{}^\gamma~.
\end{align}

\subsection{Local Lie superalgebra construction}
\label{A:SMaxwell.Local}
Let us now extend the super-Maxwell$_\infty$ algebra discussed above to negative
dimension elements. To do so, we will employ a construction due to Kac \cite{Kac:1977em}
(see also Appendix B of \cite{Bossard:2017wxl} which we follow). Its application to
our present situation is due to Jakob Palmkvist \bibnote{J. Palmkvist, private communication.}. We thank Axel Kleinschmidt for parallel comments on the
bosonic case.

Kac defined a local Lie superalgebra as a direct sum 
$\cT_{-1} \oplus \cT_0 \oplus \cT_{+1}$ of three vector
spaces, equipped with a Lie superbracket
\begin{align}\label{E:KacLocal}
[ \cT_0, \cT_{\pm 1}] = \cT_{\pm 1}~, \qquad
[\cT_{-1}, \cT_{+1}] = \cT_0~, \qquad
[\cT_0, \cT_0] = \cT_0
\end{align}
where here we use $[\cdot, \cdot]$ as a graded commutator.\footnote{Here we will be
a little sloppy and treat vector spaces and their elements interchangeably.
We are also interested in the case where odd elements are fermionic and even 
elements are bosonic, but the construction is general.}
The local Lie superalgebra can be extended to a unique superalgebra in two steps:
\begin{enumerate}
\item Freely generate all positive elements at level $k$ by nested $k$ commutators of $\cT_{+1}$ with itself, 
i.e. $\tilde \cT_{+k} = [\cT_{+1}, [ \cdots [\cT_{+1}, \cT_{+1}]]]$.
Repeat for all negative elements using $\cT_{-1}$. The maximal extension of
the local Lie superalgebra $\tilde \cT = \sum_k \tilde \cT_{+k}$
is given by the set of all such freely generated elements.

\item There is a maximal ideal $D$ among the freely generated elements.
Quotient out by it, defining $\cT = \tilde \cT / D$. In practice, this amounts
to dropping all freely generated positive elements $\tilde \cT_{+k}$ whose
commutator with $\cT_{-1}$ vanishes, and similarly for all negative elements
$\tilde \cT_{-k}$ with vanishing $\cT_{+1}$ commutator. The resulting
algebra $\cT$ is the minimal Lie superalgebra generated by the local Lie
superalgebra \eqref{E:KacLocal}. The resulting superalgebra is necessarily simple.
\end{enumerate}

This construction applies in our present case, where the positive level generators correspond to what we call the super-Maxwell$_\infty$ algebra and the non-positive level generators to $\ext\cH_L$. The level corresponds to twice the dimension, and
the local generators discussed above are
\begin{align}
\cT_{-1} = M_{\alpha a}~, \qquad
\cT_0 = M_{a b}~, \qquad
\cT_{+1} = Q_\alpha~.
\end{align}
The super Maxwell$_\infty$ algebra constructed from $Q_\alpha$ appears to be
exactly the minimal extension of the positive elements $\cT_{+k}$.
At level 2, we keep only $P_a$ in \eqref{E:FDA.QQP}, this being the only component
of $\{Q_\alpha, Q_\beta\}$ not annihilated by $M_{\alpha a}$.
The other generators uncovered 
up through level 8 (dimension 4) each have non-vanishing commutator with $M_{\alpha a}$.
It is not obvious (but seems likely) that all freely generated elements
beyond this point which are consistent with the Jacobi identity must
have non-vanishing $M_{\alpha a}$ commutator.

We require one extra element: the existence of a non-degenerate bilinear form $\eta$.
This can be constructed recursively, following the general proof given in
\cite{Bossard:2017wxl}, although in our case $\eta(\cT_i, \cT_j)$ is non-vanishing
only for $i + j = 2$. It must be non-degenerate because $\cT$ is simple. With this
in mind, we can define the elements of $\cT_{-\Delta}$ as the ``duals'' of the
elements in $\cT_{2+\Delta}$. This fixes their normalization and determines
their commutation relations with $\{Q_\alpha, P_a, \tilde Q^\alpha\}$ by reflecting
the relations from the positive level generators. The Jacobi identity then determines
their commutation relations with higher positive elements. Armed with the commutation
relations $[\cT_{-\ell}, \cT_{\ell'}] = \cT_{\ell' -\ell}$ for $\ell, \ell'>0$, one can
directly compute $[\cT_{-\ell}, \cT_{-\ell''}] = \cT_{-(\ell+\ell'')}$ for $\ell,\ell''>0$
by choosing $\ell' = 2+\ell+\ell''$ and reflecting.
This is invaluable because the positive level generators are significantly simpler to characterize.

\bibliography{library.bib}
\bibliographystyle{utphys_mod_v2}

\end{document}